\documentclass{article}

\usepackage{biblatex} 
\usepackage{authblk}
\usepackage[a4paper, total={7in, 10in}]{geometry}
\usepackage[utf8]{inputenc}
\usepackage{amsmath}
\usepackage{xcolor}
\usepackage{pgfplots}
\usepackage{graphicx}
\usepackage{url}
\usepackage{setspace}   
\usepackage{tikz}
\usetikzlibrary{positioning}
\pgfplotsset{compat=1.15}

\providecommand{\keywords}[1]
{
  \small	
  \textbf{\textit{Keywords---}} #1
}

\title{Reverse Engineering the Fly Brain Using {\it FlyCircuit} Database}

\author[1]{Yu-Tai Ching\thanks{corresponding author, correspondence please send to  ytc@nycu.edu.tw.}}
\author[2]{Chin-Ping Cho}
\author[1]{Fu-Kai Tang}
\author[1]{Yi-Chiun Chang}
\author[3]{Chang-Chieh Cheng}
\author[4]{Guan-Wei He}
\author[5]{Ann-Shyn Chang}
\author[6]{Chaochun Chuang}
\affil[1]{Department of Computer Science, National Yang Ming Chiao Tung University, Hsinchu, Taiwan}
\affil[2]{Google Taiwan Engineering Limited, Taipei, Taiwan}
\affil[3]{Information Technology Service Center, National Yang Ming Chiao Tung University, Hsinchu, Taiwan}
\affil[4]{Phison Electronics Corp., Jhunan, Miaoli, Taiwan}
\affil[5]{Brain Research Center, National Tsing Hua University, Hsinchu, Taiwan}
\affil[6]{National Center for High-performance Computing, Hsinchu, Taiwan}
\date{}

\begin{document}

\maketitle

\begin{abstract}
A method to reverse engineering of a fly brain using the {\it FlyCircuit} database is presented. 
This method was designed based on the assumption that similar neurons could serve identical functions. 
We thus cluster the neurons based on the similarity between neurons. 
The procedures are to partition the neurons in the database into groups, and then assemble the groups into potential modules. 
Some of the modules correspond to known neuropils, including Medulla were obtained. 
The same clustering algorithm was applied to analyze Medulla's structure. 
Another possible application of the clustering result is to study the brain-wide neuron connectome by looking at the connectivity between groups of neurons. 
\end{abstract}

\doublespacing
\section{Introduction}

Reverse Engineering of Integrated Circuits (ICs) has been a technique in the electronic industry for a significant period. 
This process involves capturing images of the internal structure of an IC and deciphering its components and their functionality. 
To understand the functionality, experienced experts are needed to recognize modules in the IC and the relationships between the modules. 
In June 2008, an ambitious project at Janelia Farm introduced the reverse engineering of a fly brain to the Electrical Engineering and Computer Science Society community through IEEE Spectrum\cite{4531462}. 
The project entailed acquiring electron microscopy volume data of a female fly brain and developing computer methods for analyzing volume images\cite{zheng2018complete}.

Reverse engineering a fly brain is much more difficult than reverse engineering an IC. 
The internal structure of an IC consists of wires. 
The intersection of two wires is a transistor. 
A subsystem is usually a block in the IC containing many transistors. 
The wires in an IC are generally rectilinear paths and in the same layer or several layers, considered in 2D or 2.5D space. 
Furthermore, an IC is not designed from scratch, instead many previously built modules (blocks) are used. 
Knowing some modules helps reduce the functionality of the IC. 
Neurons in the fly brain are truly 3D furthermore, there is little information about the ``modules'' in the fly brain. 

In this context, we present an alternative approach to reverse engineering the fly brain. 
We utilized a database containing light microscopic images of 27K neurons provided by the Brain Research Center at National Tsing Hua University in Hsinchu, Taiwan\cite{FlyCircuit}. 
The neurons were traced\cite{Ching12}, their skeletons presented, and subsequently mapped onto a representative brain model\cite{chiang11}. 

While recognizing modules in the fly brain is difficult, we present to neuron-cluster to identify potential modules.  
Assume that the similar neurons could be functionally identical.   
We first put similar neurons together to form groups, and we then assemble the groups to find potentially meaningful modules. 
Groups were identified using a user-assistant method, neurons were classified into 154 groups. 
After assembling, groups are organized into 17 clusters.
Notably, some of these clusters corresponded to previously identified structures in the fly brain, including the AL (antennal Lobe), LH (lateral horn), VLP (ventral lateral protocerebrum), MB, and the Visual system. 
Among the visual system, two clusters pertained to Med (medulla) and Lop (lobula plate) respectively, while the remaining two corresponded to Lob (lobula complex).
We thus consider each cluster a possible module. 
Each module can further be partitioned into smaller components to study the internal structure of the module. 
In the manuscript, we study the structure of the Medulla in the visual system of the fly brain. 


A visualization tool was implemented for the fly brain reverse engineering task. 
The tool enables the visualization of a single neuron, a group of neurons, the fly brain's hull, and the neuropils. 
Additionally, it supports the visualization of user-defined groups, described through a JSON file. 
This feature supports visually evaluating computational results, for example, the result of neuron-clustering. 


In the following, we describe the data set used in Section~\ref{data}. The method is in Section~\ref{method}. 
In Section~\ref{method}, we first define the similarity between two neurons, we then present the algorithm to obtain the 17 potential modules. 
Analysis of the structure of the Medulla is then presented in Section~{\ref{MedStruct}}. 
We have the conclusions and discussions in Section~\ref{Conclusion}. 

\section{The Data Set\label{data}} 
\begin{table}
\begin{center}
\caption {Female and male neurons involved in this study. Number of neurons for different genes in the database are listed.  }{\label{dataTable} }
\begin{tabular}{|c|c|c|c| }
\hline
Gene product or source & gene name & Female & Male \\ \hline
choline O-acetyltransferase &{\it Cha} & 3375 & 0 \\  \hline
vesicular glutamate transporter &{\it VGlut} & 6001 & 0 \\  \hline
glutamic acid decarboxylase (1) &{\it Gad} & 7967 & 0 \\  \hline
5-hydroxytryptamine receptor 1A &{\it 5HT1A} & 227 & 191 \\  \hline
5-hydroxytryptamine receptor 1B &{\it 5HT1B} & 41 & 57 \\  \hline
tyrosine hydroxylase &{\it TH} & 515 & 379 \\  \hline
thyrotropin-releasing hormone &{\it Trh} & 995 & 1188 \\  \hline
tyrosine decarboxylase 2 &{\it Tdc2} & 264 & 0 \\  \hline
neuropeptide F &{\it npf} & 159 & 200 \\  \hline
source: Chiang's lab &{\it E0585} & 193 & 0 \\  \hline
source: Chiang's lab  &{\it GH164} & 1 & 6 \\  \hline
source: Chiang's lab  &{\it G0239} & 12 & 0 \\ \hline
fruitless &{\it fru} & {{2755}} & {{3373}} \\  \hline \hline
 Total & & 22505 & 5388\\ \hline
\end{tabular}
\end{center}
\end{table}

Neurons used in this study are from the FlyCircuit database (http://www.flycircuit.tw/v1.1/)\cite{FlyCircuit}. 
27,893 neurons were involved in this study. 
Neuron images were acquired using the confocal microscope. 
Each neuron is labeled by one of the genes among {\it Cha}, {\it VGlut}, {\it Gad}, {\it 5HT1A}, {\it 5HT1B}, {\it TH}, {\it Trh}, {\it Tdc2}, {\it npf}, {\it E0585}, {\it GH164}, {\it G0239}, and {\it fru}. 
Table~\ref{dataTable} lists the number of neurons labeled by different genes in both female and male brains respectively. 

\begin{figure}
\begin{center}
\begin{tabular}{ |c | c | }
\hline
        \includegraphics[width=6cm]{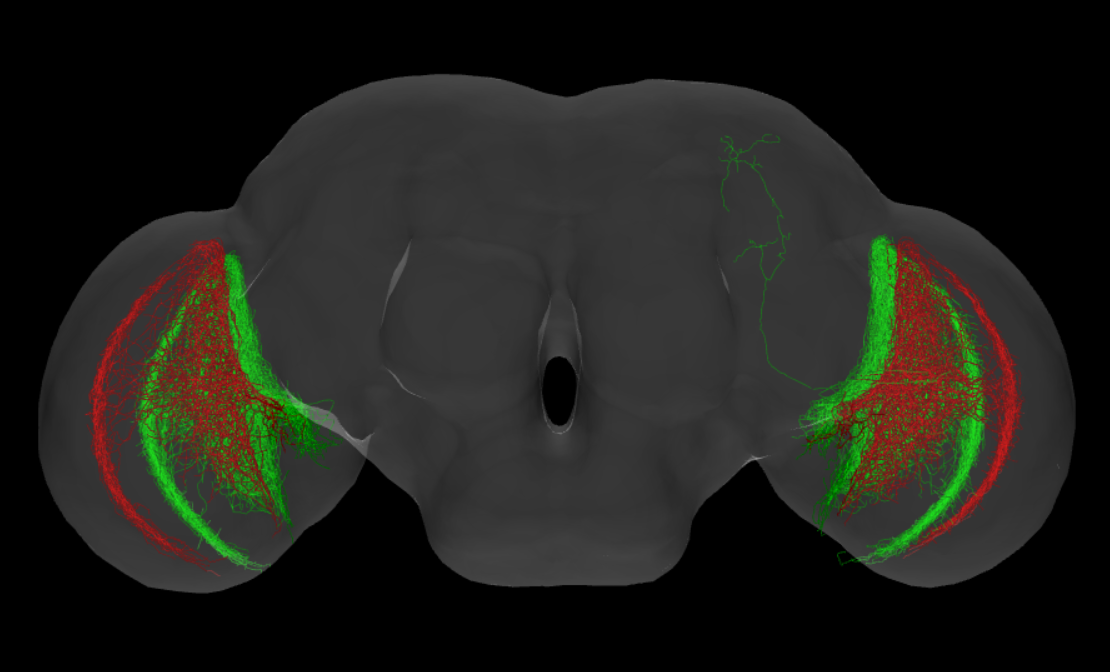} 
    & 
        \includegraphics[width=6cm]{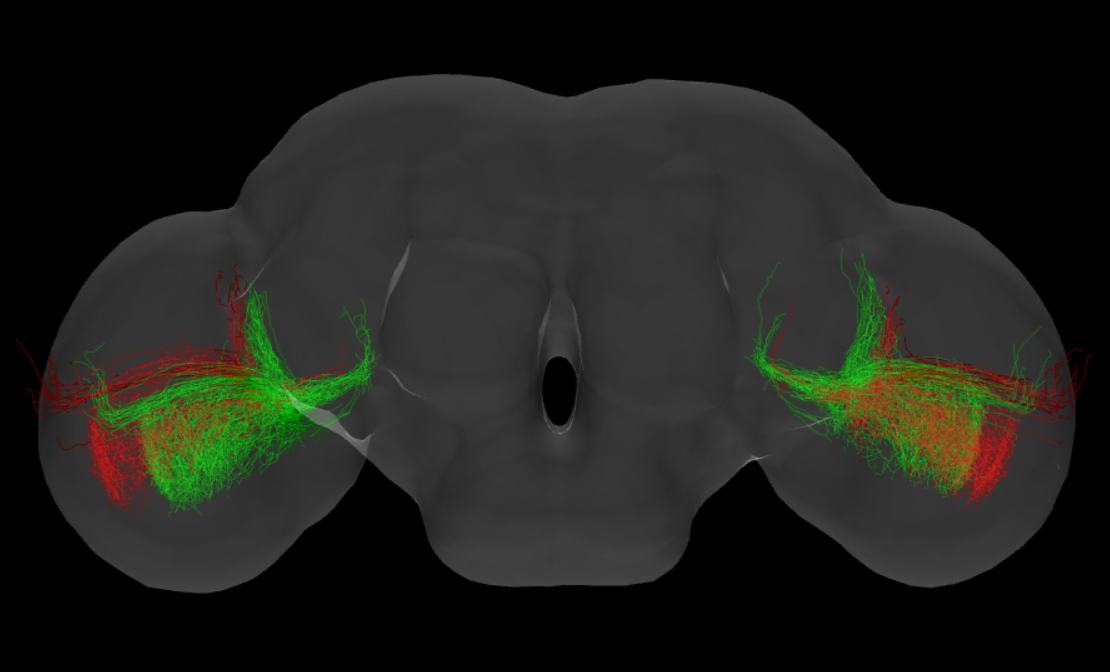} 
    \\ \hline

\end{tabular}
\caption{The same type of neurons in different gender fly brains are not aligned after warping. Red and Green neurons are respectively female and male neurons. 
But each of the female and male neurons are aligned well. }\label{FvsM}
\end{center}
\end{figure}

Neurons were traced and presented as their skeleton. 
All neurons including the male neurons were mapped into a representative female fly brain. 
The same type of neurons of both genders are not aligned in the representative brain, but similar neurons from the same gender are aligned well as shown in Figure~\ref{FvsM}. 
In Figure~\ref{FvsM}, the female and male neurons are respectively red and green. 

22505 female neurons and 5388 male neurons are in the database. 
Assume that these 22505 neurons are randomly sampled from the brain-wide neurons in the female fly brain. 
We use female neurons to study the brain-wide connectome of the fly brain. 
The major 5388 male neurons are the {\it fru} neurons. 
The number of female and male {\it fru} neurons is about the same (respectively 2755 and 3373). 
We can compare the distribution of the {\it fru} neurons in the female and male fly brains.  

The skeleton of a neuron is a rooted tree stored as a list of 3D points $(x_i,y_i,z_i)$ and a list of edges $(i,j)$, $i$ and $j$ are the $i$'th and the $j$'th point in the point list.  
The first point is the center of the soma. 
Since all of the neurons were mapped into a representative brain, all of the neurons are in the same coordinate system enclosed in a $839 \times 457 \times 234$ rectangular box. 
Each voxel is a 0.3-micrometer cube. 
The origin of the coordinate system is the center of the box. 
Because neurons are in the same coordinate system, the Euclidean distance between two points or between a point and a line segment of two neurons can be computed without further alignment. 

\section{Method\label{method}}
\subsection{Similarity between Neurons}

\begin{figure}
    \centering
    \includegraphics[width=15cm]{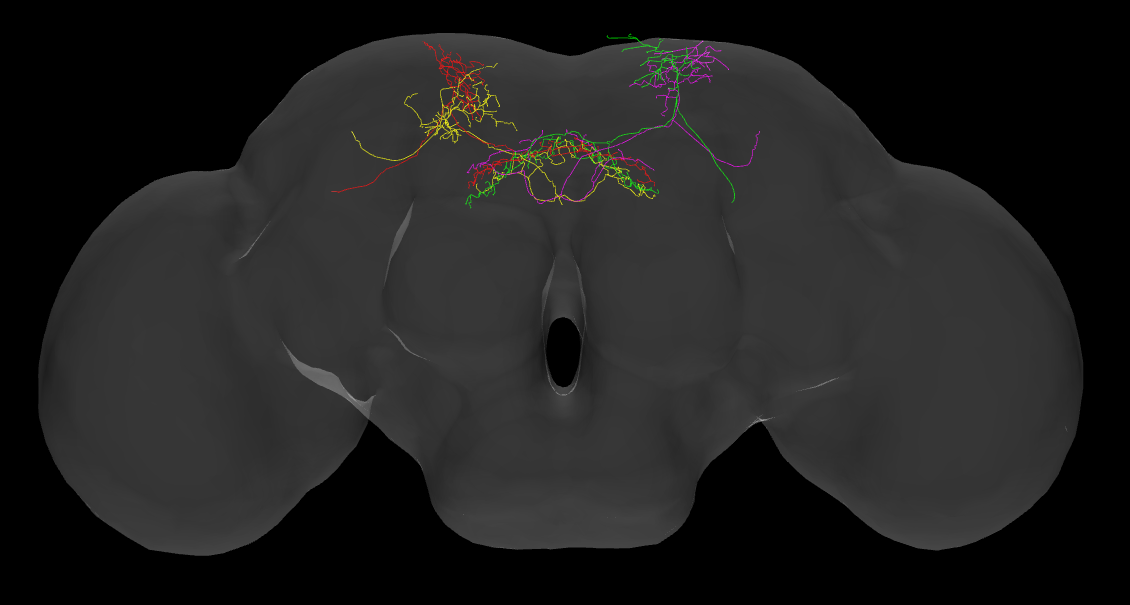}
    \caption{An example shows the similar neurons.  }
    \label{SimilarNeurons}
\end{figure}

\begin{figure}
\begin{center}
\begin{tikzpicture}
\draw[gray, dashed, <->] (4,4) -- (-4,-4);
\draw[gray, thick, <->] (-2,-2) -- (1,1);
\filldraw[black] (0,0) circle (2pt) node[anchor=west]{$p_1'$};
\draw[gray, dashed] (-2,2) -- (0,0);
\filldraw[black] (-2,2) circle (2pt) node[anchor=west]{$p_1$};
\filldraw[black] (1,1) circle (2pt) node[anchor=west]{$q_1$};
\filldraw[black] (-2,-2) circle (2pt) node[anchor=west]{$q_2$};

\draw[gray, dashed] (-5,-1) -- (-3,-3);
\filldraw[black] (-5,-1) circle (2pt) node[anchor=west]{$p_2$};
\filldraw[black] (-3,-3) circle (2pt) node[anchor=west]{$p_2'$};

\end{tikzpicture}
\caption{The distance between a point and a line segment. The distance between $p_1$ and $\overline{q_1,q_2}$ is $\overline{|p_1,p_1'|}$. The distance between $p_2$ and $\overline{q_1,q_2}$ is $\overline{|p_2,q_2|}$} 
\label{DisP2Line}
\end{center}
\end{figure}
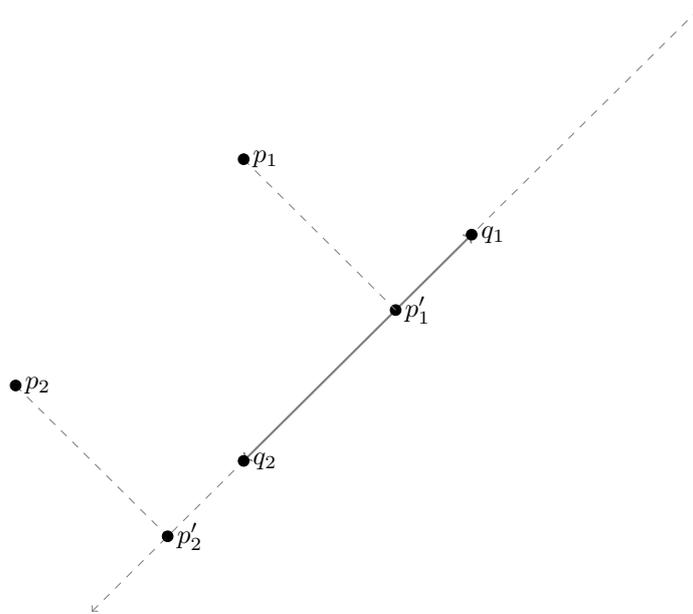

Figure~\ref{SimilarNeurons} shows examples of some similar neurons. 
Observe that these neurons look the same, occupying the same region, and left/right symmetric. 
The similarity between a pair of neurons is defined based on this observation. 
If neurons $a$ and $b$ are on the same side of the brain. 
The neuron $a$ to neuron $b$ similarity is the percentage of points of $a$ that are ``close enough'' to the line segments of the $b$. 
The closeness is the Euclidean distance between a point and a line segment. 
Given a point $p$ and a line segment $\overline{q_1,q_2}$, the distance $d$ between $p$ and $\overline{q_1,q_2}$ is the perpendicular distance between $p$ and the straight line $\overleftrightarrow{q_1,q_2}$ if the closest point on $\overleftrightarrow{q_1,q_2}$ to $p$ is between $q_1$ and $q_2$; otherwise the $d = \min\{|\overline{p,q_1}|, |\overline{p,q_2}| \}.$
In Figure~\ref{DisP2Line}, the distance between $p_1$ and $\overline{q_1,q_2}$ is $\overline{|p_1,p_1'|}$; the distance between $p_2$ and $\overline{q_1,q_2}$ is $\overline{|p_2,q_2|}$. 
If the distance $d$ is less than a given threshold, we say that $p$ is close enough to $\overline{q_1,q_2}$. 
In this study, the threshold is 10. 
The similarity of $a$ to $b$ is the percentage of the points in $a$ that are close enough to any line segment in $b$. 
The similarity is a real number over the range $[0,1]$. 
Note that, the similarity relationship between two neurons is not symmetric, i.e., the similarity of $a$ to $b$ is not equal to the similarity of $b$ to $a$. 
We use $s_{(a,b)}$ to denote the similarity of $a$ to $b$. 
To ensure two neurons $a$ and $b$ are similar, we use the mutual similarity $d_{(a,b)}$ = $\min(s_{(a,b)},s_{(b,a)})$ in this study. 

The left-right similarity of neurons is also considered.  
This study defines the mirror image as the $yz$ plane passing through the center of the $839 \times 457 \times 234$ box. 
The coordinates of the center point are (420, 229,117). 
For a neuron $a$, let $a'$ be the mirror image of $a$ respective to the symmetric plane.
Let $b$ be a neuron such that $d_{(a',b)}$ is greater than 0.8, the mutual similarity between $a$ and $b$ is defined to be
$d_{(a',b)}$.  

We shall partition the neurons into groups based on the similarity of neurons to find some groups. 
This task needs a visualization tool which is stated in the following. 
We shall then present the module-finding methods. 

\subsection{Visualization Tool}
\begin{figure}
    \centering
    \includegraphics[width=15cm]{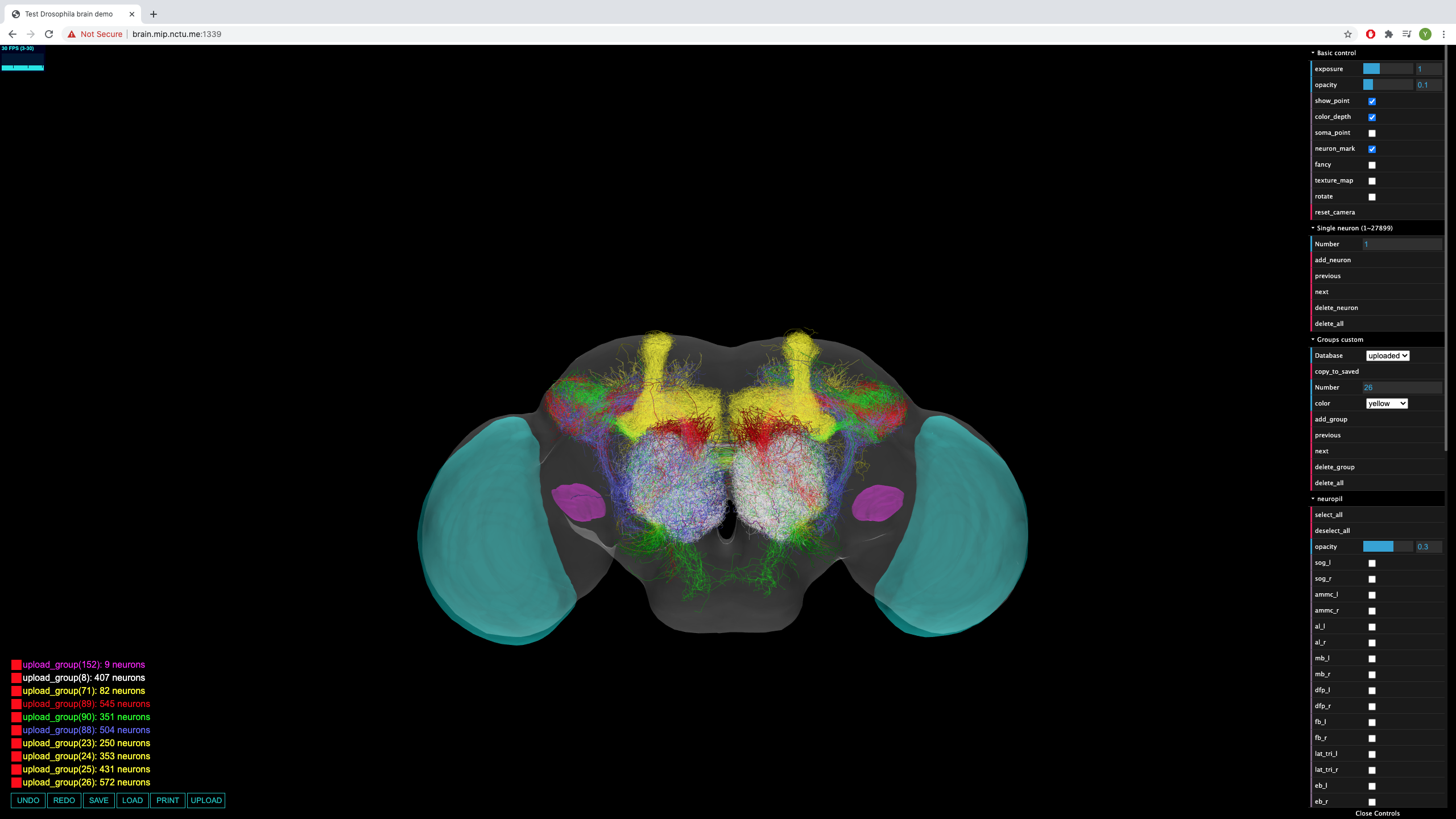}
    \caption{The visualization tool for studying the fly brain. }
    \label{VisualizationWebPage}
\end{figure}

The visualization tool is implemented to display a single neuron, sets of neurons, and selected neuropils for visually examining computational results.   
The tool is a server/client system. 
The server computer stores the hulls of the fly brain, the neuropils, and the databases of the set of 27k neurons. 
Users are on the client end and connect to the database through a web browser. 
Rendering and displaying images are on the client's computer. 
An example is shown in Figure~\ref{VisualizationWebPage}. 
Some neuropils and groups of neurons are shown. 
On the bottom-left are the groups of neurons rendered on the screen. 
The red squares on the left are the buttons, clicking a button turns on or off the corresponding group. 
On the right of the screen, user interfaces to the server are shown. 
Users can select to render neurons, groups of neurons, and neuropils. 
Users can also choose the way of rendering, for example, transparent rendering. 
Rotating the rendered result is also controlled by a button on the right.
Zoom in and zoom out are controlled by the mouse. 

A group of neurons can be rendered in the color specified by the user or assigned by the system. 
8 colors can be selected. 
Users can select many groups, and render them all together on one screen in different colors.  
When there are many groups rendered in one session, 
8 colors may not be enough if many groups are rendered, users can turn on or off some groups to prevent the region of interest been blocked. 
Operations such as rotation, zoom-in and zoom-out, and transparent rendering support 3D visualization of many groups. 

Rendering of the hulls of the fly brain and neuropils are supported. 
These objects are stored in the boundary representation (triangle patches as the boundary). 
The fly brain is by default drawn in the middle of the window. 
The neuropils can be rendered by selecting from a pop-out menu on the right. 

\subsection{Finding the Potential Modules}
There are two steps, the first step is to cluster the neurons into groups. 
The second step is to assemble the groups to find the potential modules. 
Both steps employ graph algorithms. 

\subsubsection{Finding Groups}
We define the graph $G_{\rm neurons}$ for group finding. 
Let $G_{\rm neurons} = (V_{\rm neurons}, E_{\rm neurons})$ be a graph where $V_{\rm neurons}$ is a set of vertices representing the set of 27K neurons, and $E_{\rm neurons}$ is a set of weighted edges $(u,v)$, $u, v\in V_{\rm neurons}$.
The weight associated with the edge $(u,v)$ is the mutual similarity $d_{(u,v)}$. 

Groups are found under user guidance. 
Given a target neuron $s$ as the seed neuron and a threshold $th$ for the mutual similarity, we compute the connected component $C$ that every vertex $u$ in $C$ can be reached from $s$ using a path where the weights associated with edges on the path are greater than or equal to a given threshold $th$.  
Finding $C$ can be implemented using graph algorithms such as the breadth-first search or the Union/Find operation\cite{Cormen3rd}. 
Both algorithms produce the same result. 
A JSON file is generated for the neurons in $C$ and fed into the visualization tool for visual inspection of the group. 
Different threshold $th$ resulting different sizes of $C$, there are more neurons if a low $th$ is applied. 
Too low $th$ could cause $C$ to merge with other connected components. 
The most appropriate $th$ is hard to define and thus determined by iteratively trying different $th$, inspecting the resulting connected component, and then chosen by a user. 
The most appropriate $th$ is the one that one more relaxation of $th$ causes the size of $C$ to increase significantly and the shape of $C$ change. 
Once a connected component is found, the neurons in the connected component are marked used, and never involved in the future group-finding process.
The location of somas of neurons in the same groups could be different. 
Further partitioning of groups based on the location of soma proceeds. 
The $k$-means method was applied to each group's soma location. 
$k$ was selected by trying $k=2$ up to $k=10$. 
Users decided the most appropriate $k$. 
Finally, 212 groups found.
Among these 212 groups, some contained very few neurons or were neither symmetric nor at the center. 
These groups were removed from the further analysis. 
The remaining 154 groups consist of 23,625 neurons in the database. 
A group is named by a number in the sequence the group found. 

\subsubsection{Finding Potential Modules}
A fly brain should be similar to an IC, consisting of modules. 
Other than the mushroom body, the olfactory, and the visual systems, people do not know much about the modules in the fly brain.  
A method is designed to assemble the groups to find potential modules. 
The method is a heuristic algorithm in which we compute the maximum spanning tree of a complete graph, then we partition the Maximum Spanning Tree into subtrees. 
Each subtree corresponds to a potential module. 

A weighted graph $G_{\rm group} = (V_{\rm group}, E_{\rm group})$ is constructed for the 154 groups obtained previously. 
$V_{\rm group}$ is a set of 154 vertices representing the 154 groups. 
$E_{\rm group}$ is a set of weighted edges $(i,j)$ where $i$ and $j$ are two groups and the weight associated the edge $(i,j)$ is computed as the following. 

Each group $i$ is converted to a set of voxels occupied by the neurons in group $i$.
The set of voxels occupied by a neuron is obtained by:
\begin{enumerate}
    \item Compute the set $S$ of voxels intersecting with the skeleton of the neuron.
    \item For each voxel $v$ in $S$, a $5\times  5\times 5$ volume centered at $v$ are marked occupied by the neuron.
\end{enumerate}
The voxel occupied by group $i$ is the union of the sets of voxels occupied by all the neurons in group $i$, denoted ${\bf v}_i$. 
Let ${\bf v}_{i,j}$ be the intersection of ${\bf v}_i$ and ${\bf v}_j$.
The weight associated with the edge $(i,j)$ denoted $w(i,j)$ is 
\begin{equation}
\max (\frac{|{\bf v}_{i,j}|} { |{\bf v}_i| }, \frac{ |{\bf v}_{i,j}| } { |{\bf v}_j|}), \label{EdgeWeights}
\end{equation}  
where $|{\bf v}|$ is the number of voxels in ${\bf v}$.  

\begin{figure}
    \centering
    \includegraphics[width=15cm]{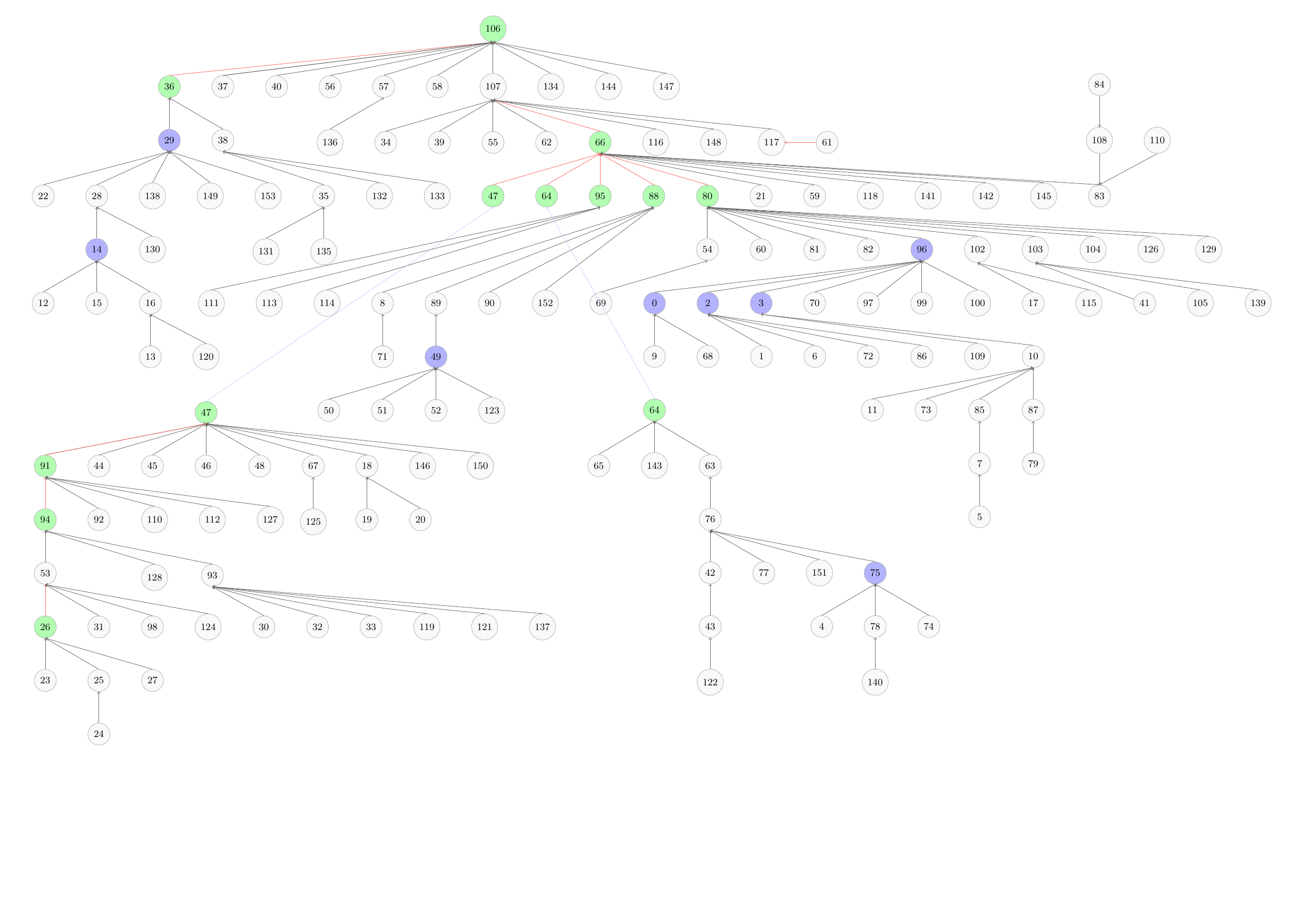}
    \caption{The maximum spanning tree for finding modules.}
    \label{MaximumSpanning}
\end{figure}

\begin{figure}
    \centering
    \includegraphics[width=7cm]{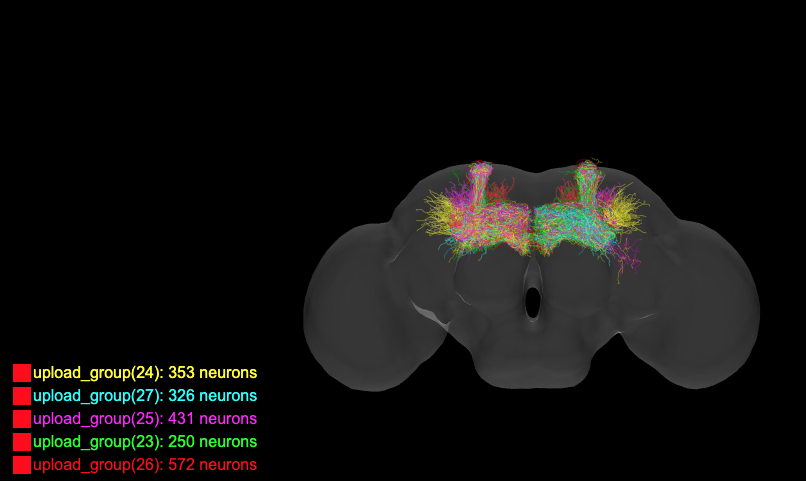}
    \includegraphics[width=7cm]{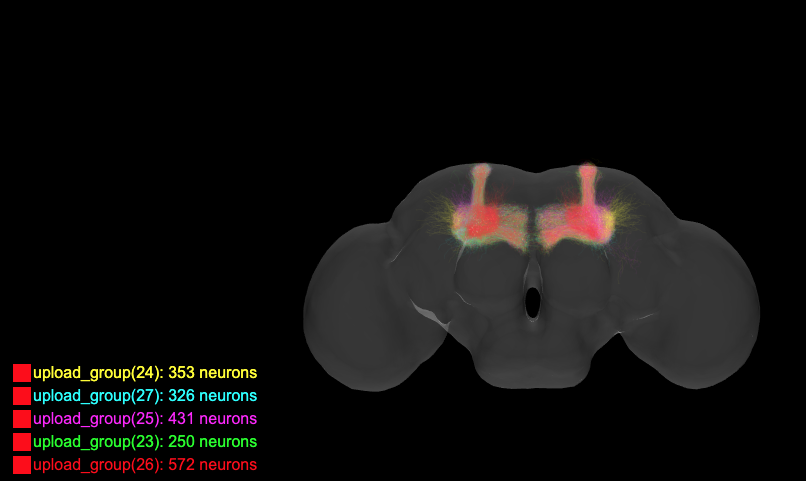}
    \caption{{\tt MB}, the subtree rooted 26 is the Mushroom Body}
    \label{moduleMB}
\end{figure}

\begin{figure}
    \centering
    \includegraphics[width=7cm]{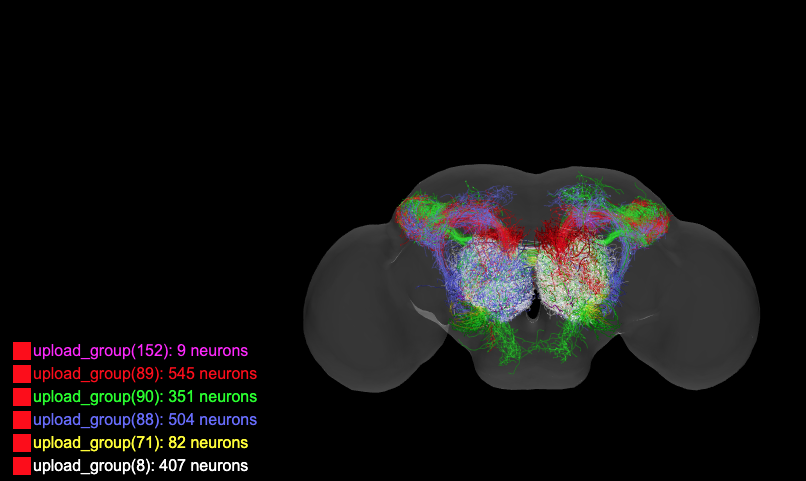}
    \includegraphics[width=7cm]{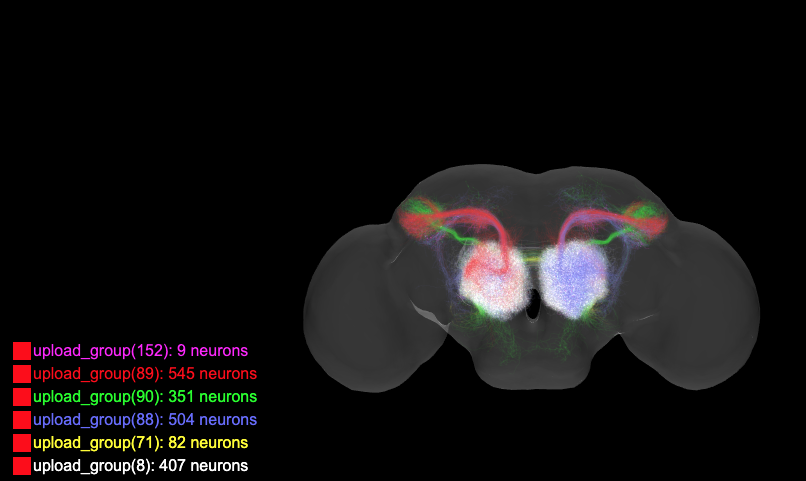}
    \caption{{\tt AL}, the subtree rooted 88 corresponds to the Antennal Lobe.}
    \label{moduleAL}
\end{figure}

\begin{figure}
    \centering
    \includegraphics[width=7cm]{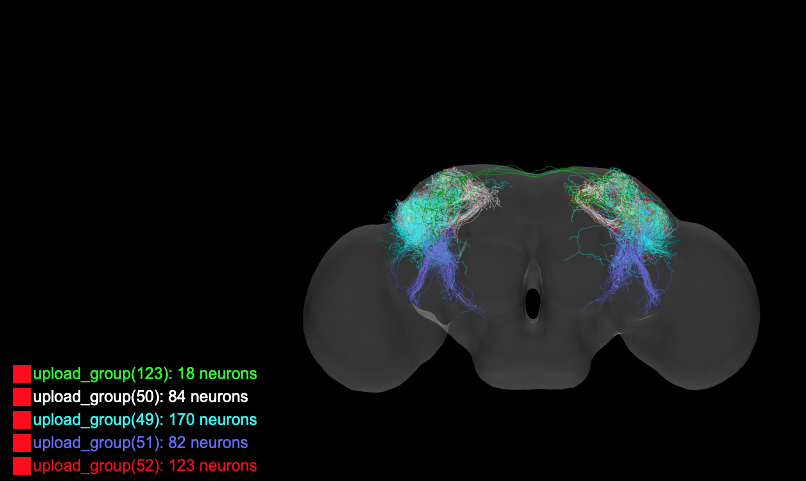}
    \includegraphics[width=7cm]{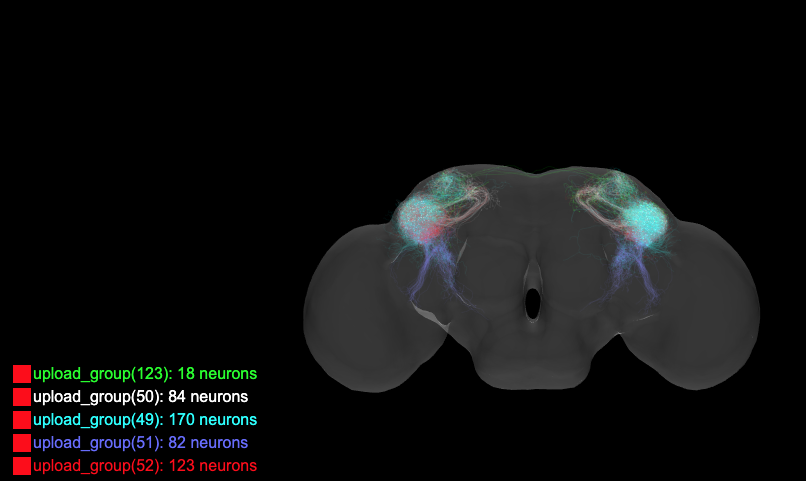}
    \caption{{\tt LH}, the tree rooted 49 is the Lateral Horn. }
    \label{moduleLH}
\end{figure}

\begin{figure}
    \centering
    \includegraphics[width=7cm]{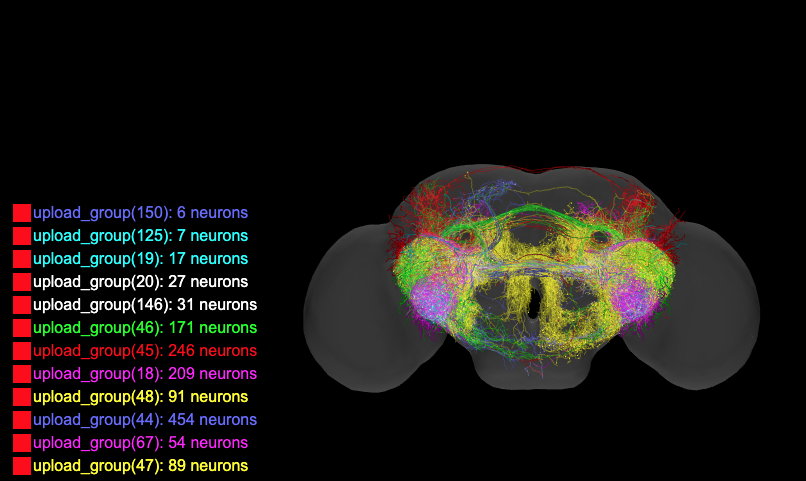}
    \includegraphics[width=7cm]{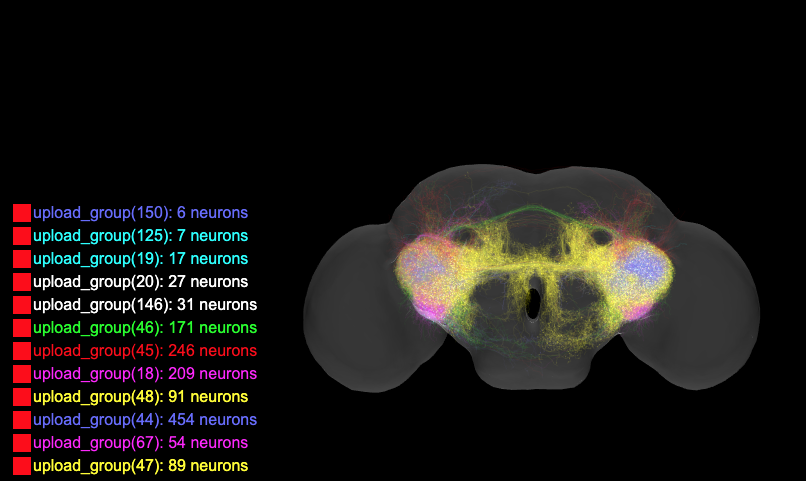}
    \caption{{\tt VLP} the tree rooted 47 is the VLP.}
    \label{moduleVLP}
\end{figure}

\begin{figure}
    \centering
    \includegraphics[width=7cm]{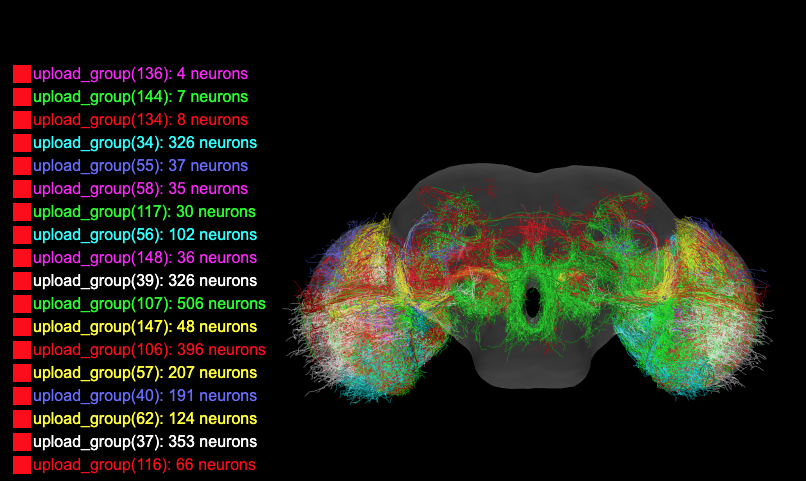}
    \includegraphics[width=7cm]{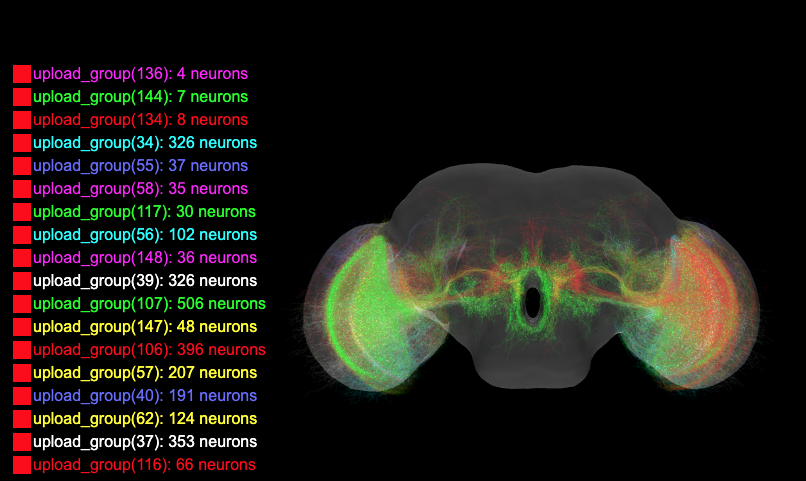}
    \caption{The tree rooted 106, one of the four subtrees of the Visual system, named {\tt Visual1}. Neurons in this subtree mainly occupy the Medulla.}
    \label{moduleVisual1}
\end{figure}

\begin{figure}
    \centering
    \includegraphics[width=7cm]{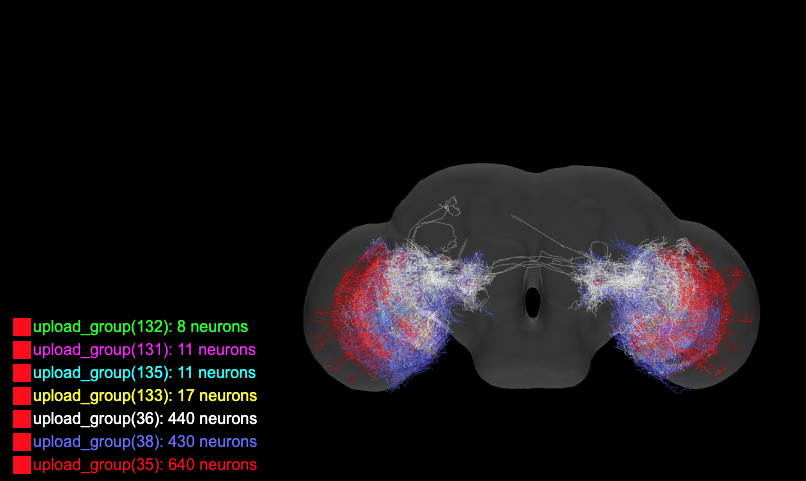}
    \includegraphics[width=7cm]{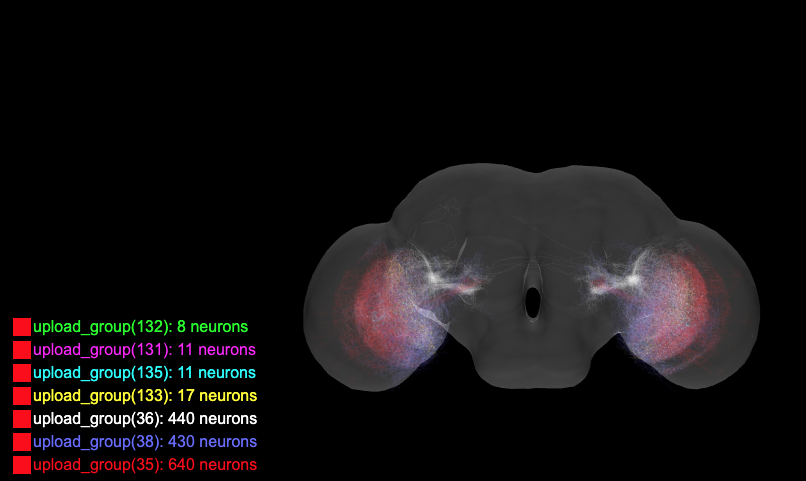}
    \caption{The tree rooted 36, one of the four subtrees of the Visual system, named {\tt Visual2}. Neurons occupy Lop and innervate Med and Lob.}
    \label{moduleVisual2}
\end{figure}

\begin{figure}
    \centering
    \includegraphics[width=7cm]{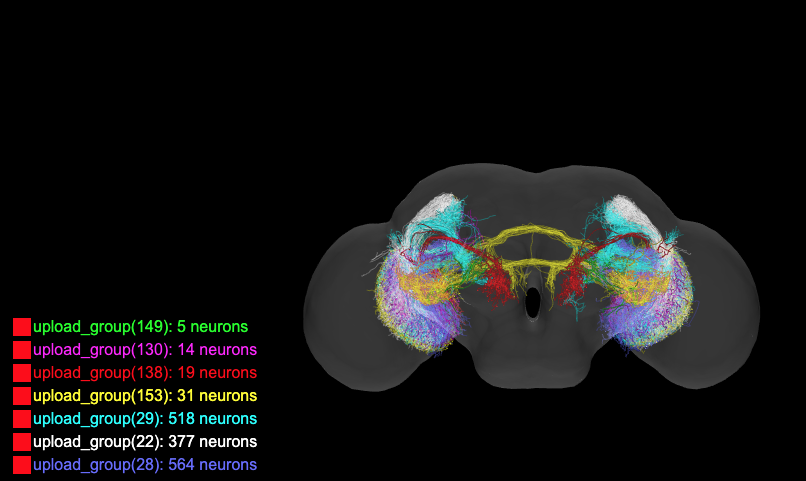}
    \includegraphics[width=7cm]{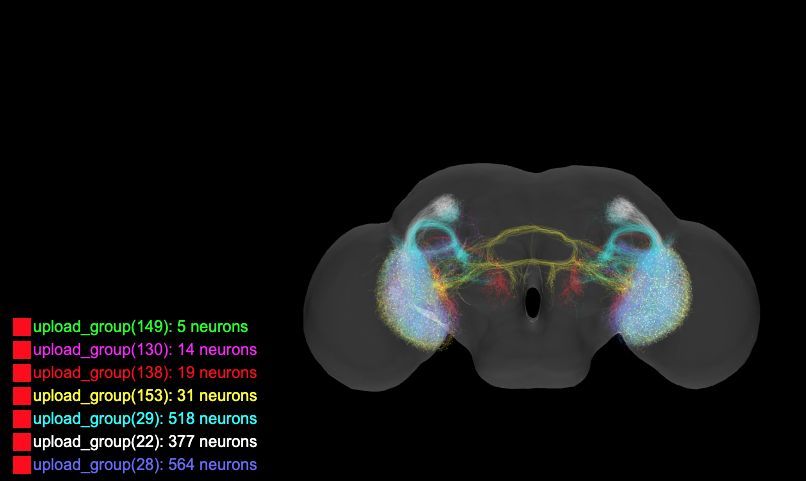}
    \caption{The tree rooted 29, one of the four subtrees of the Visual system, named {\tt Visual3}. Neurons occupy Lob.}
    \label{moduleVisual3}
\end{figure}

\begin{figure}
    \centering
    \includegraphics[width=7cm]{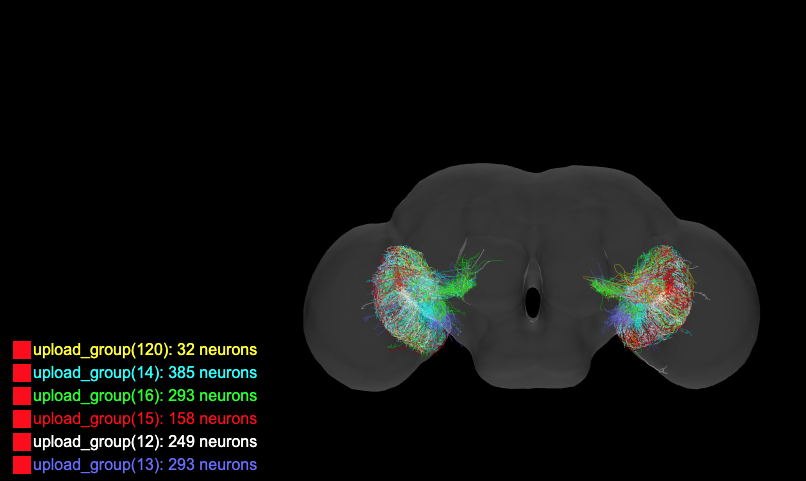}
    \includegraphics[width=7cm]{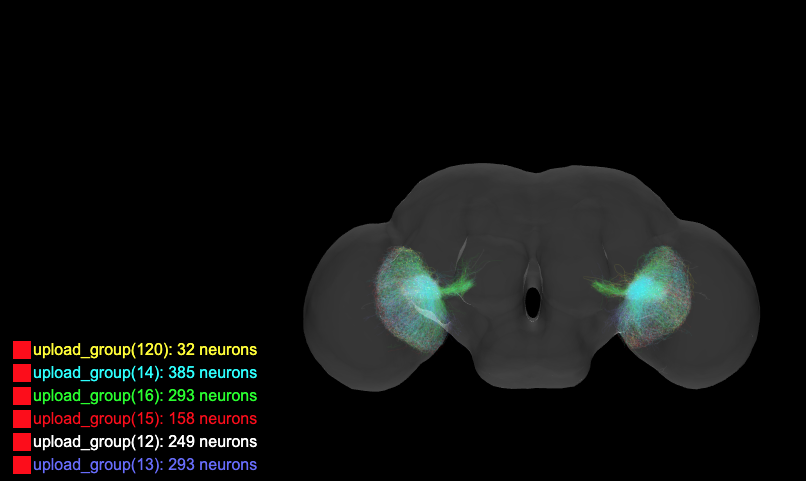}
    \caption{The tree rooted 14, one of the four subtrees of the Visual system, named {\tt Visual4}. Neurons occupy Lob.}
    \label{moduleVisual4}
\end{figure}

\begin{figure}
    \centering
    \includegraphics[width=7cm]{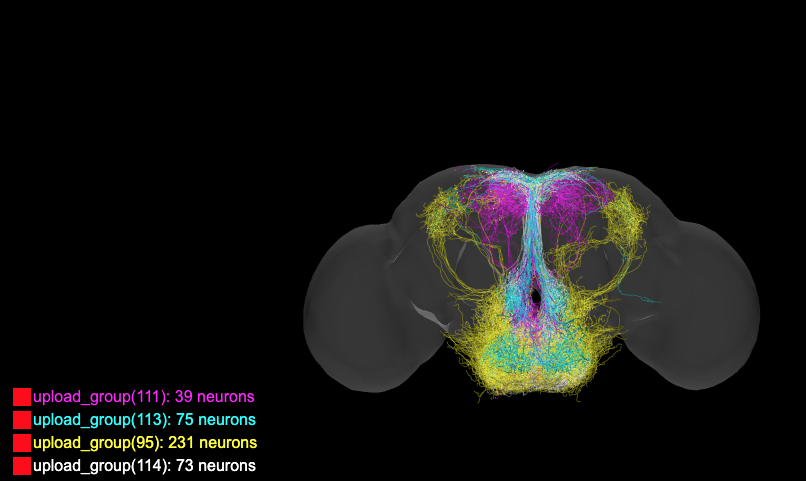}
    \includegraphics[width=7cm]{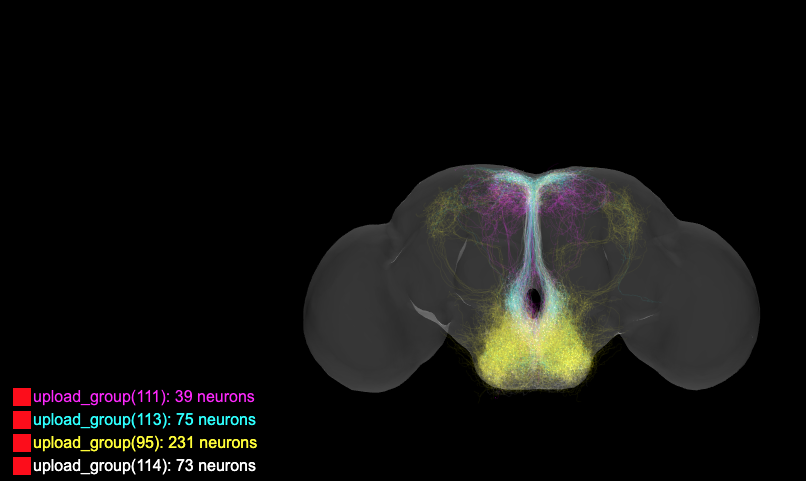}
    \caption{The tree rooted 95 is named {\tt Esophagus95}}
    \label{moduleEsophagus1}
\end{figure}

\begin{figure}
    \centering
    \includegraphics[width=7cm]{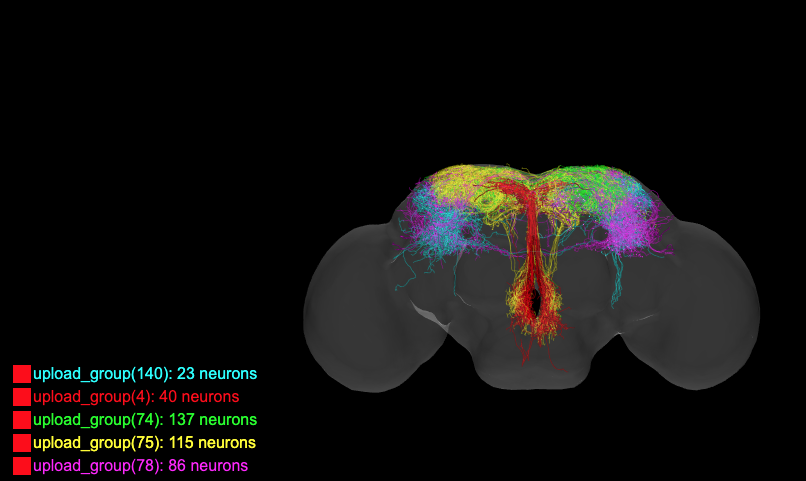}
    \includegraphics[width=7cm]{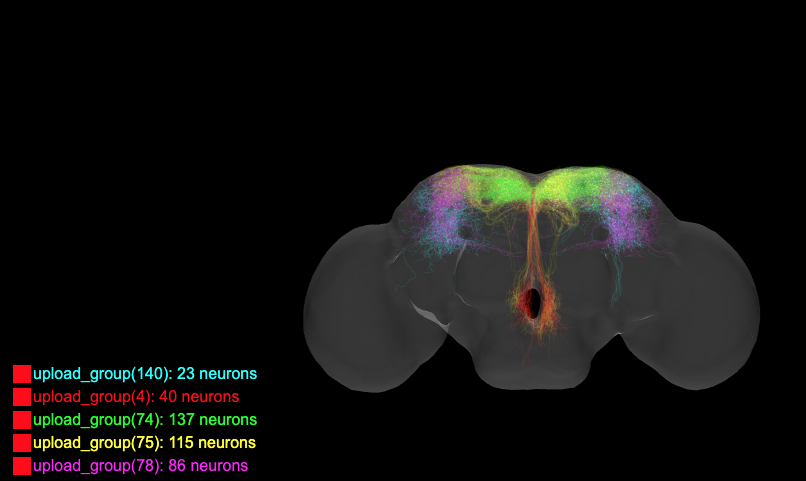}
    \caption{The tree rooted 75 is named {\tt Esophagus75}.}
    \label{moduleEsophagus2}
\end{figure}

\begin{figure}
    \centering
    \includegraphics[width=7cm]{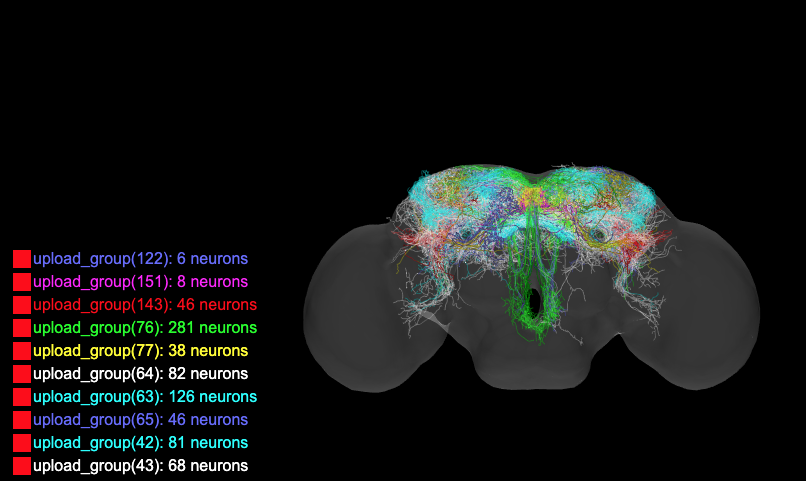}
    \includegraphics[width=7cm]{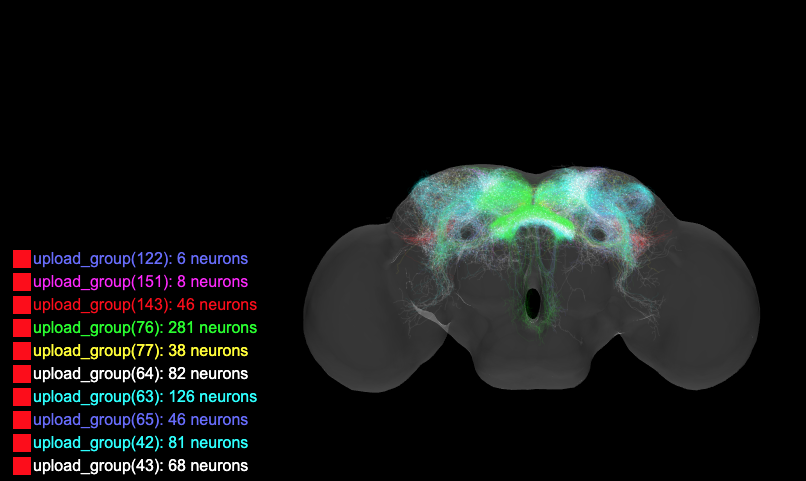}
    \caption{The tree rooted 64 is named {\tt Superior64}. }
    \label{moduleSuperious64}
\end{figure}

\begin{figure}
    \centering
    \includegraphics[width=7cm]{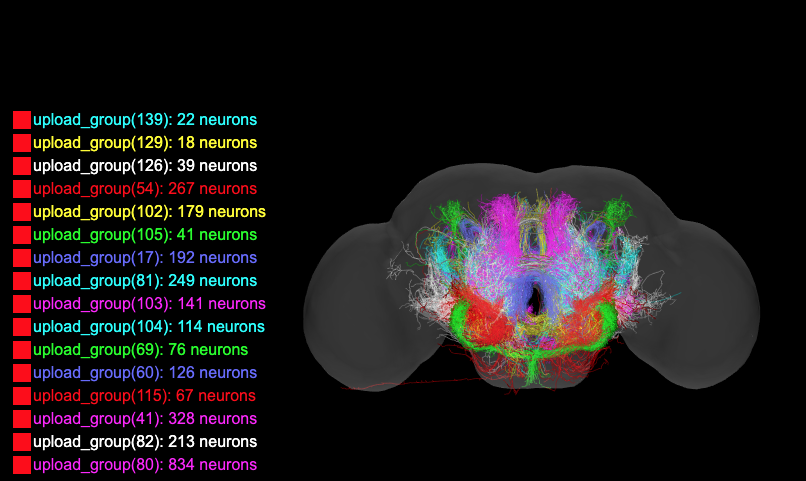}
    \includegraphics[width=7cm]{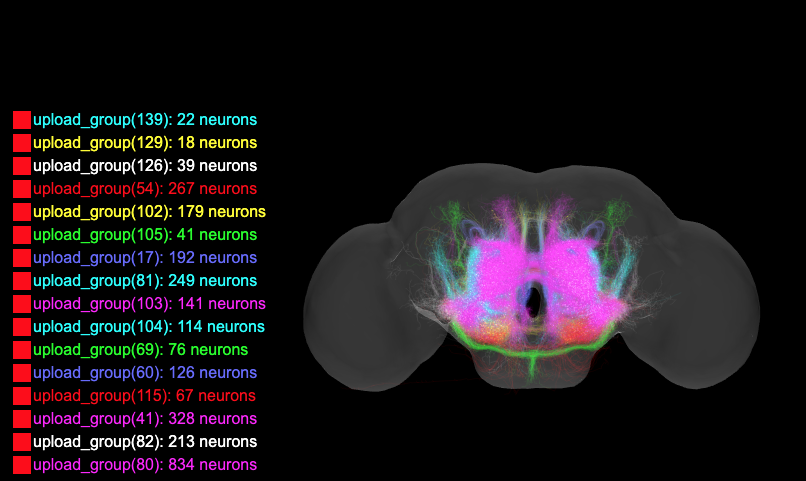}
    \caption{The tree rooted 80 is named {\tt Inferior80}.}
    \label{moduleInferior80}
\end{figure}

\begin{figure}
    \centering
    \includegraphics[width=7cm]{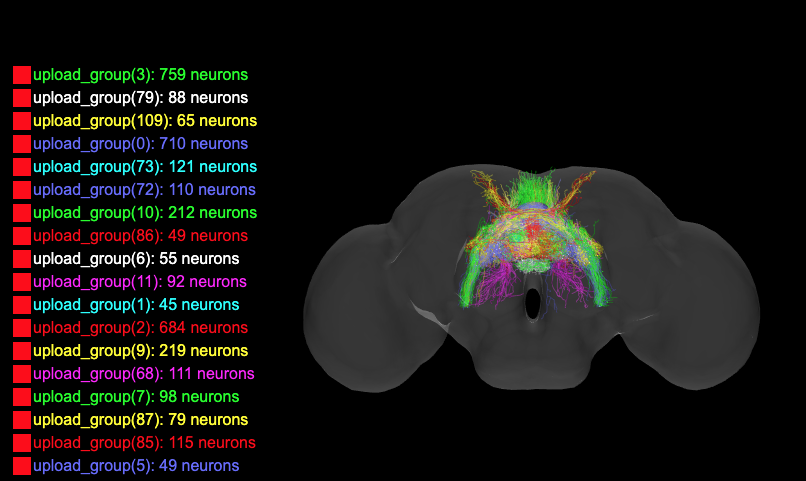}
    \includegraphics[width=7cm]{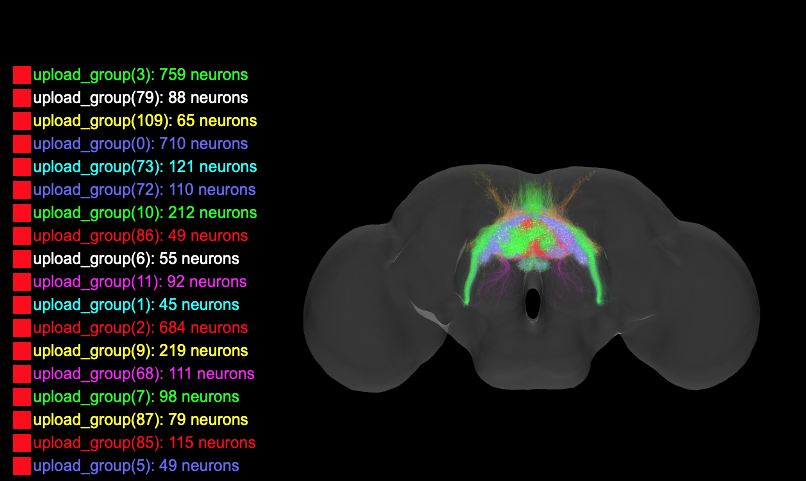}
    \caption{The trees rooted 0, 2, 3 together is named {\tt Core1}.}
    \label{moduleCore1}
\end{figure}

\begin{figure}
    \centering
    \includegraphics[width=7cm]{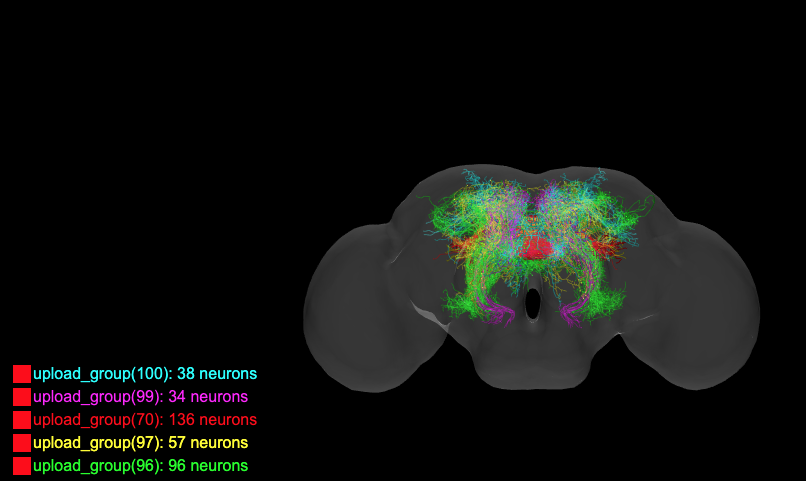}
    \includegraphics[width=7cm]{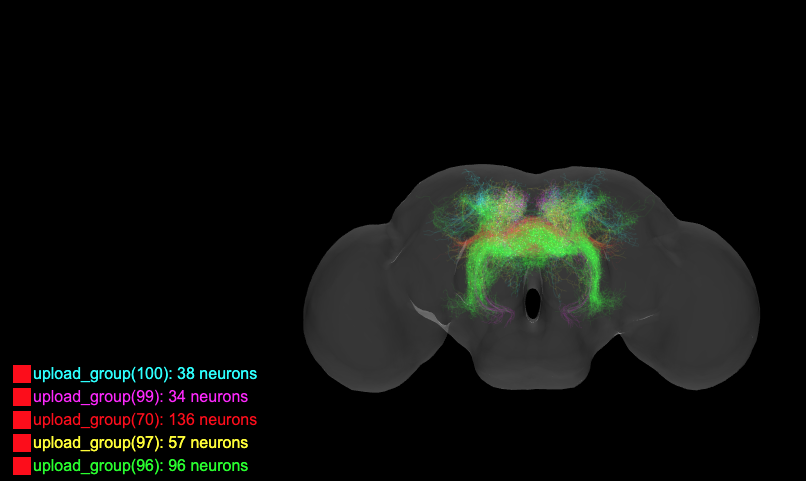}
    \caption{The tree rooted 96 is named {\tt Core2}.}
    \label{moduleCore2}
\end{figure}

\begin{figure}
    \centering
    \includegraphics[width=7cm]{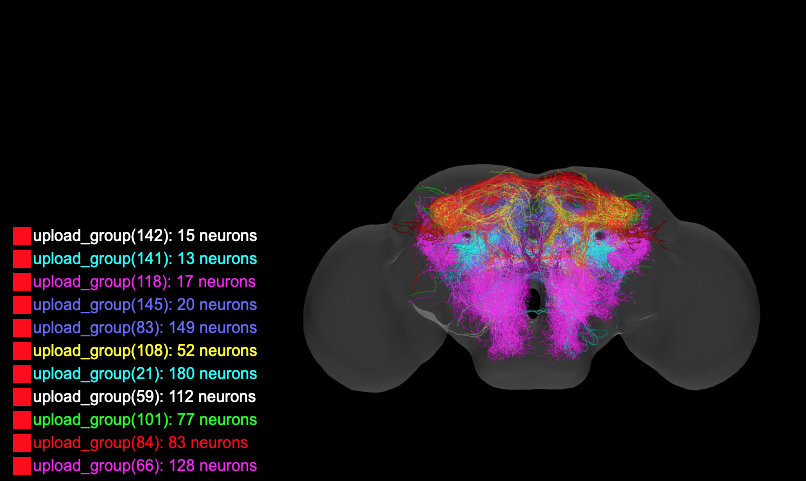}
    \includegraphics[width=7cm]{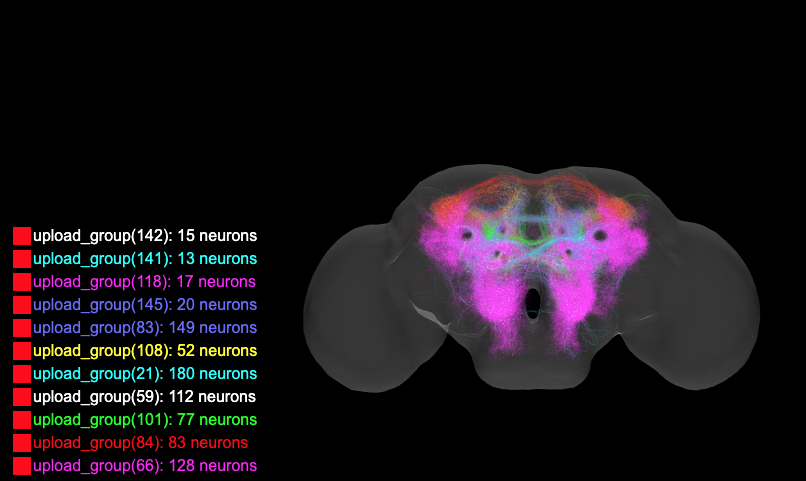}
    \caption{The tree rooted 66 is named {\tt Center}.}
    \label{moduleCenter}
\end{figure}

\begin{figure}
    \centering
    \includegraphics[width=7cm]{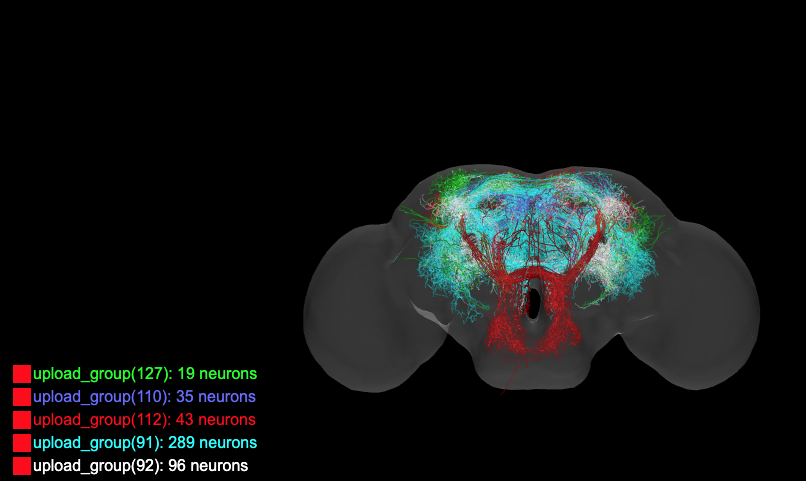}
    \includegraphics[width=7cm]{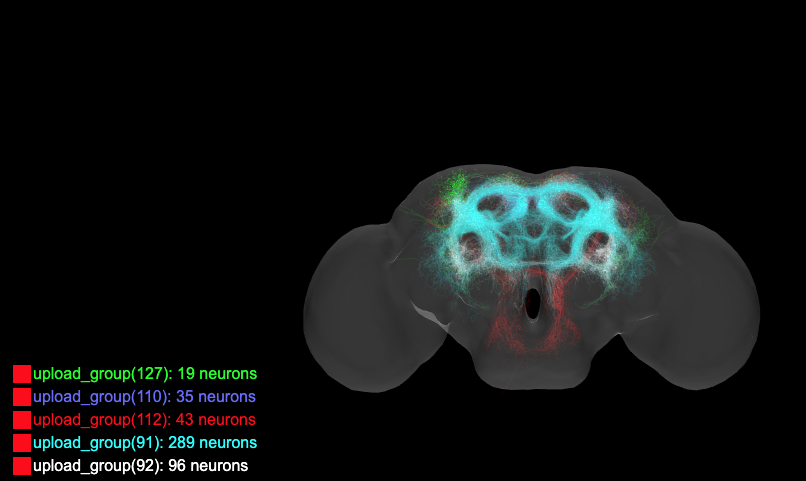}
    \caption{The tree rooted 91 is named {\tt Superior91}.}
    \label{ModuleSuoerior91}
\end{figure}

\begin{figure}
    \centering
    \includegraphics[width=7cm]{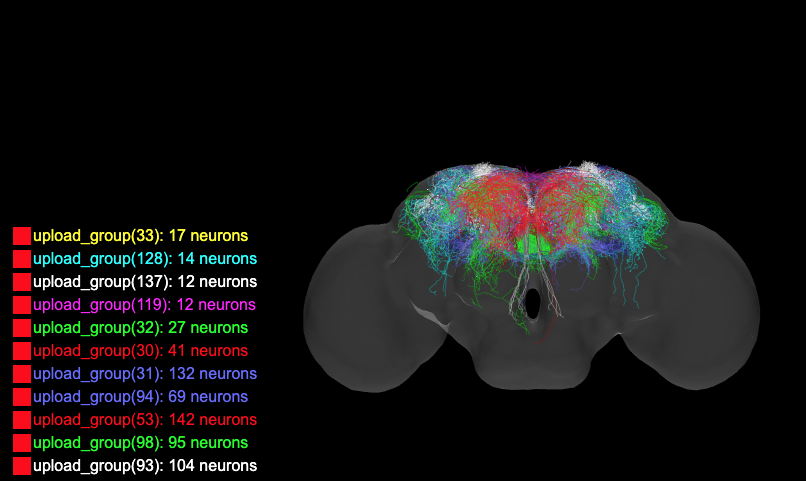}
    \includegraphics[width=7cm]{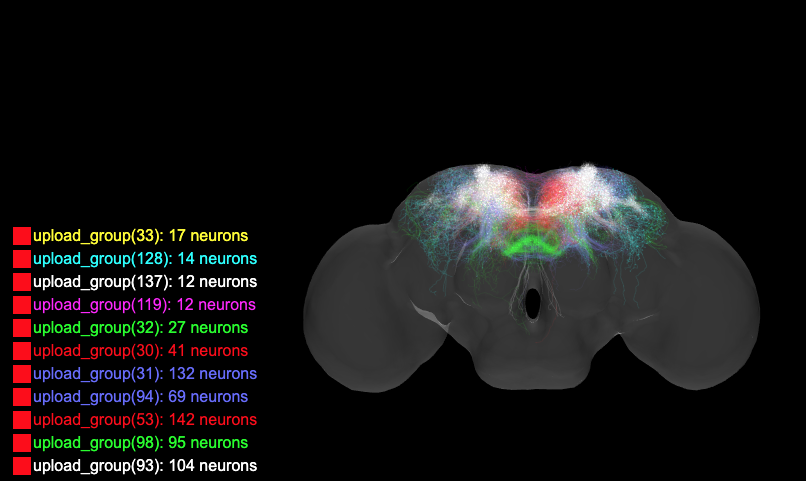}
    \caption{The tree rooted 94 is named {\tt Superior94}.}
    \label{moduleSuperious94}
\end{figure}

The Maximum Spanning Tree $T$ of $G_{\rm group}$ is computed. 
Note that $|V_{\rm group}|$ is 154, there are 153 edges to connect the 154 vertices (groups), and the sum of the weights of the edges is maximized.  
There are light tree edges in the $T$; these light edges are likely connecting two modules. 
Groups connected by heavy tree edges are likely in the same module.  
In how we defined the weights of tree edges in Eq.~(\ref{EdgeWeights}), groups in a module tend to have an edge connecting to the largest group in the module. 
The resulting maximum spanning tree has many groups connected to group 106 and group 107 in the visual system. 
These two groups are large thus the edges incident to these two groups are heavy. 
We restrict groups not in the visual system, except group 66, connect to groups 106 and 107, and use the light tree edges to select some nodes (the green nodes) as roots of subtrees. 
The resulting maximum spanning tree is Figure~\ref{MaximumSpanning}. 
In Figure~\ref{MaximumSpanning}, the red edges (light edges) have weights less than 0.3. 
The green nodes are the root of subtrees in which the groups are in a potential module. 
The blues nodes are selected to partition a group with the help of the visualization tool. 
These potential modules are described as follows. 
Some subtrees correspond to known modules, 
\begin{itemize}
\item {\tt MB}, the subtree rooted 26, the Mushroom Body (Figure~\ref{moduleMB}).  
\item {\tt AL}, the subtree rooted 88, the Antennal Lobe (Figure~\ref{moduleAL}).
\item {\tt LH}, the subtree rooted 49, the Lateral Horn (Figure~\ref{moduleLH}).
\item {\tt VLP}, the subtree rooted 47, neurons are mainly in VLP (Figure~\ref{moduleVLP}).
\item Four subtrees in the Optical Lobe
    \begin{itemize}
        \item {\tt Visual1}, the subtree rooted 106, neurons are mainly in Medulla (Figure~\ref{moduleVisual1}). 
        \item {\tt Visual2}, the subtree rooted 36, neurons are mainly in Lop (Figure~\ref{moduleVisual2}), and some neurons innervate Med and Lob. 
        \item {\tt Visual3}, the subtree rooted 29 (manually selected), neurons are mainly in Lob (Figure~\ref{moduleVisual3}). 
        \item {\tt Visual4}, the subtree rooted 14 a manually selected node, neurons are mainly in Lob as well) (Figure~\ref{moduleVisual4}). 
    \end{itemize}
\end{itemize}
Other than the modules corresponding to the known neuropils, the potential modules are described in the following. 
Group 66 connects to node 107 in {\tt Visual1}, node 47 in {\tt VLP}, node 88 in {\tt AL} by red edges. 
It also connects nodes 64, 95, and 80 by red edges. 
Subtrees rooted 64 and 95 are respectively named {\tt Superior64} (Figure~\ref{moduleSuperious64}) and {\tt Esophagus95} (Figure~\ref{moduleEsophagus1}),  
many neurons in this module are around the Esophagus. 
Another module that has this property is the subtree rooted 75, named {\tt Esophagus75} (Figure~\ref{moduleEsophagus2}). 
The subtree rooted 80 is a big subtree.
In this subtree, we manually choose node 96, a subtree rooted 96 consisting of neurons in the center of the fly brain including the neurons in EB (group 70). 
Trees rooted 0, 2, 3 consist of dense neurons in the center, a potential module named {\tt Core1} (Figure~\ref{moduleCore1}). 
Tree rooted 96 not including {\tt Core1} named {\tt Core2} (Figure~\ref{moduleCore2}). 
Tree rooted 80, not including {\tt Core1} and {\tt Core2} named {\tt Inferior80} (Figure~\ref{moduleInferior80}). 
The subtree rooted 66, not including {\tt VLP}, {\tt Superior64}, {\tt AL}, {\tt Esophagus95}, and {\tt Inferion80} is {\tt Center} (Figure~\ref{moduleCenter}). 
The last two subtrees roote 91 and 94 are named {\tt Superior91} (Figure~\ref{ModuleSuoerior91}) and {\tt Superior94}
(Figure~\ref{moduleSuperious94}). 

\begin{figure}
    \centering
    \includegraphics[width=15cm]{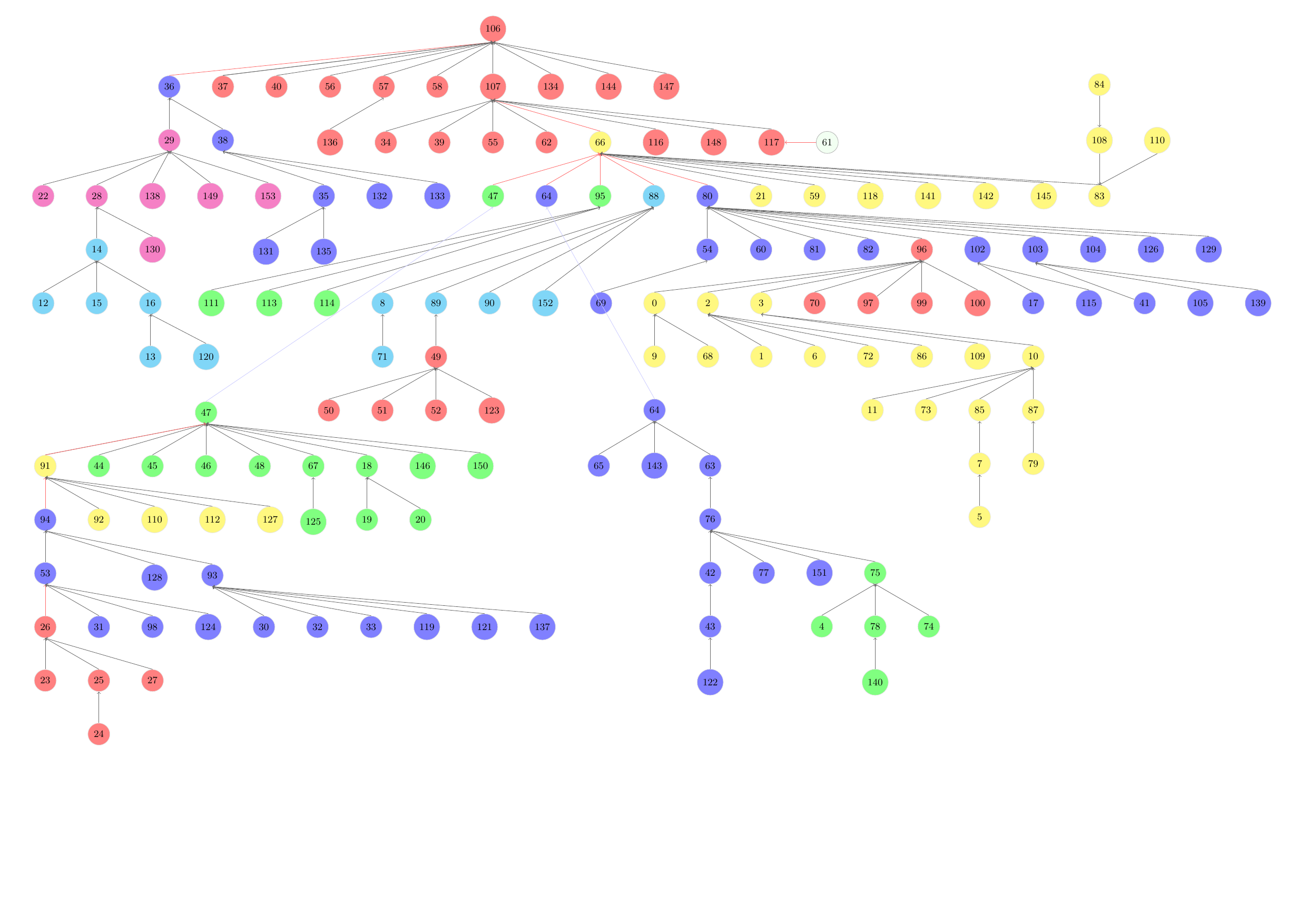}
    \caption{The same maximum spanning tree as Figure~\ref{MaximumSpanning}. Each module is rendered in the same color. }
    \label{MaximumSpanningTreeColor}
\end{figure}

In the Maximum Spanning Tree, {\tt MB} connects to {\tt Superior94} by a red edge. 
This observation can be interpreted as that the {\tt MB} might have the most strong connection to {\tt Superior94}. 
Using the visualization tool, we can see that many neurons in {\tt Superior94} are around the $\alpha$-arm of the {\tt MB}. 
Redraw maximum spanning such that nodes in the subtree correspond to a potential module in the same color, we get Figure~\ref{MaximumSpanningTreeColor}.

Table~\ref{fruRatio} lists the ratio of the male and female {\it fru} neurons (last column in the table) in the modules. 
In some modules the female has more {\it fru} neurons than the male, for examples in {\tt Esophagus95}, {\tt Visual2}, {\tt Visual3}, {\tt LH}, {\tt VLP}, and {\tt Superior94}. 
The modules {\tt Visual1}, {\tt Center66}, and {\tt Superior91}, male has more {\it fru} neurons than female. 
Observe these modules using the visualization tool, modules {\tt Visual1}, {\tt Center66}, and {\tt Superior91} are closer to the peripheral than modules {\tt Esophagus95}, {\tt Visual2}, {\tt Visual3}, {\tt LH}, {\tt VLP}, and {\tt Superior94}. 
If labeling the neurons by the {\it fru} gene is random, the male fly has more {\it fru} neurons close to the peripheral than the female fly. 

\begin{table}
\begin{center}
\caption {Clusters and fru neurons distribution. }\label{fruRatio}
\begin{tabular}{|c|c|c|c|c| }
\hline
   & Total neurons & Female neurons & Male neurons & Male/Female \\ \hline
 {\tt Core1} & 3661 & 194/0.070 & 208 / 0.062 & 0.886\\ \hline 
 {\tt Core2} & 361 & 16/0.006 & 17/0.005 & 0.883\\ \hline
 {\tt Esophagus95} & 418 & 67/0.0243 & 38/0.0113 & 0.465 \\ \hline 
 {\tt Esophagus75} & 401 & 80/0.029 & 134/0.040 & 1.379\\  \hline
 {\tt MB} & 1932 & 342/0.124 & 442/0.131 & 1.056 \\ \hline
 {\tt Visual1} & 2802 & 195/ 0.070 & 359/0.106 & 1.514 \\ \hline
 {\tt Visual2} & 1557 & 150/0.054 & 97 / 0.029 & 0.537 \\ \hline 
 {\tt Visual3} & 1528 & 171/0.062 & 126/0.037 & 0.597\\ \hline
 {\tt Visual4} & 1410 & 0 & 0 & \\ \hline
 {\tt AL} & 1898 & 72/0.026 & 84/0.025 & 0.962\\ \hline
 {\tt LH} & 477 & 26/0.009 & 15 /0.004 & 0.444 \\ \hline
 {\tt VLP} & 1402 & 189/0.069 & 125/0.037 & 0.536\\ \hline
 {\tt Center} & 804 & 34/0.0123 & 93/0.0276 & 2.244 \\  \hline
 {\tt Superior64} & 782 & 61/0.022 & 81/0.024 & 1.091\\  \hline
 {\tt Superior91} & 482 & 105/0.038 & 255/0.076 & 2.0\\ \hline 
 {\tt Superior94} & 609 & 39/0.014 & 27/0.008 & 0.571 \\ \hline 
 {\tt Inferior80} & 2906 & 603/0.219 & 769/0.228 & 1.041 \\ \hline

\end{tabular}
\end{center}
\end{table}

Possible applications of the modules are studying the structure of a module and studying the connectivity of neurons to explore the connectome of neurons in the fly brain. 
In the next section, we study the structure of the Medulla (the module {\tt Visual1}). 

\section{The Structure of the Medulla\label{MedStruct}}
We demonstrate an application using the modules found to analyze the structure of the Medulla. 
The maximum spanning tree technique is employed again, but fewer user assistants are involved. 
Neurons in ({\tt Visual1}) corresponding to the Medulla are used. 

Three sets, $S_{\rm MedF}$, $S_{\rm MedFfru}$, and $S_{\rm MedMfru}$, of neurons in {\tt Visual1} were collected. 
$S_{\rm MedF}$ consists of the neurons in the female fly brain but does not include the female {\it fru} neurons. 
The female and male {\it fru} neurons are in $S_{\rm MedFfru}$ and $S_{\rm MedMfru}$ respectively. 
The sets $S_{\rm MedF}$ and $S_{\rm MedFfru}$ are employed for analyzing the structure of a female fly brain. 

Given a set of neurons, a weighted graph $G=(V, E)$ is constructed. 
$V$ is the set of vertices representing the neurons, and each pair of vertices $u$ and $v$, $(u,v)$ is an edge in edge in $E$. 
The weight associated with $(u, u)$ is the mutual similarity of neuron $u$ and $v$. 
Given a threshold $th$, a graph $G'=(V,E')$ is obtained by removing the edges $e$ in $E$ where the weight of $e$ is less than the threshold $th$. 
$G'$ consists of many connected components. 
The threshold is set high so that many small components are obtained. 
This group partition algorithm is applied to the three sets of neurons, $S_{\rm MedF}$, $S_{\rm MedFfru}$, and $S_{\rm MedMfru}$. 

The maximum spanning tree method is applied to assemble the small components. 
We use the neurons in $S_{\rm MedF}$ as an example to explain the assembling algorithm. 
A graph, $G_{\rm MedF}=(V_{\rm MedF}, E_{\rm MedF})$, is constructed from the small components where $V_{\rm MedF}$ is the set of vertices representing the small components and $E_{\rm MedF}$ is the set of weighted edges connecting two small components in $V_{\rm MedF}$.
The weight associated with the edge $e \in E_{\rm MedF}$ $e=(g_i,g_j)$ is the highest mutual similarity $d(u, v)$, $u\in g_i$ and $v\in g_j$, $g_i$ and $g_j$ are two small components. 
Let $T$ be the maximum spanning tree of the graph $G_{\rm MedF}$. 
An edge $e=(g_i,g_j)$ in $T$ is a {\it bridge} if the weights of all of the other edges incident to $g_i$ and $g_j$ are greater than the weight of $e$. 
After removing all the bridges, the resulting connected components will be candidate units. 
Among the candidates, we select the big enough connected components or the small connected components that contain symmetric structures as the units in the Medulla. 
Unit selection was carried out with the help of the visualization tool. 

\begin{figure}
\begin{center}
\begin{tabular}{ |c | c | c| }
\hline 
Group  & Female & Male \\ 
\hline

 0  & 
        \includegraphics[width=6cm]{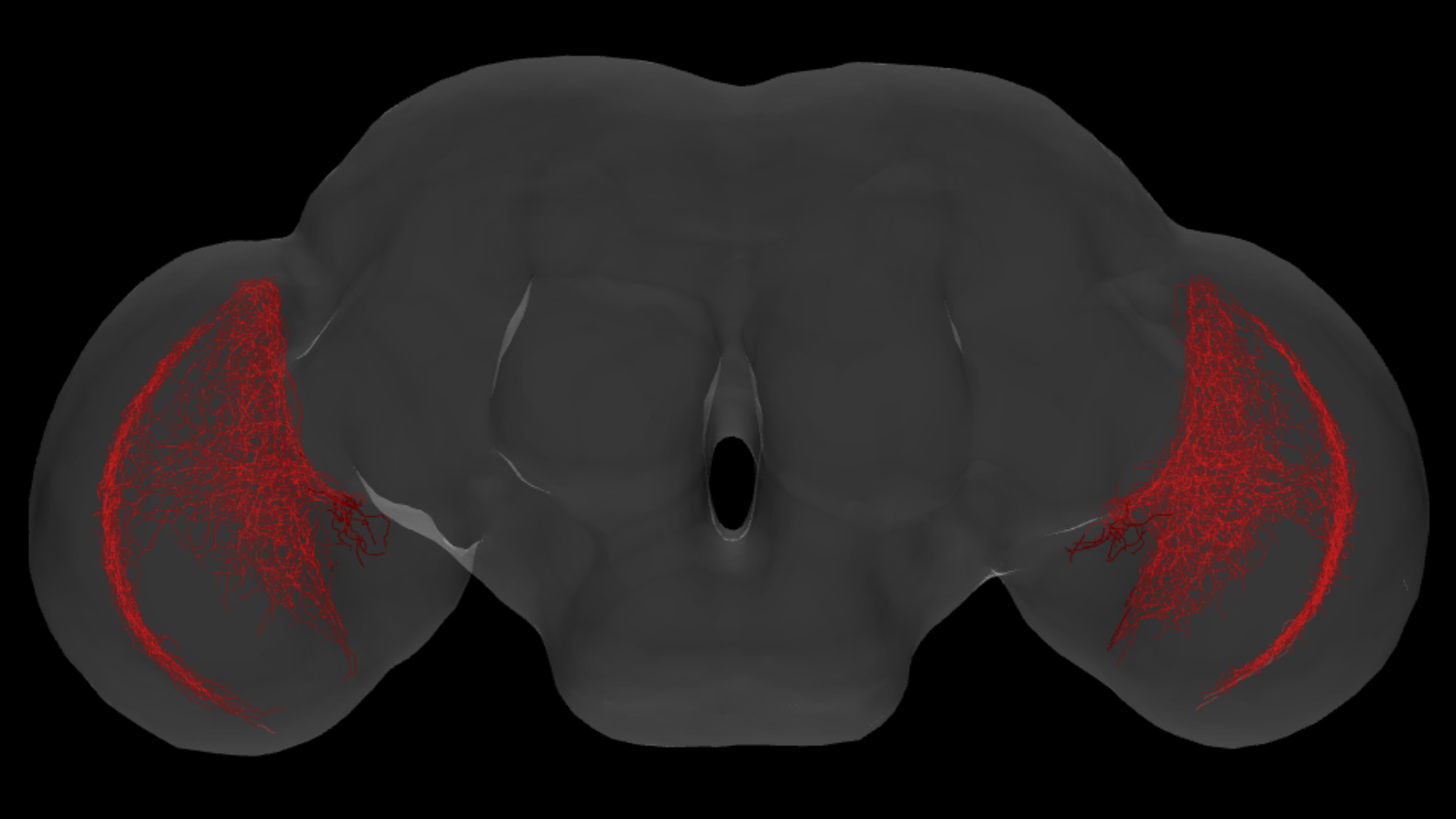} 
    & 
        \includegraphics[width=6cm]{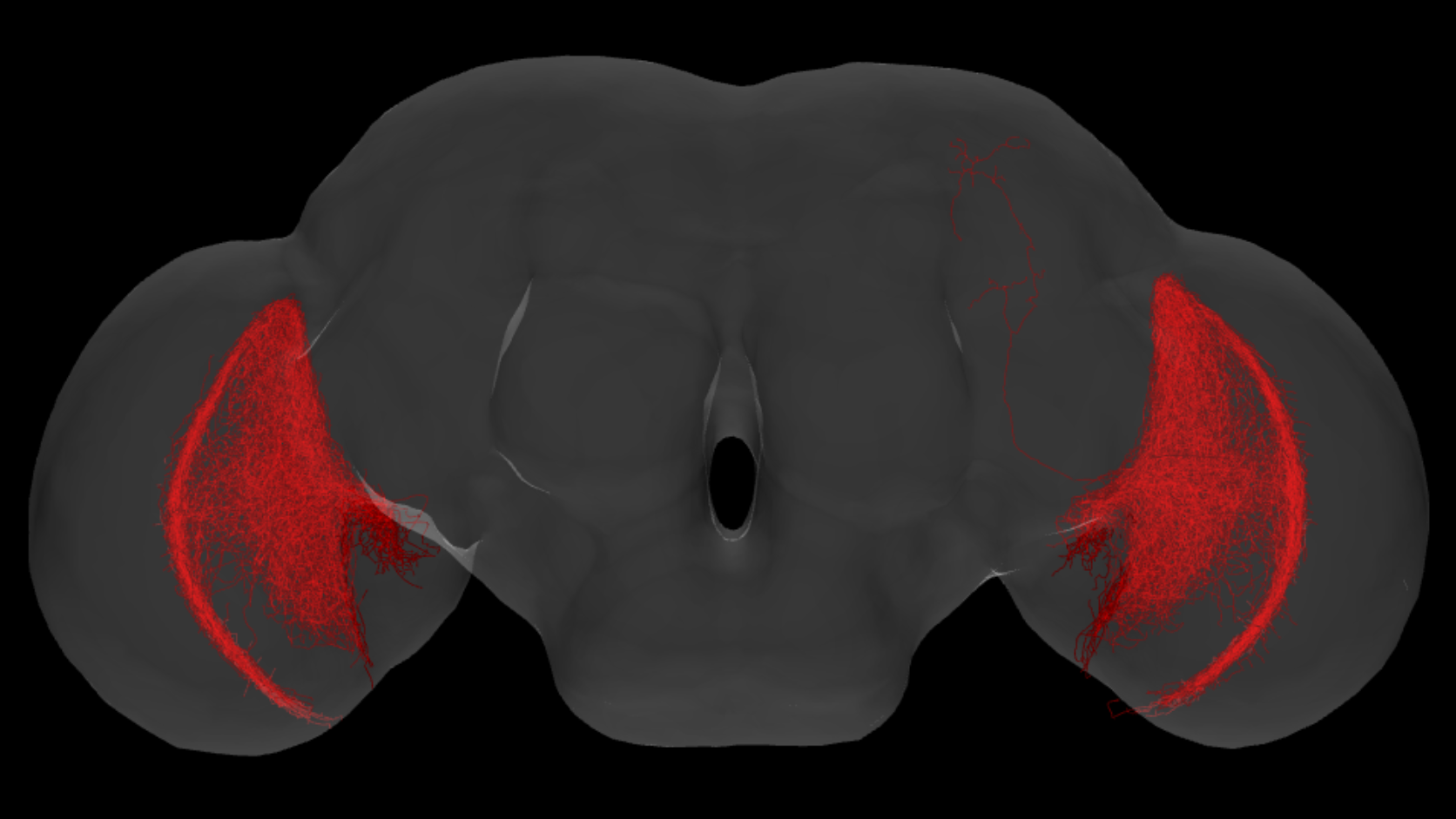} 
    \\ \hline
 1  & 
        \includegraphics[width=6cm]{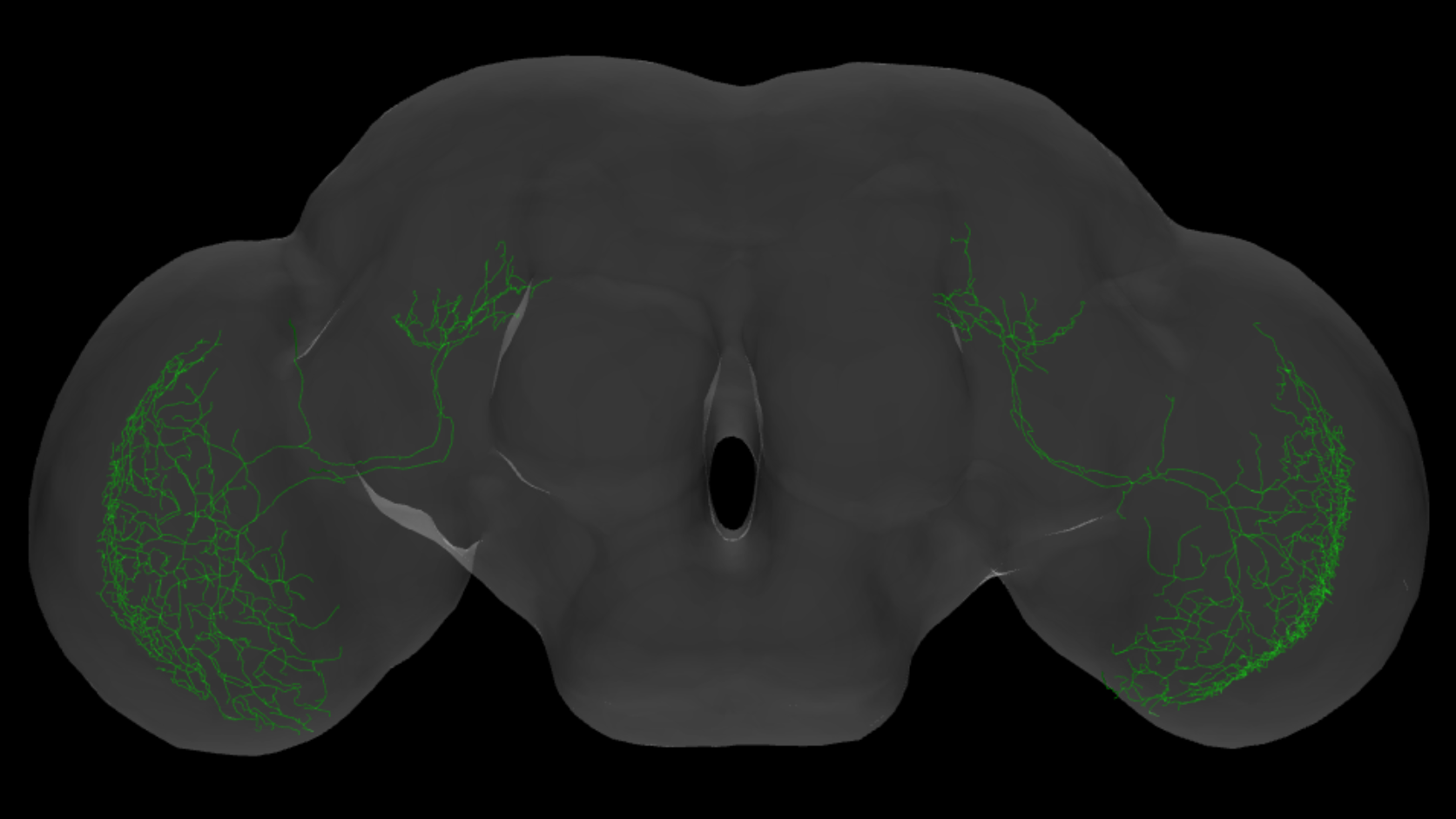} 
    & 
        \includegraphics[width=6cm]{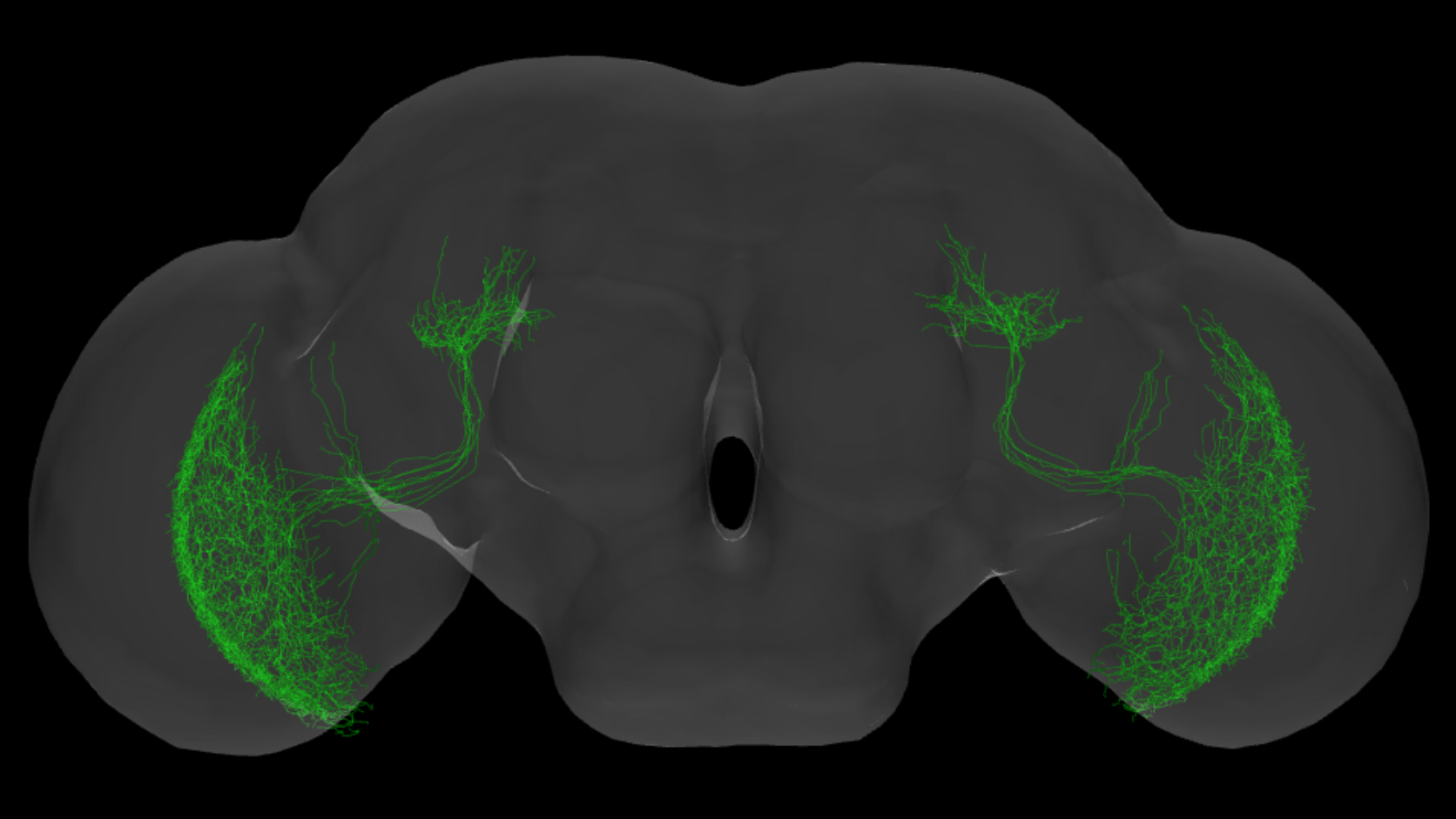} 
    \\ \hline
 2  & 
        \includegraphics[width=6cm]{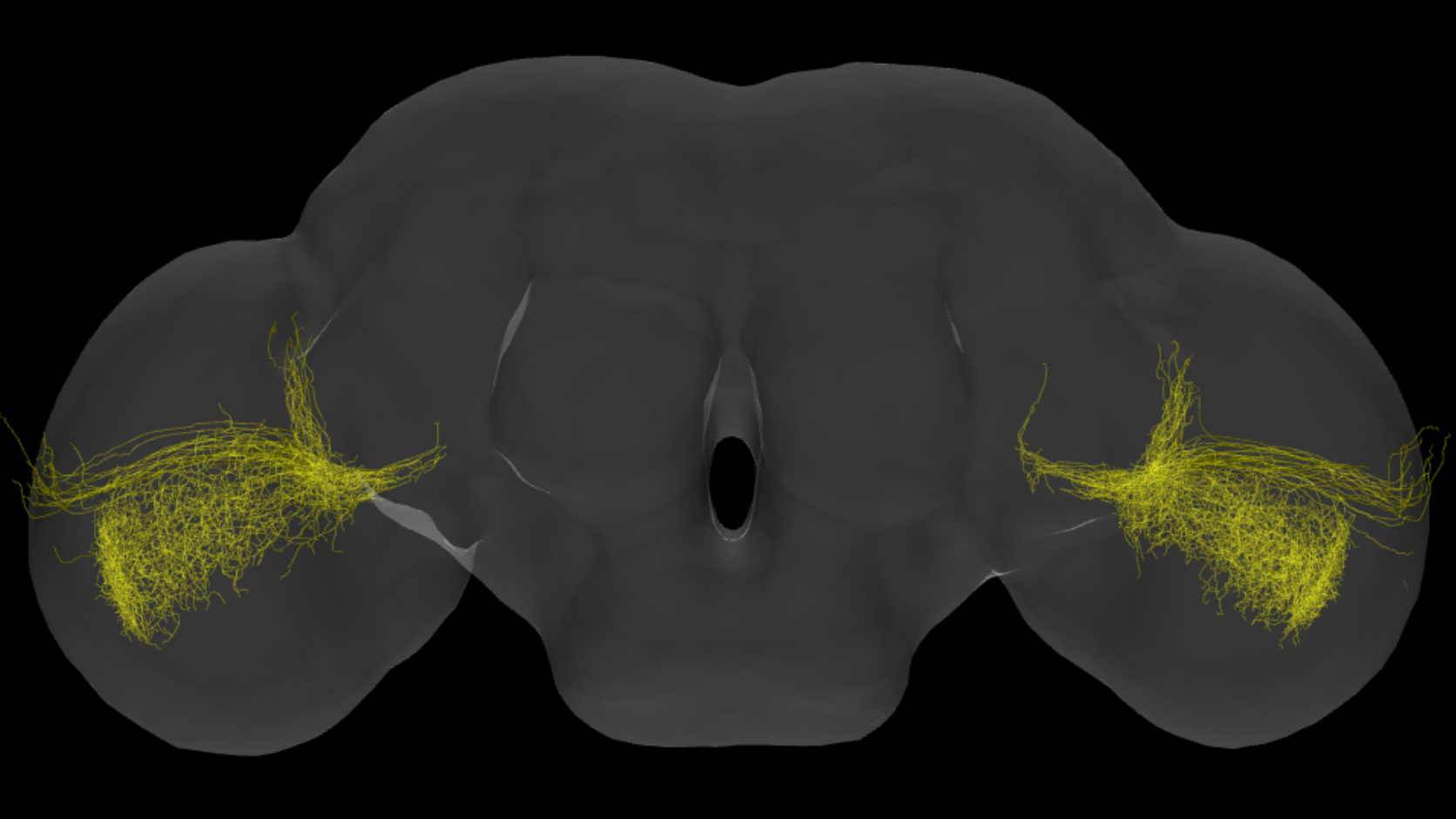} 
    & 
        \includegraphics[width=6cm]{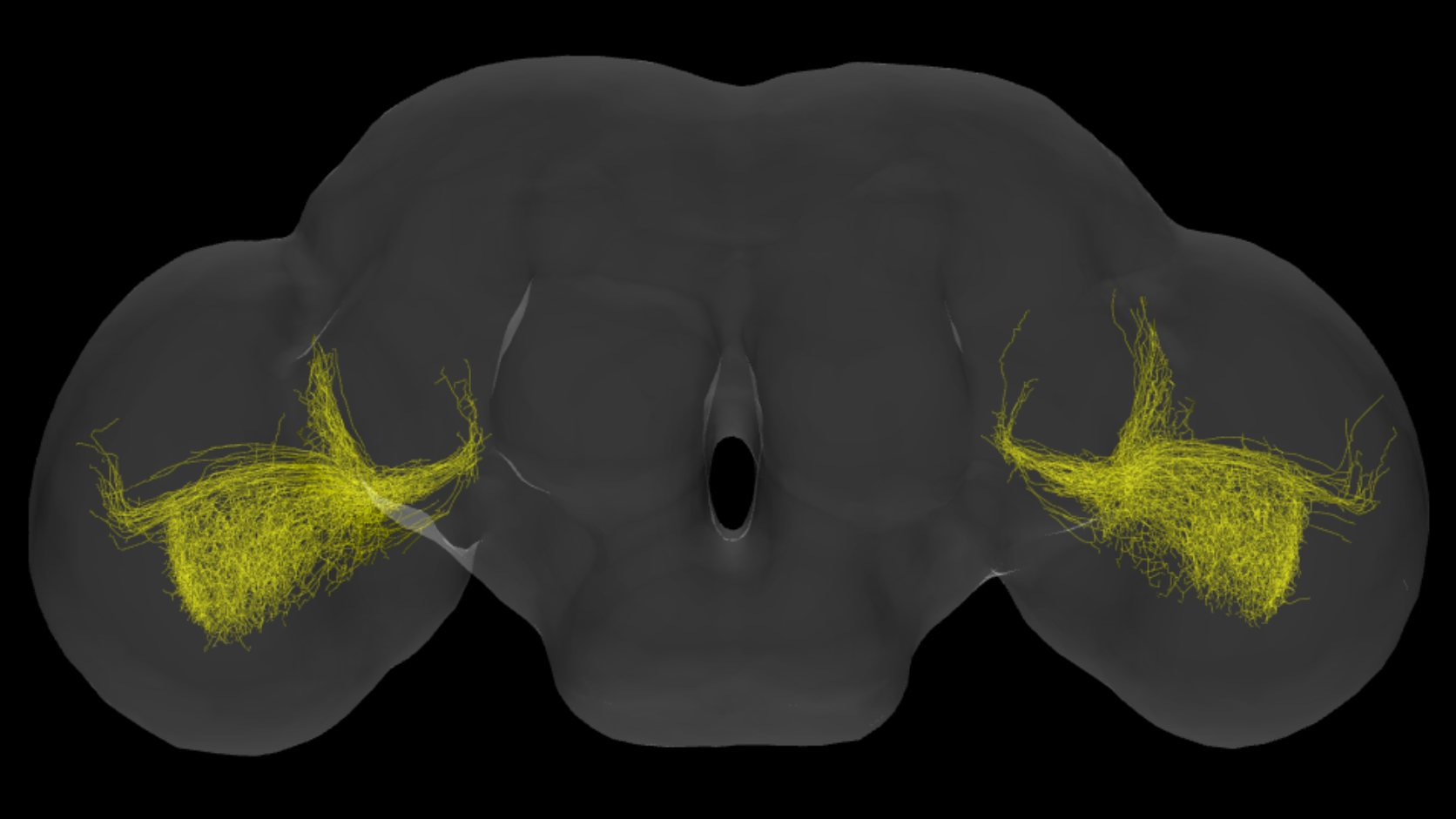} 
    \\ \hline
 3  & 
        \includegraphics[width=6cm]{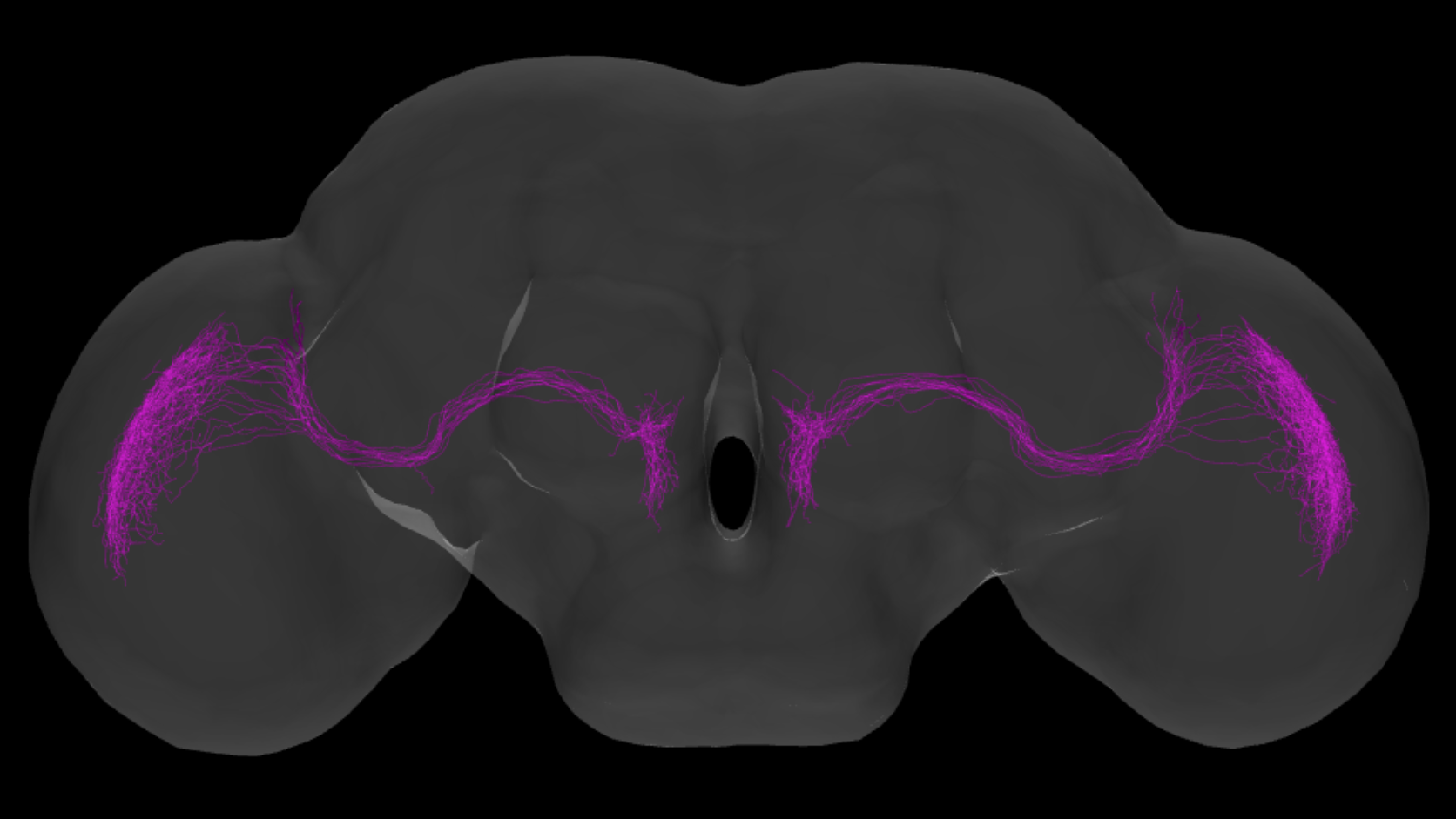} 
    & 
        \includegraphics[width=6cm]{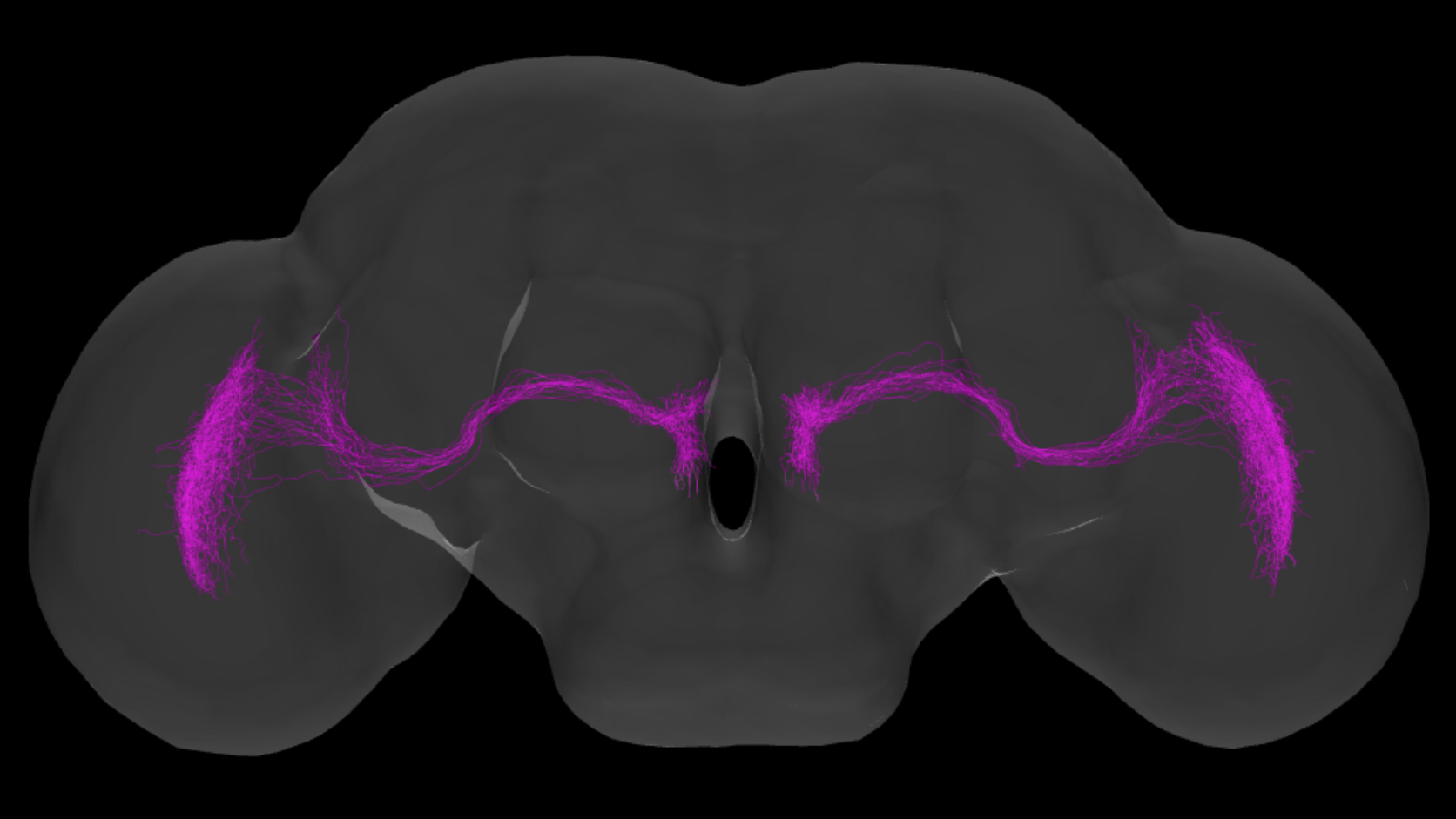} 
    \\ \hline
 4  & 
        \includegraphics[width=6cm]{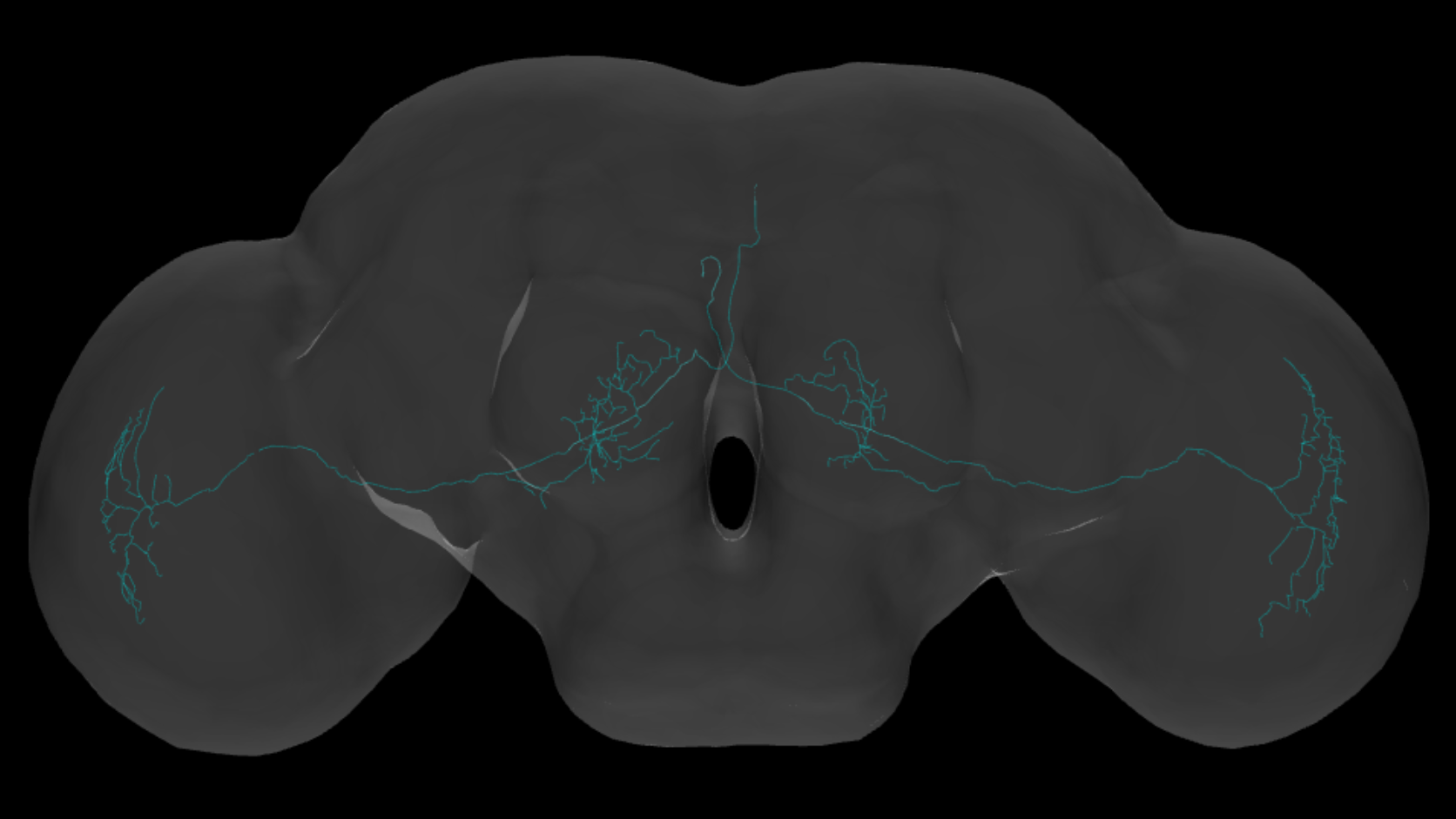} 
    & 
        \includegraphics[width=6cm]{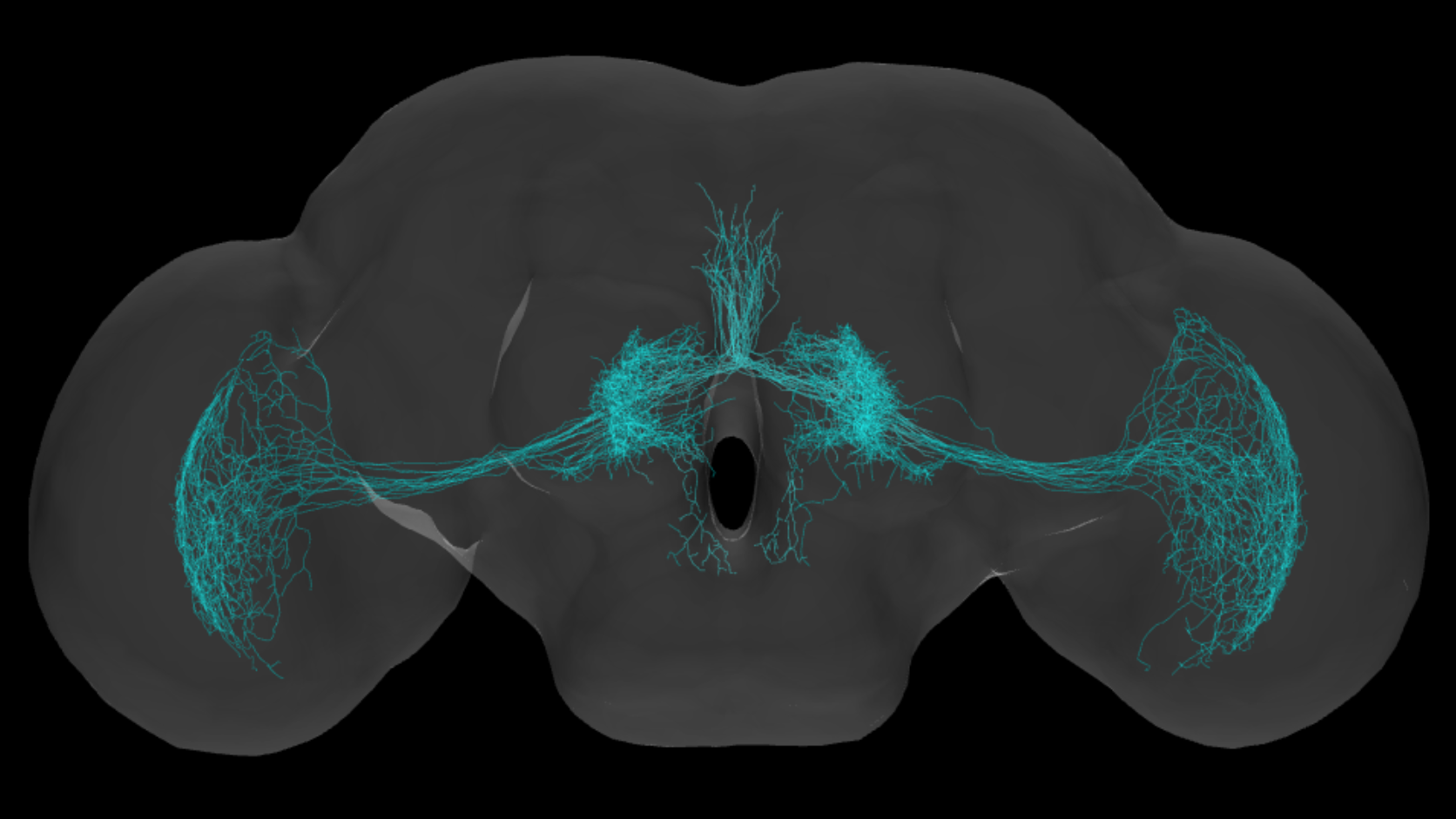} 
    \\ \hline
 5  & 
        \includegraphics[width=6cm]{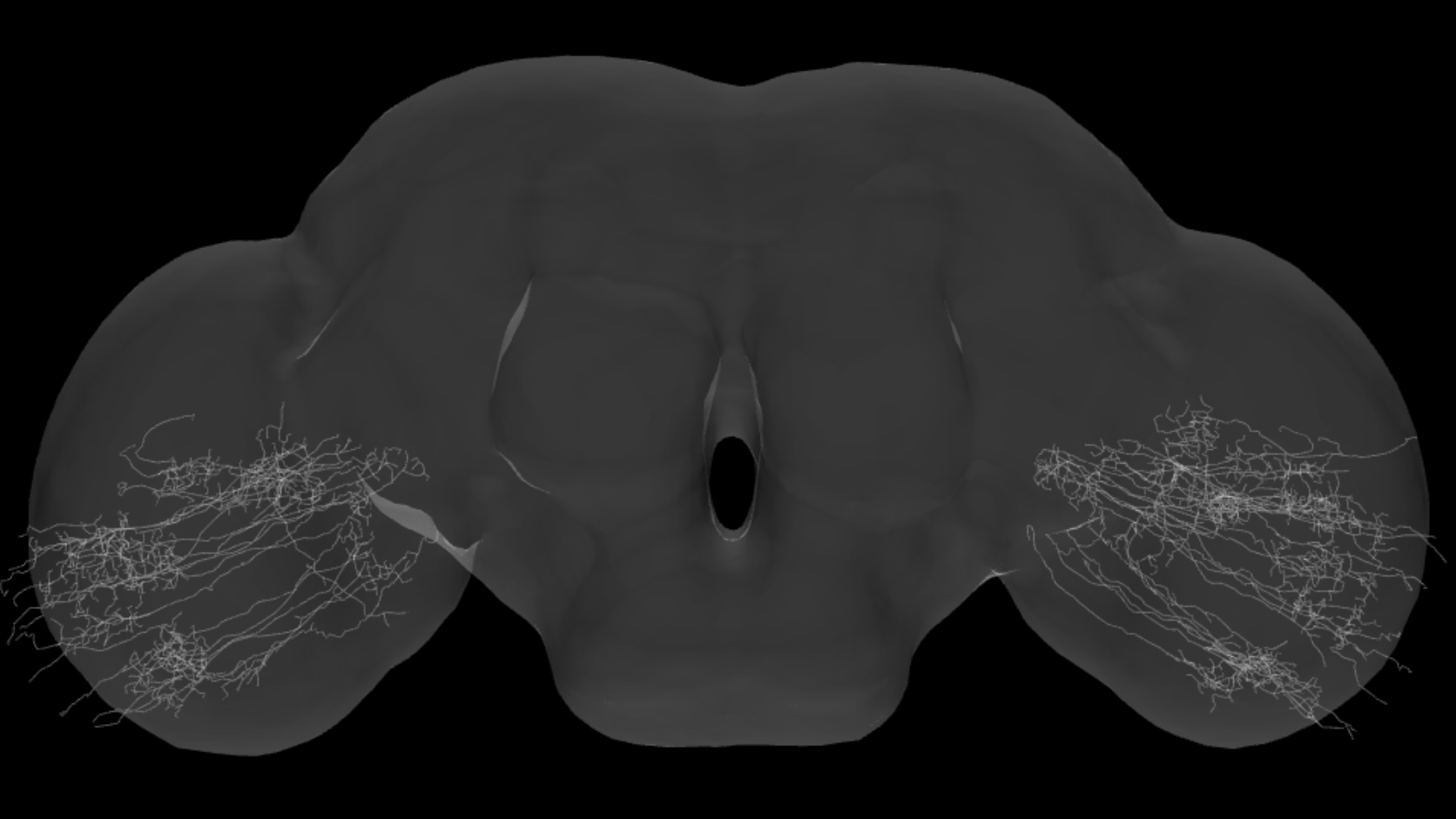} 
    & 
        \includegraphics[width=6cm]{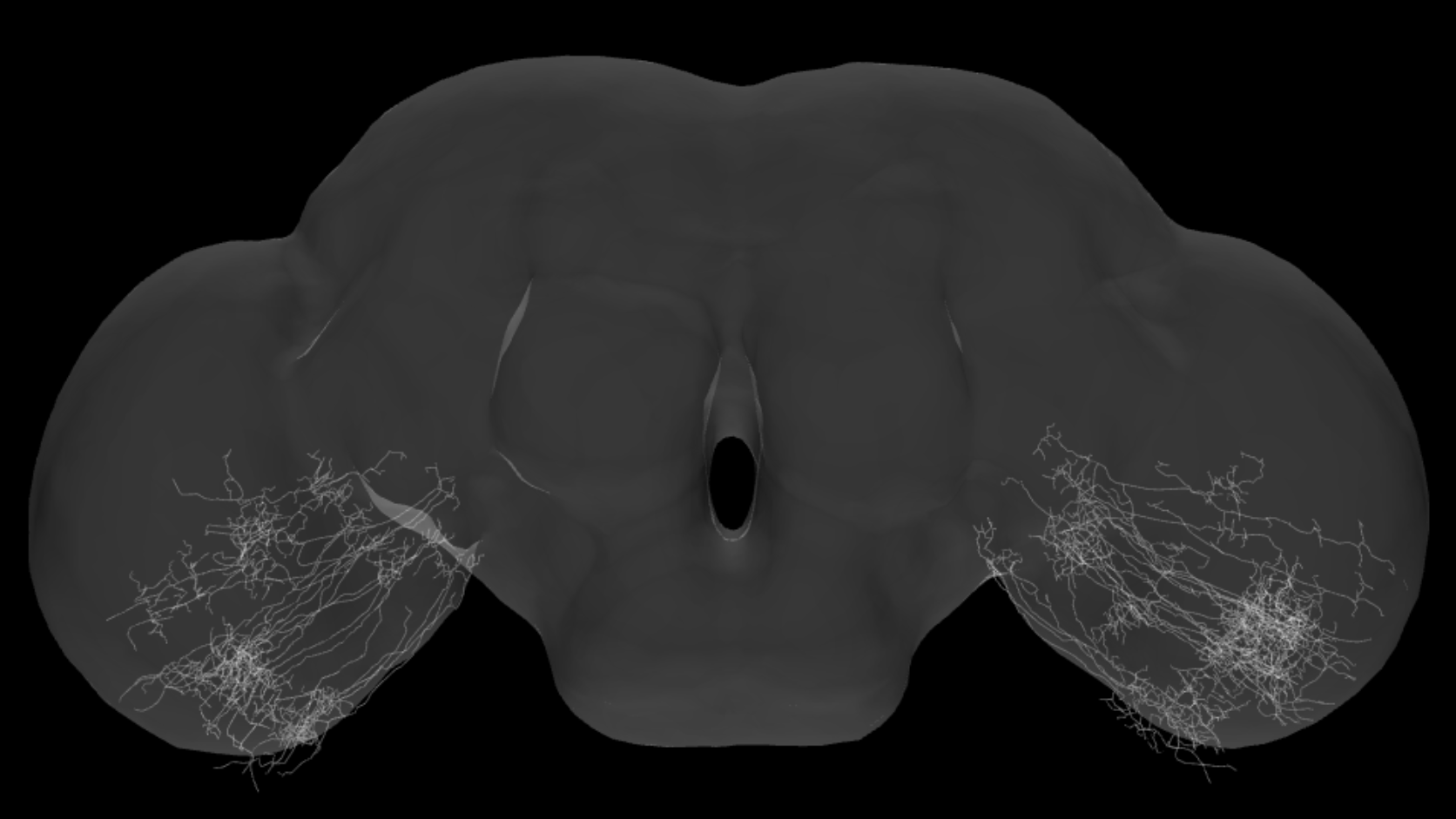} 
    \\ \hline

\end{tabular}
\caption{Comparing the {\it fru} neurons in female and male Medulla. The maximum spanning tree algorithm obtained six main components in female and male Medulal. Each row shows a component. }\label{fruMedFM}
\end{center}
\end{figure}

This almost automatic clustering procedure is applied to the three sets of neurons $S_{\rm MedF}$, $S_{\rm MedFfru}$, and $S_{\rm MedMfru}$.
Since there is no ground truth for the clustering results validation, we compare the results obtained to the sets $S_{\rm MedFfru}$ and $S_{\rm MedMfru}$. 
Six major components for both female and male {\it fru} neurons (Figure~\ref{fruMedFM}) were obtained. 
In Figure~\ref{fruMedFM}, each row presents the same component in both female and male Medulla.  
Since Male flies have smaller brains, the same structures of different genders are mapped to different locations in the female-typical brain. 
However, we get the same components from processing the neurons of different genders' brains.  
We are confident that the results obtained using the maximum spanning tree algorithm are reliable. 

\begin{figure}
\centering
    \includegraphics[width=12cm]{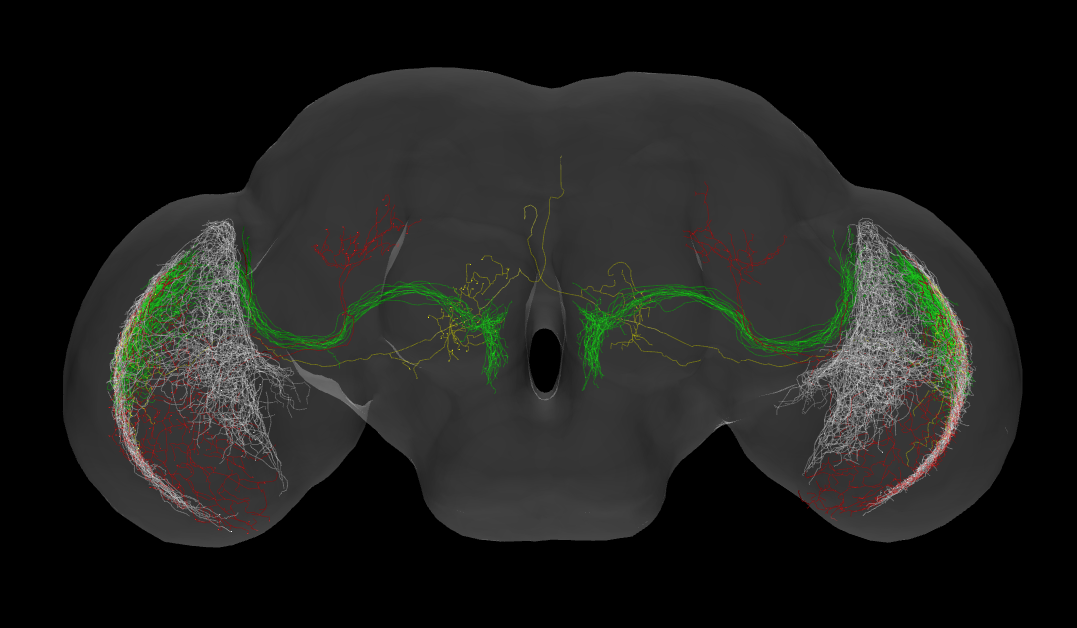} \\
    (a) The Base Layer of {\it fru} neurons in the female Medulla. The four components in different colors are rendered in the same image. \\
    \includegraphics[width=12cm]{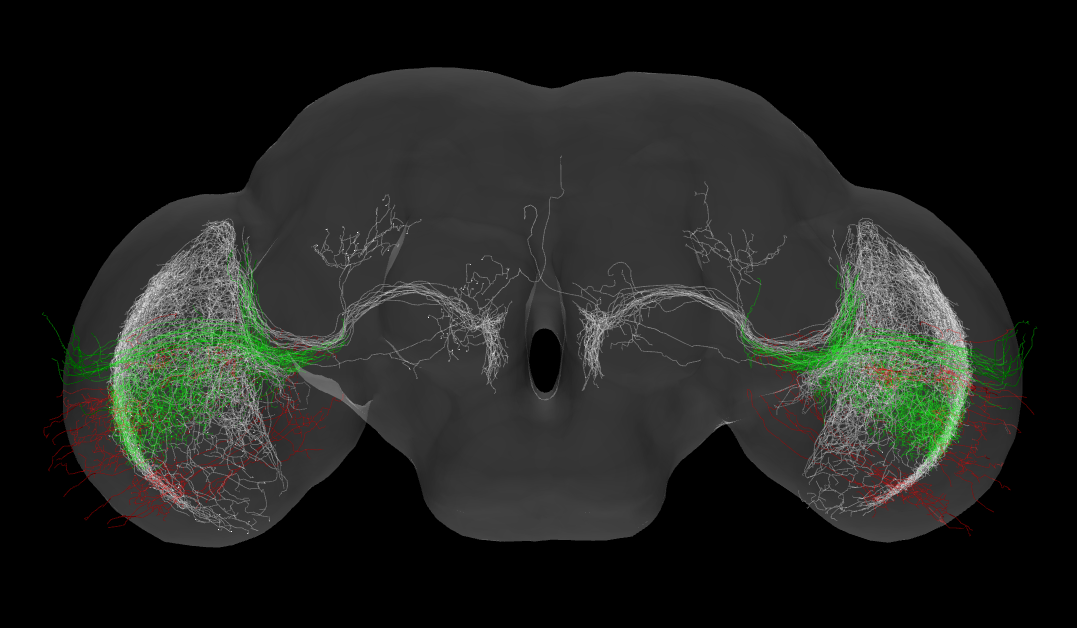} \\
    (b) The Base Layer (white), the Fan-Shape layer (green), and the Via (red) in the same image. \\
    \includegraphics[width=12cm]{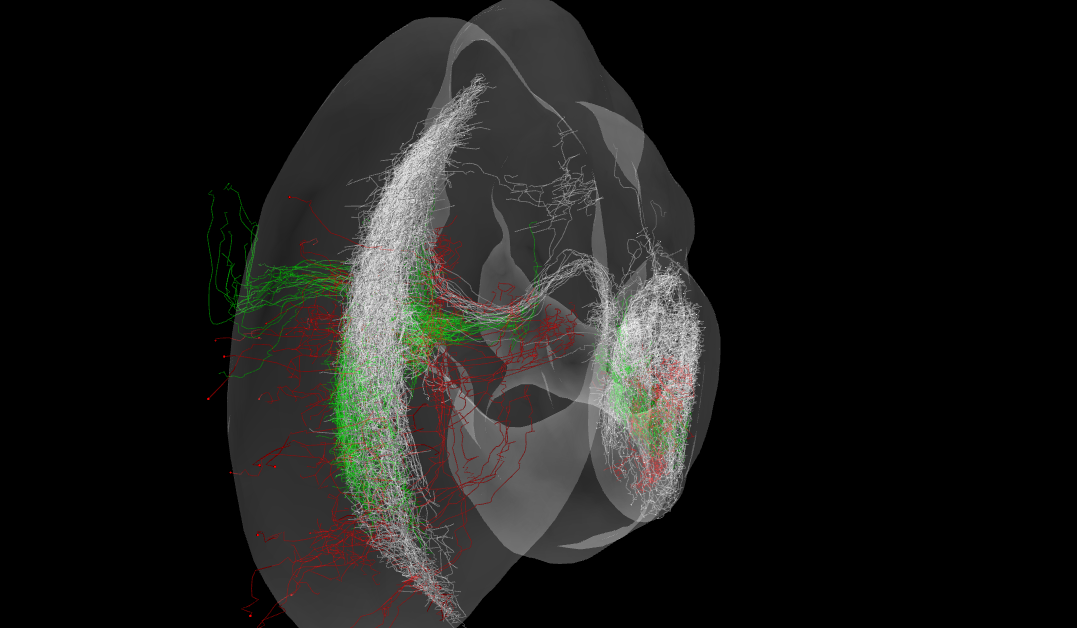} \\
    (c) Different view of the image in (b). 
    \caption{The structure of the {\it fru} neurons in the female fly Medulla. }
    \label{fruFemaleMedulla}
\end{figure}

Only the female neurons (including the female {\it fru} neurons) are used to analyze the structure of the Medulla. 
To describe the Medulla, we start with the smaller and simpler structure of the {\it fru} neuron components. 
In Figure~\ref{fruMedFM} the Female column, rows 0 (red), 1 (green), 3 (magenta), and 4 (blue) are the ``Base-Layer''. 
Row 2 is the ``Fan-Layer'' because the neurons in this structure cover a fan-shaped area of the Base-Layer, and some other fibers form a thin tract that looks like the {\it equator} of the Medulla. 
Row 5 is the ``Via'' because neurons in this structure pass through the Base-Layer and the ``Two-Ply components'' (shall be shown later). 
In Figure~\ref{fruFemaleMedulla} (a) the four components of the Base-Layer are shown in 4 colors. 
Figure~\ref{fruFemaleMedulla} (b) has all the components in one image, where the Base-Layer is white, Fan-Layer is green, and the Via is red. 
Figure~\ref{fruFemaleMedulla} (c) is from the other view direction to view the three components.  

\begin{figure}
\begin{center}
\begin{tabular}{ | c | c| }
\hline 
        \includegraphics[width=7cm]{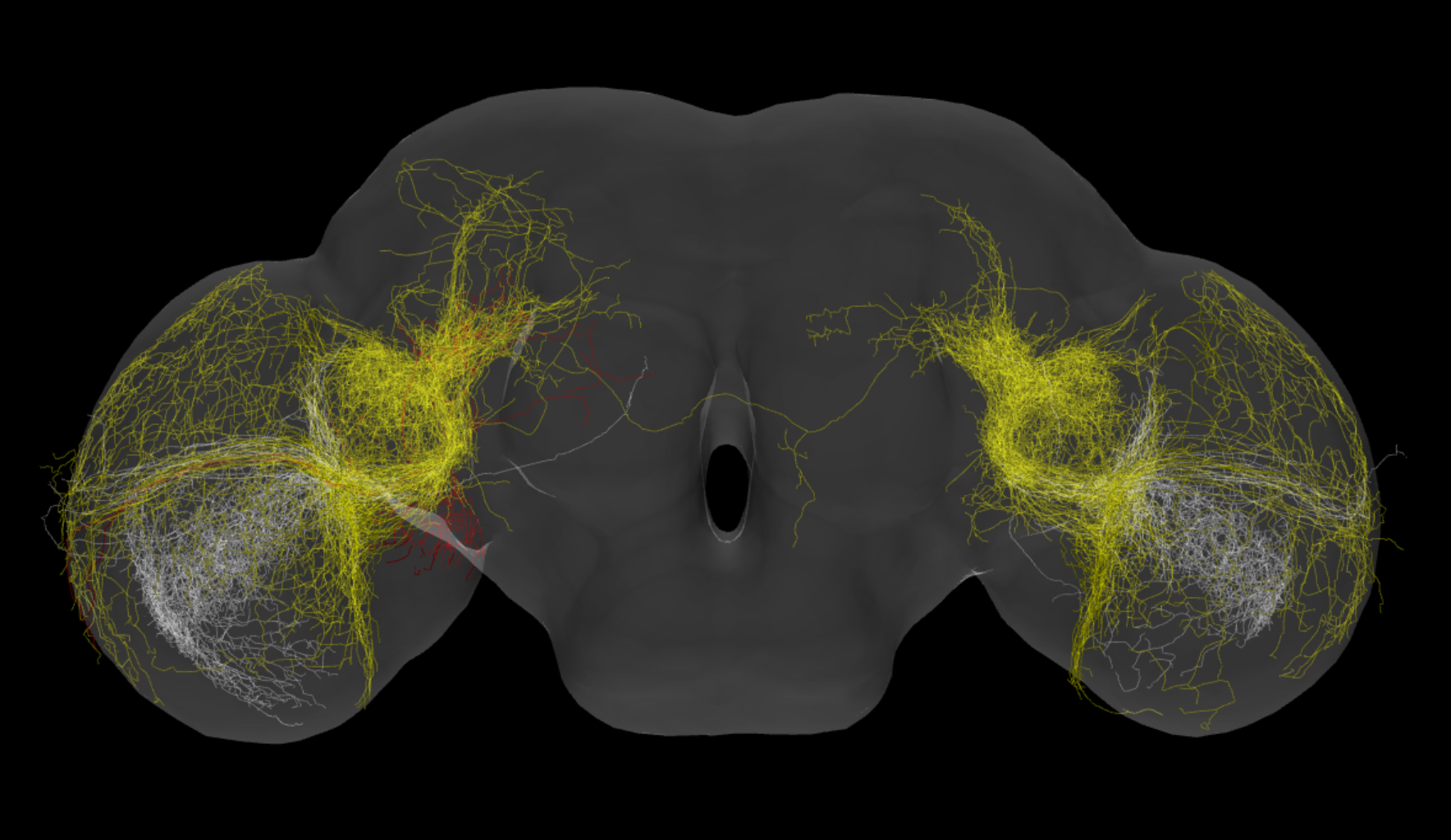} 
    &
        \includegraphics[width=7cm]{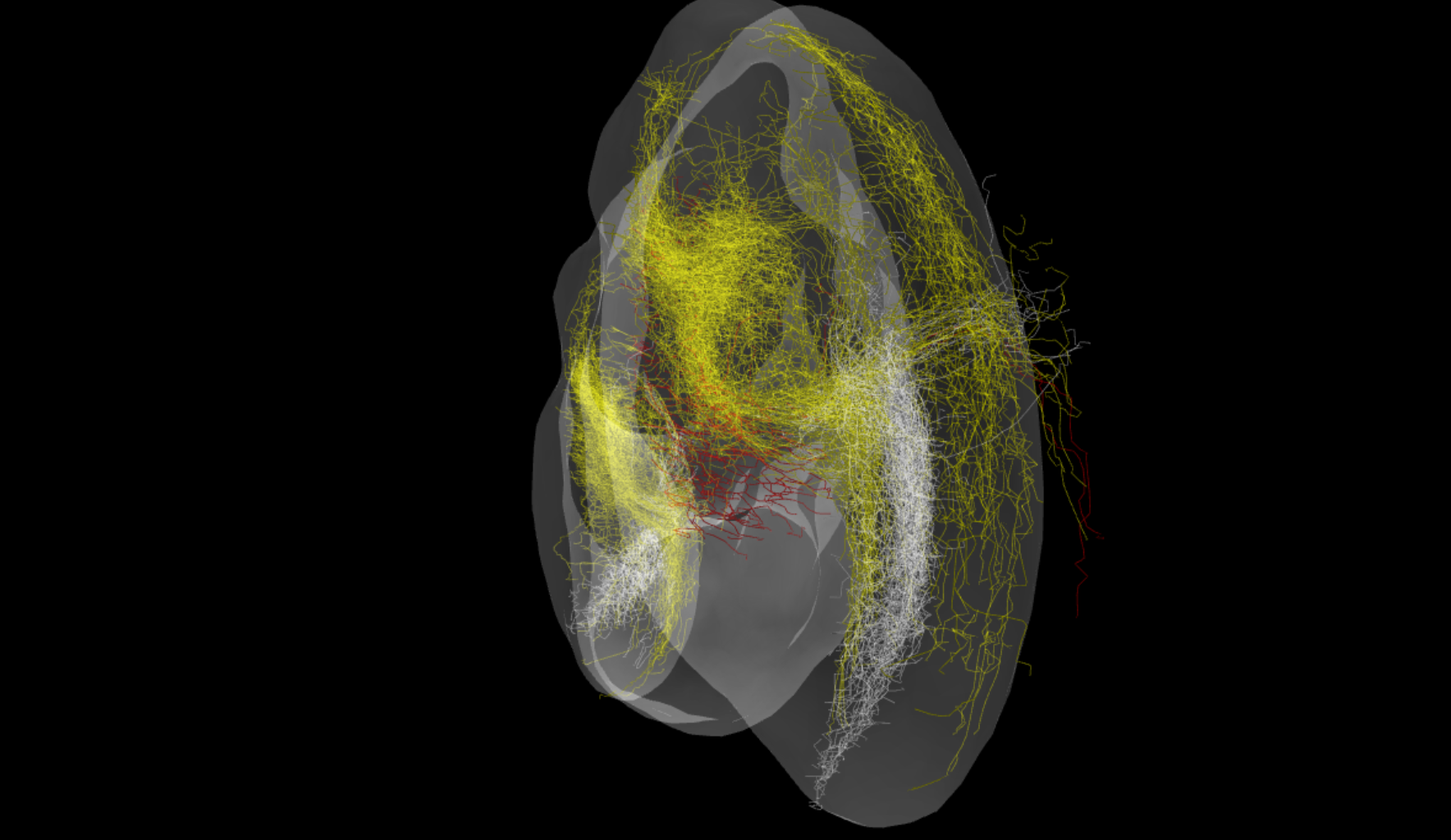} \\

        \includegraphics[width=7cm]{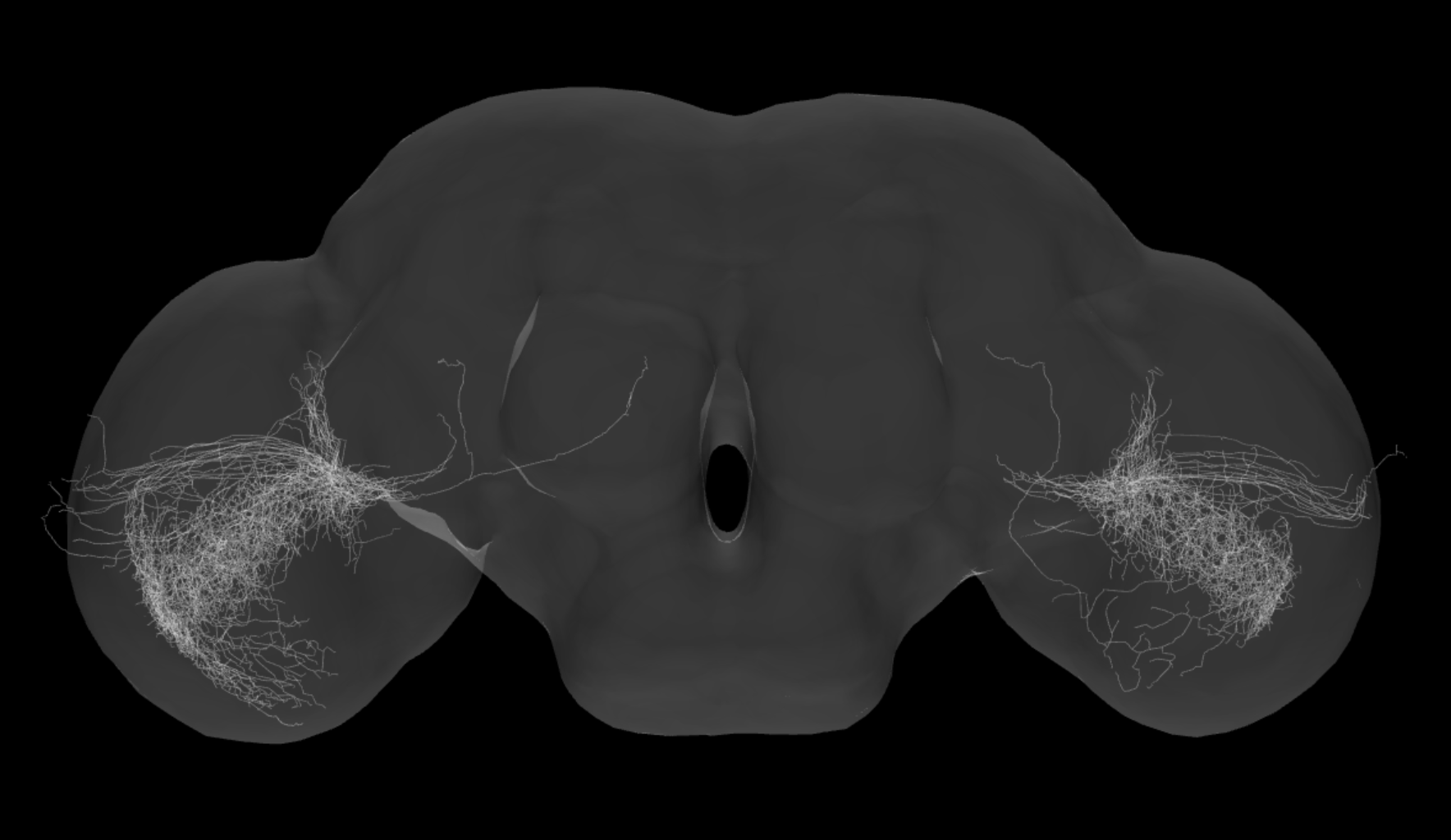} 
    & 
        \includegraphics[width=7cm]{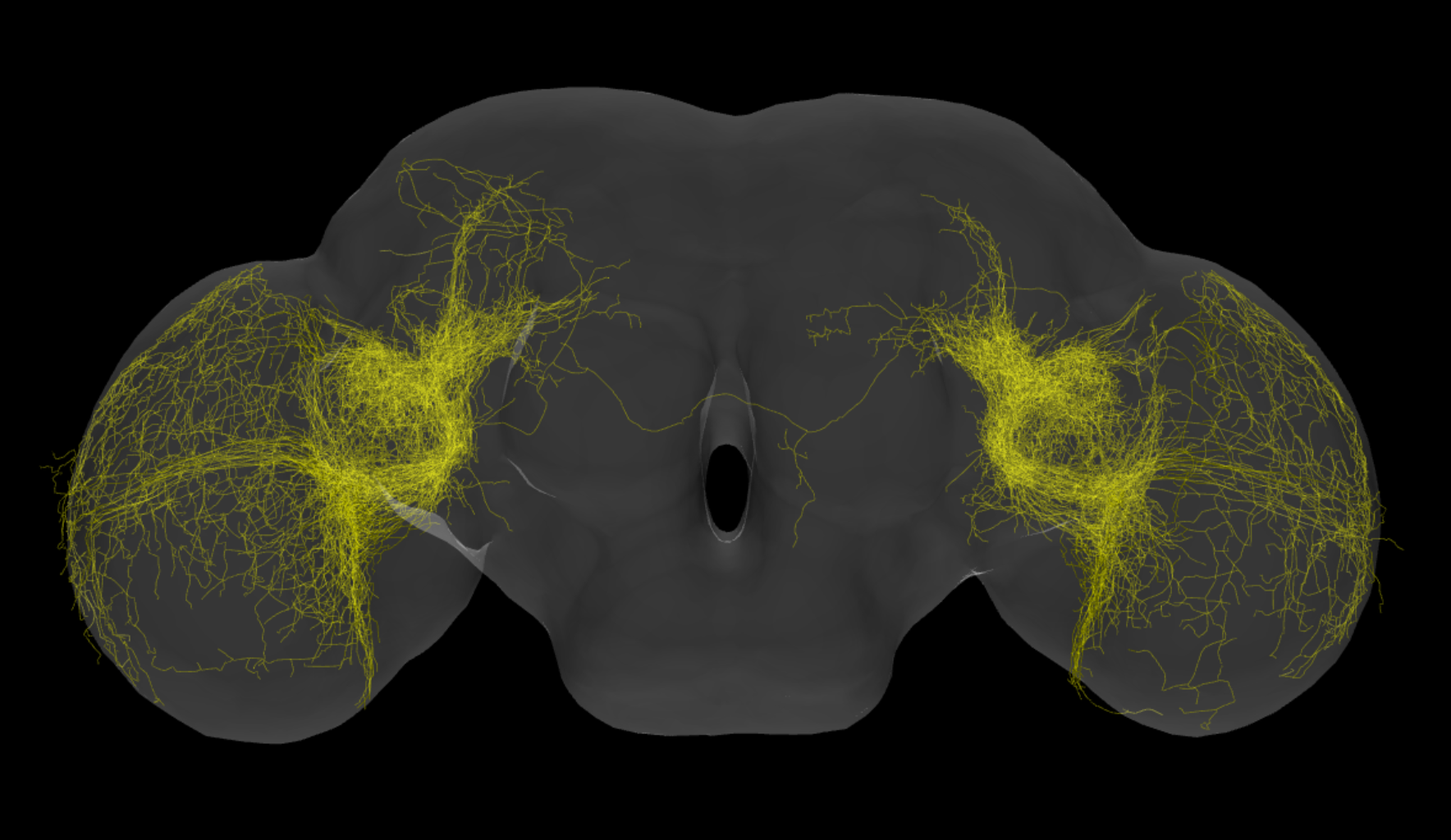} 
    \\ \hline
        \includegraphics[width=7cm]{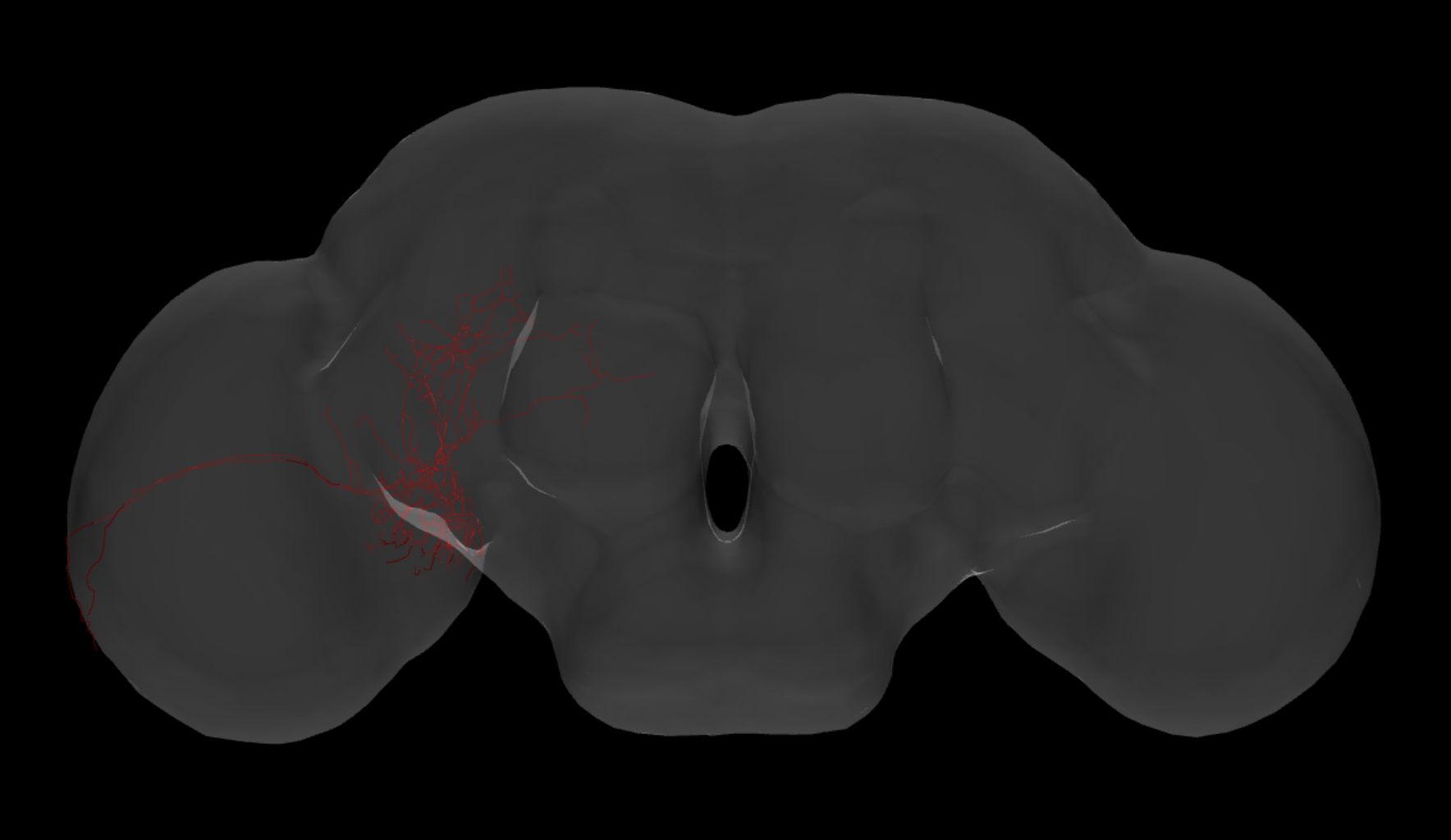} 
    & 
    \\ \hline

\end{tabular}
\caption{Fan-shaped structure groups of the neurons in Medulla of female fly brain. }\label{MedFFan}

\end{center}
\end{figure}

The structure of the Medulla is more complex than the structure of {\it fru} neurons in Medulla.
There are the Base-Layer, the Fan-Layer, the Two-Ply, a set of Via neurons, and many small components distributed over some layers. 
The Fan-Shape Layer is shown in Figure~\ref{MedFFan}. 
Two images in the first row are the 3 components obtained from the Maximum Spanning Tree algorithm.
Each of the three components is shown in the second (white and yellow) and the third row (red).  
The red one on the third row shows only the ``equator''. 
The white one is the same as the Fan-Shape Layer of the {\it fru} neuron. 
The yellow one covers a larger area of the Medulla. 
Note that the white and the yellow ones do not share the same layer (the image on the right of the first row) but their ``equators'' are coincide with the red one. 

\begin{figure}
\begin{center}
\begin{tabular}{ | c | c| }
\hline 
        \includegraphics[width=7cm]{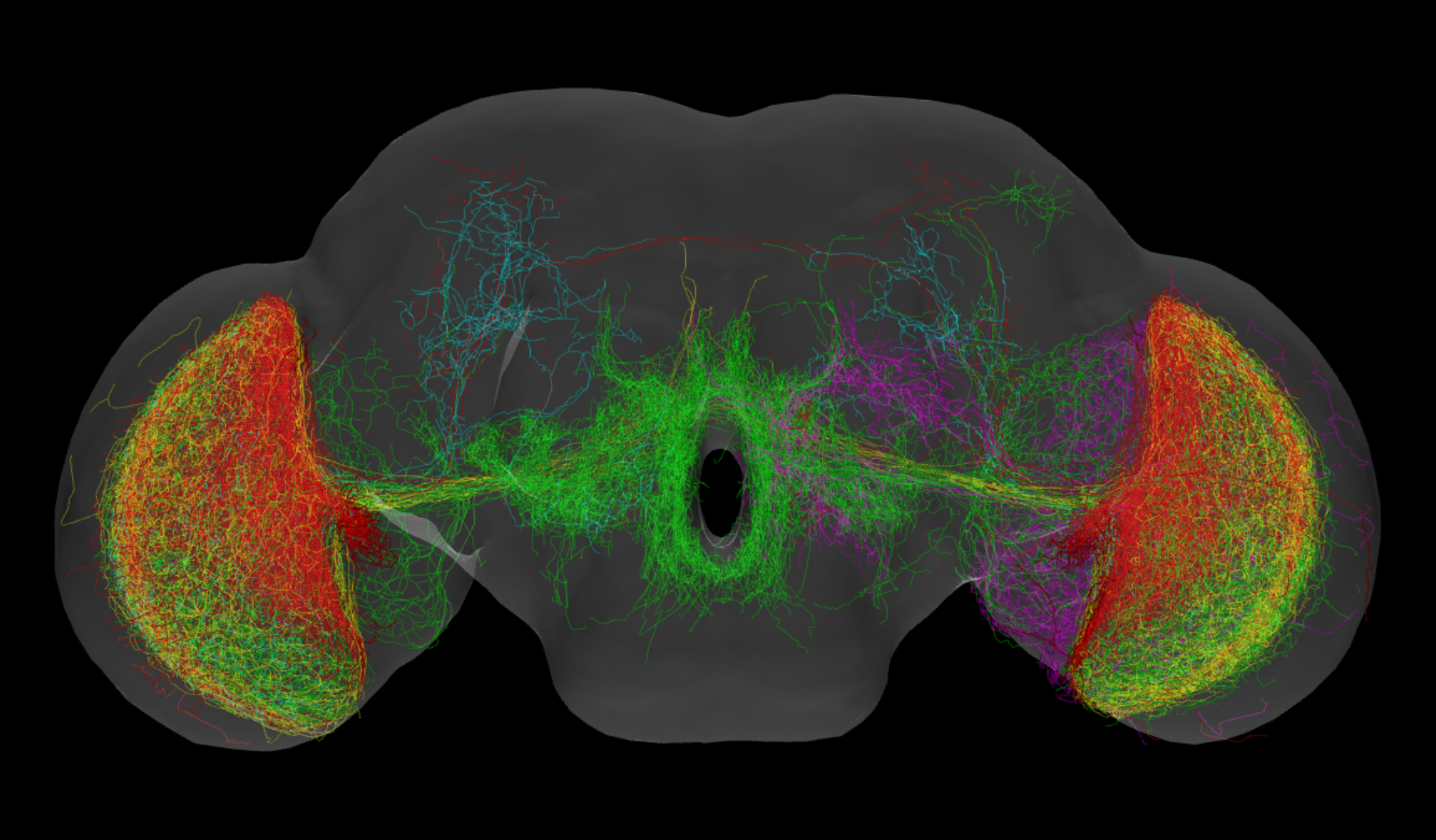} 
    &
        \includegraphics[width=7cm]{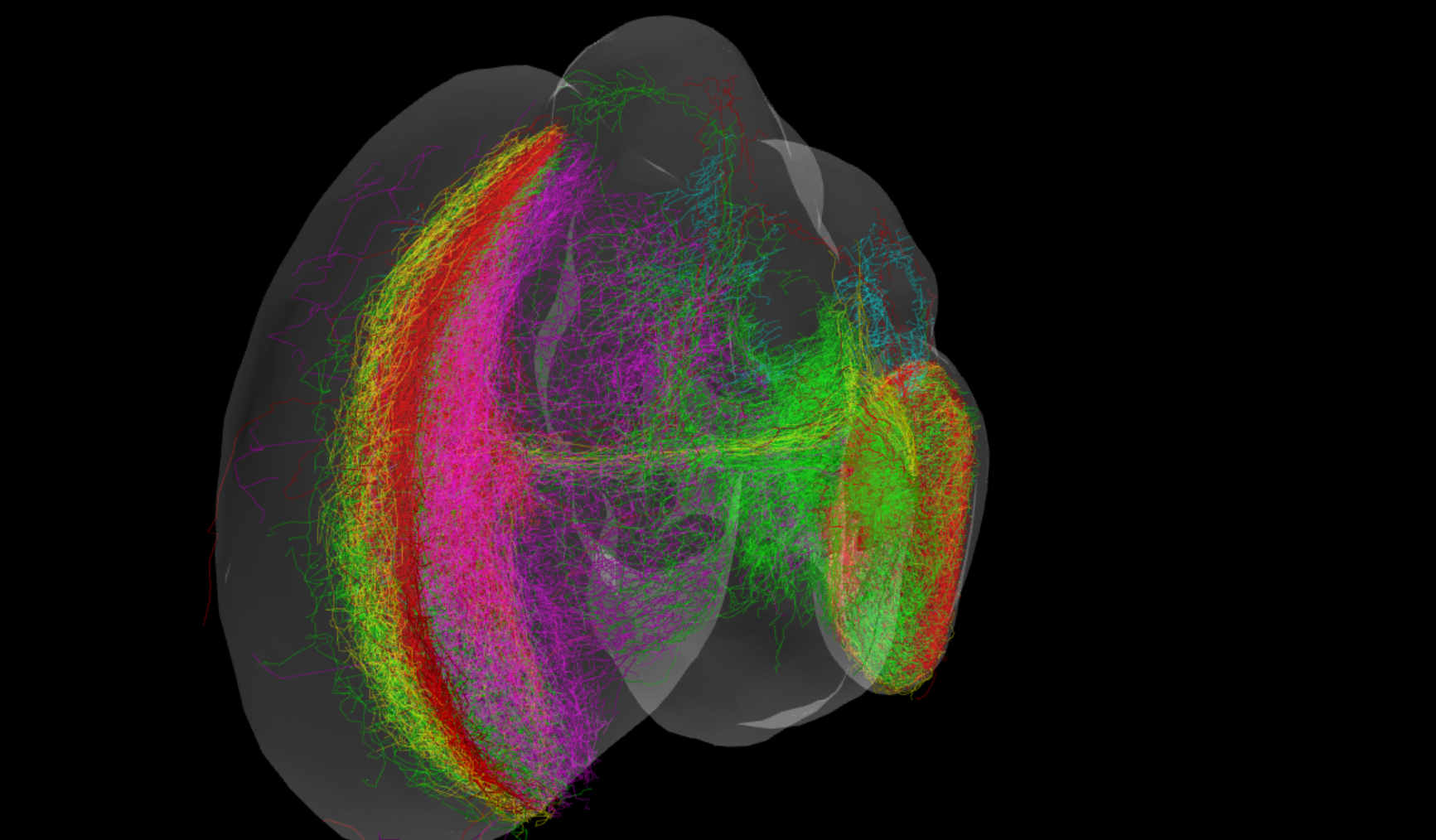} \\

        \includegraphics[width=7cm]{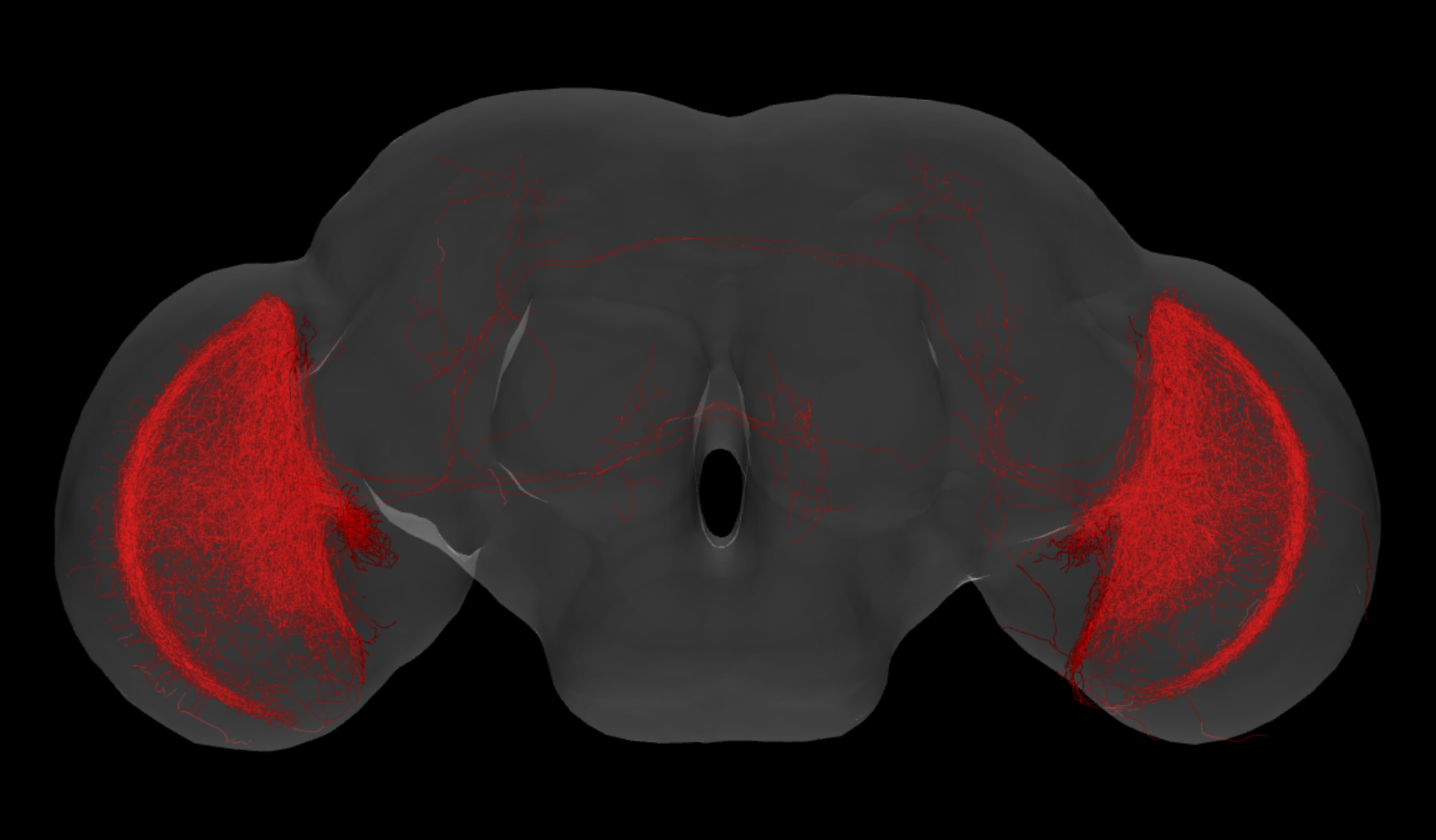} 
    & 
        \includegraphics[width=7cm]{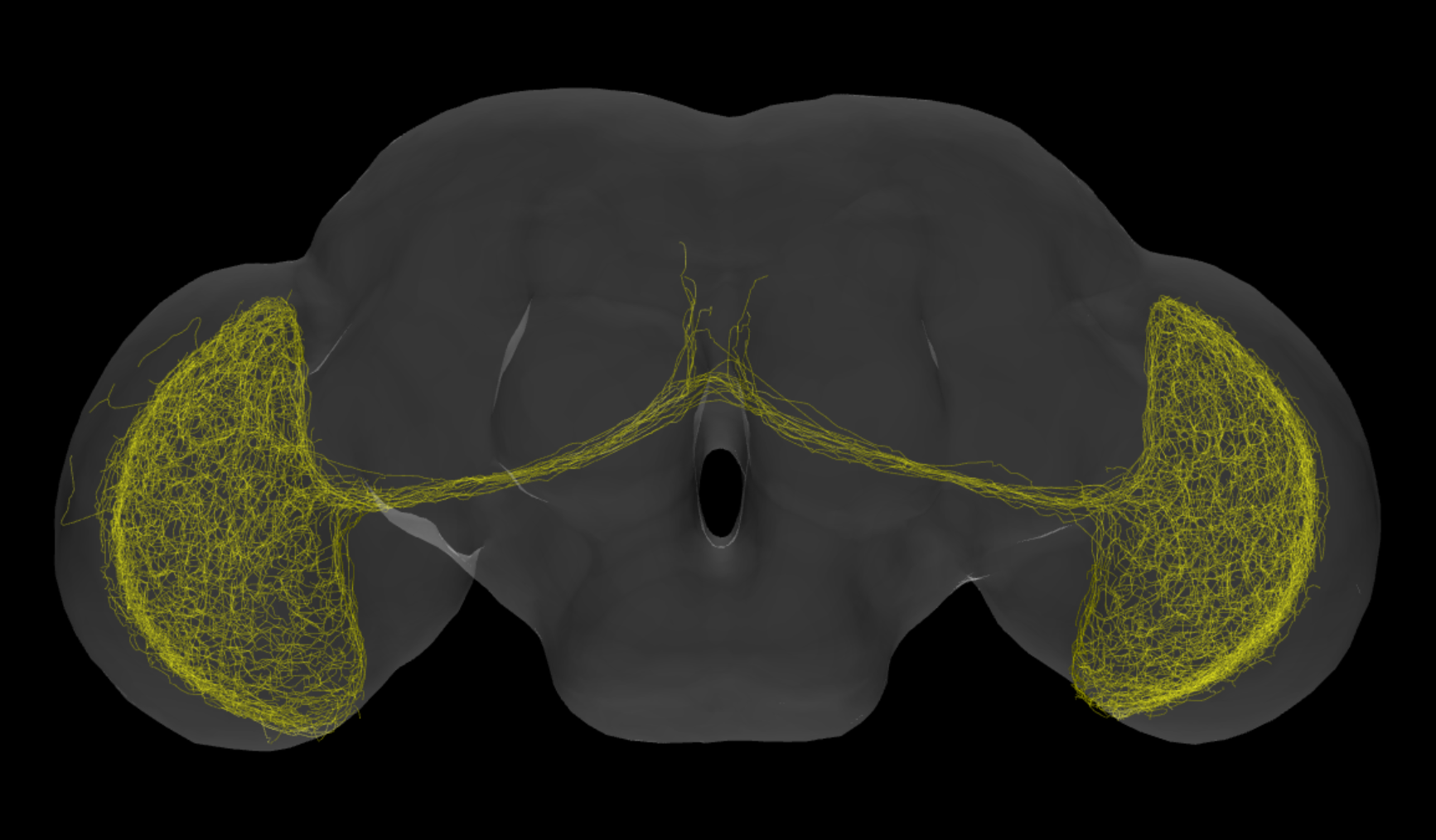} \\
     \hline
 
        \includegraphics[width=7cm]{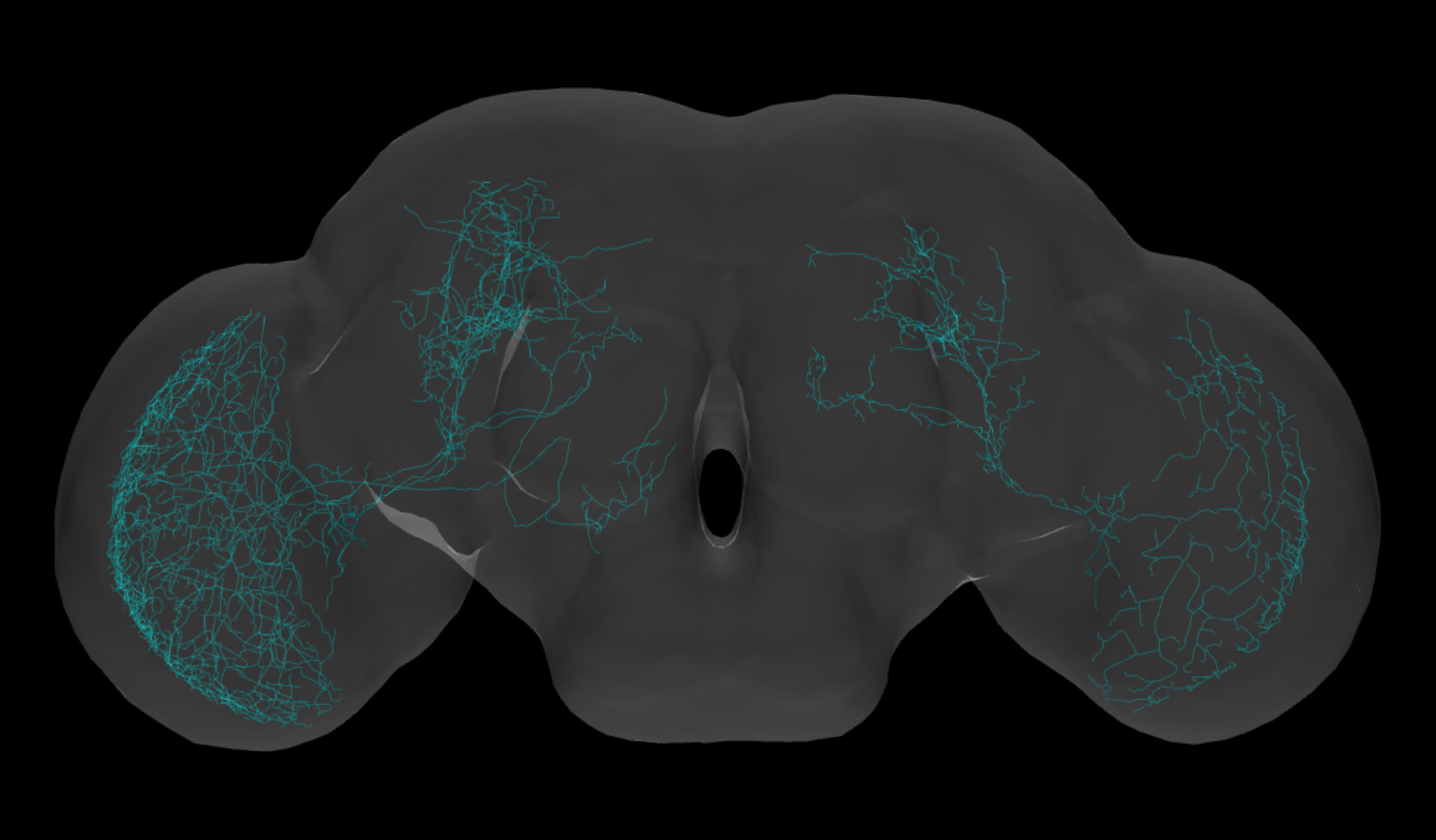} 
    & 
        \includegraphics[width=7cm]{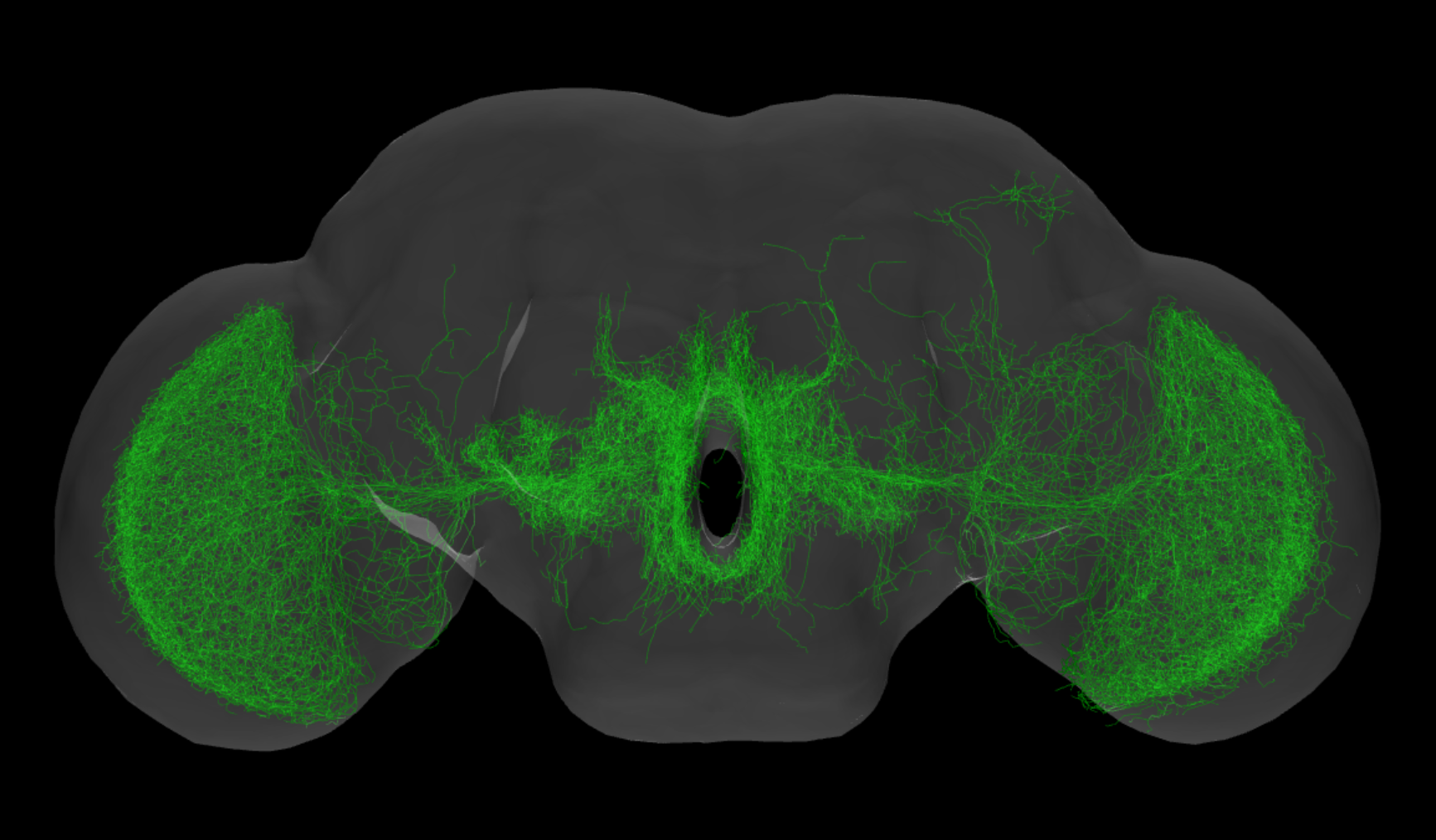} \\
     \hline
 
        \includegraphics[width=7cm]{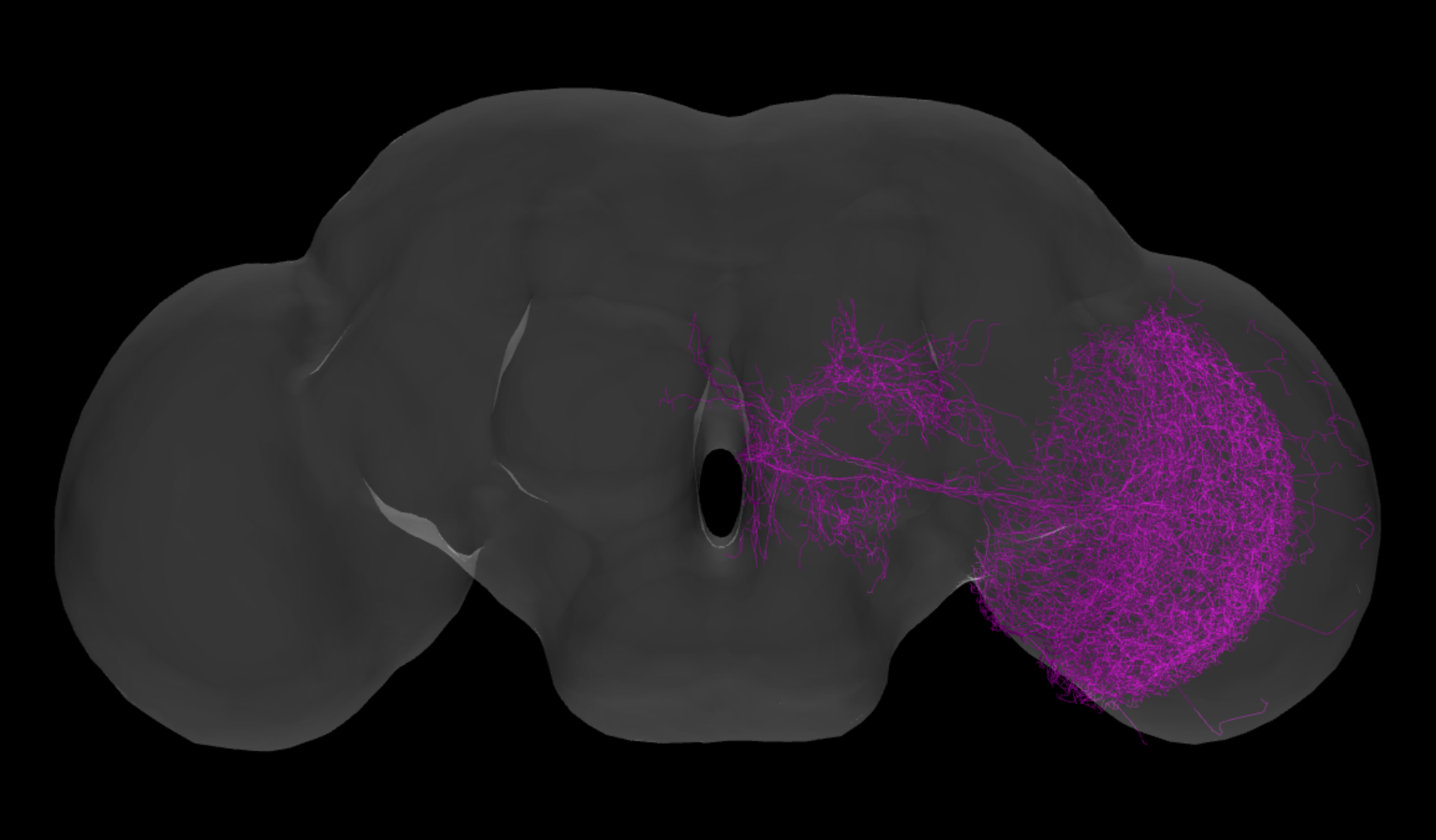} 
    & 
         \\
     \hline

    \\ \hline

\end{tabular}
\caption{Base-layer components in the Medulla of the female fly brain. }\label{MedBaseLayer}
\end{center}
\end{figure}

\begin{figure}
    \centering
    \includegraphics[width=5cm]{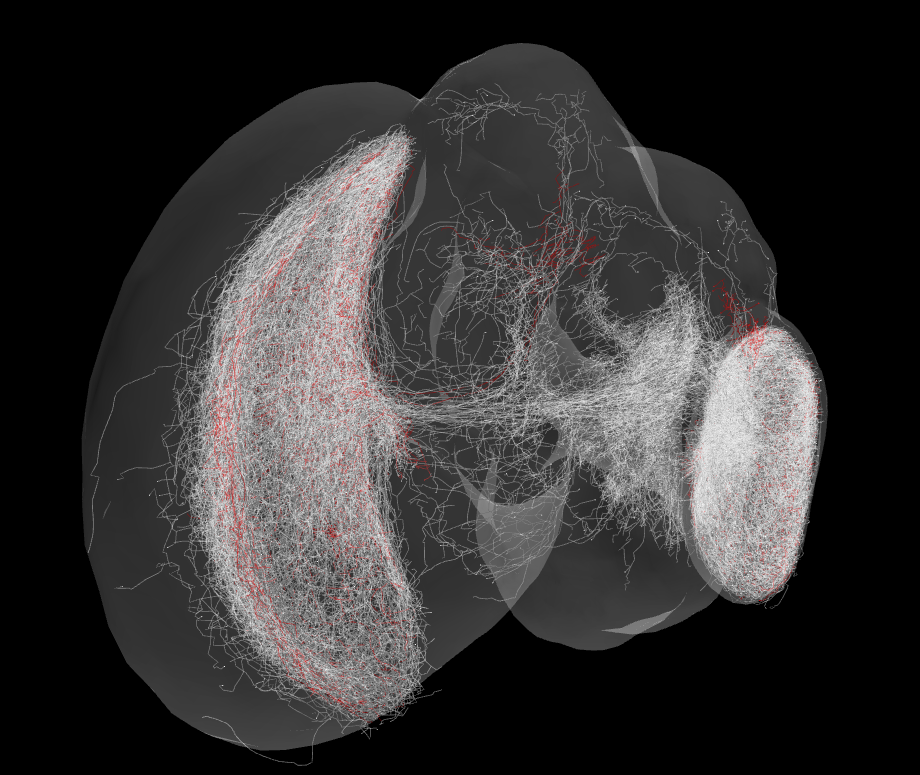}
    \caption{The Base-Layer of the {\it fru} neurons (red) is a part of the Base-Layer of the female non-{\it fru} neurons (white). }
    \label{fruOnBaseLayer}
\end{figure}

Five components were found for the Base-Layer in Medulla.
The five components are shown in the first row in Figure~\ref{MedBaseLayer} in different colors, the picture on the left is the front view and the image on the right is the side view. 
It seems the components form sub-layer of the Base-Layer.
Each component is shown in the following rows in Figure~\ref{MedBaseLayer}. 
The {\it fru} Base-Layer (red) is among the Layers of the female Base-Layer (white) shown in Figure~\ref{fruOnBaseLayer}.

\begin{figure}
\begin{center}
\begin{tabular}{ | c | c| }
\hline 
        \includegraphics[width=7cm]{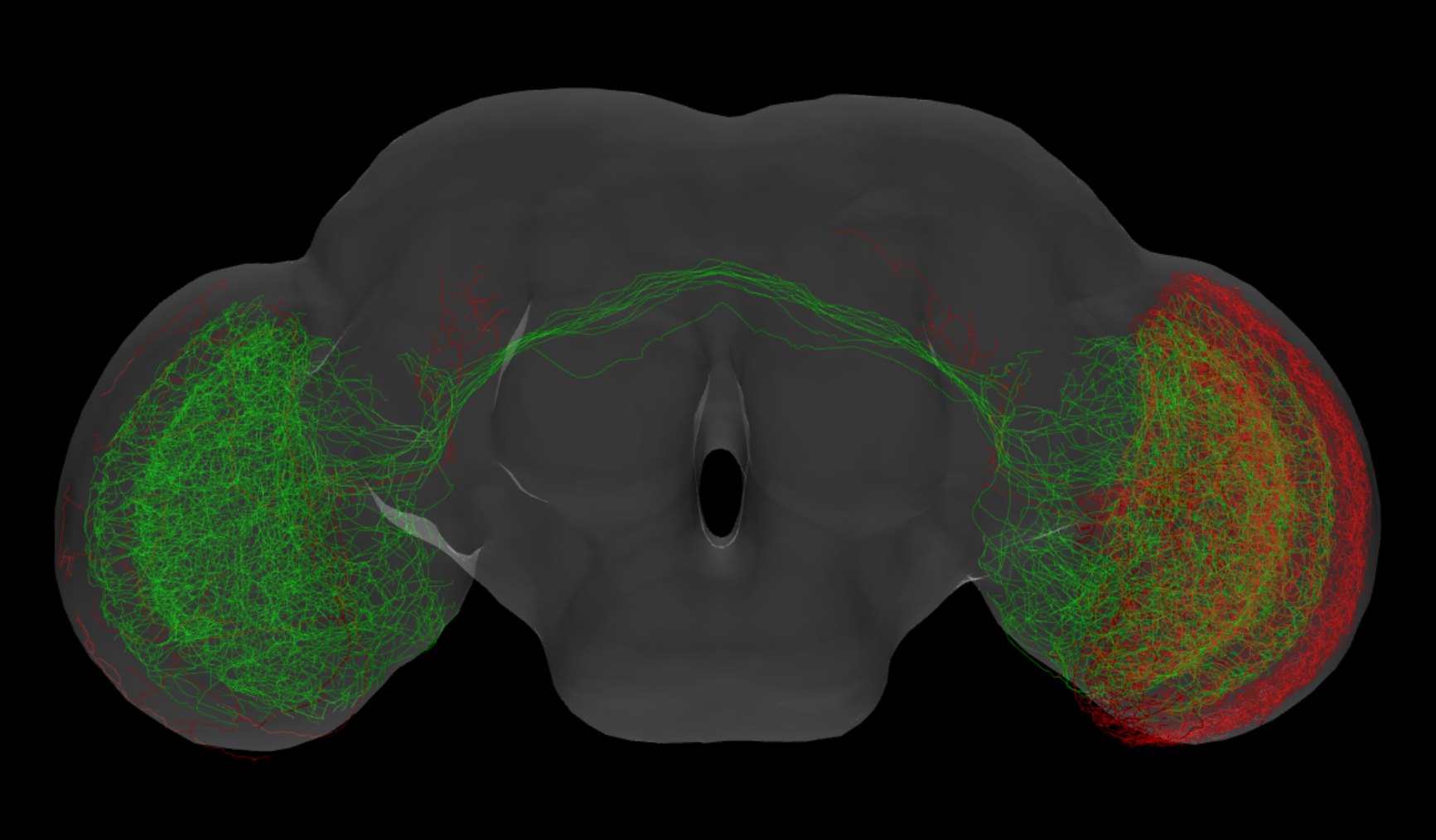} 
    &
        \includegraphics[width=7cm]{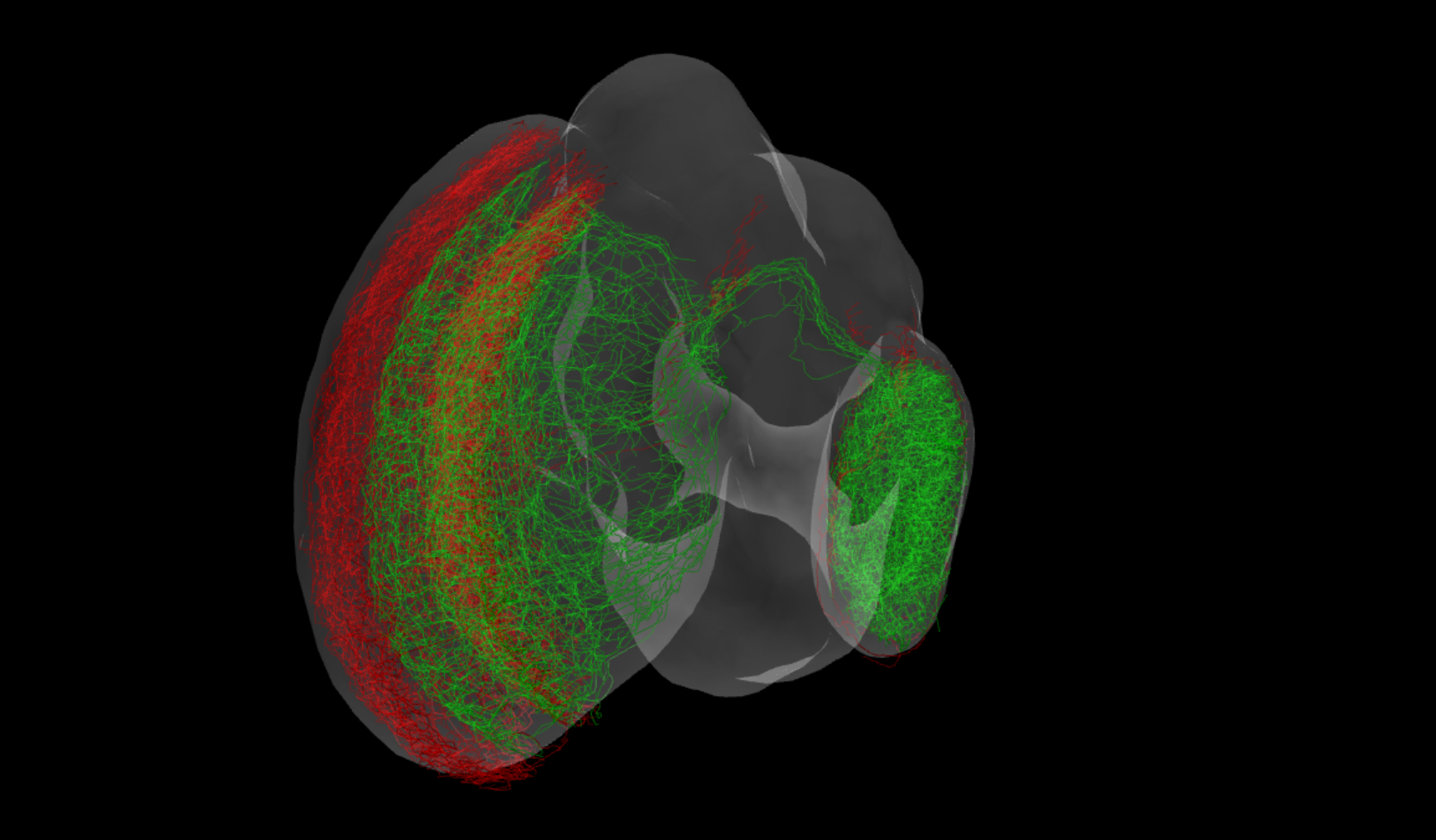} \\

        \includegraphics[width=7cm]{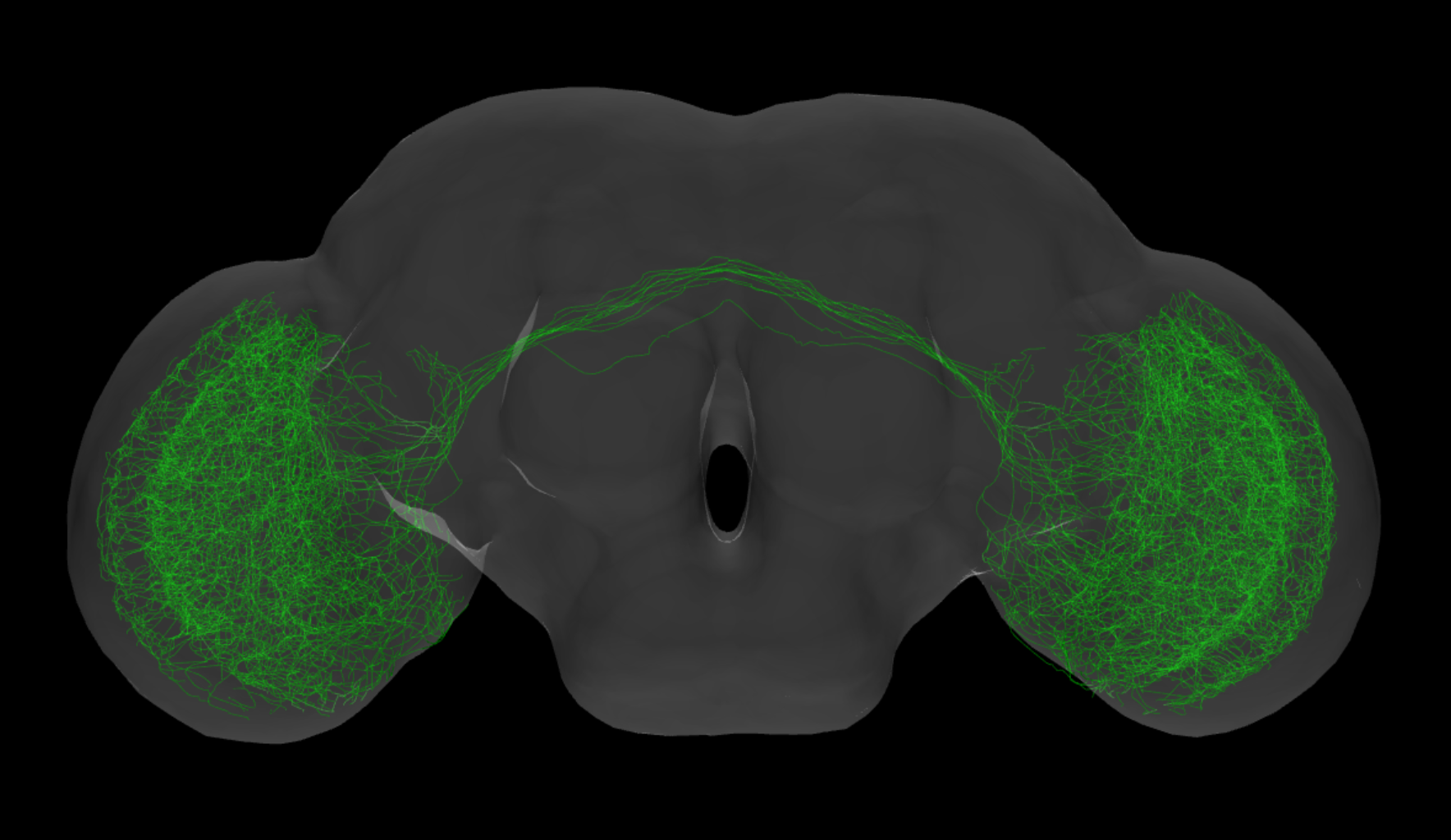} 
    & 
        \includegraphics[width=7cm]{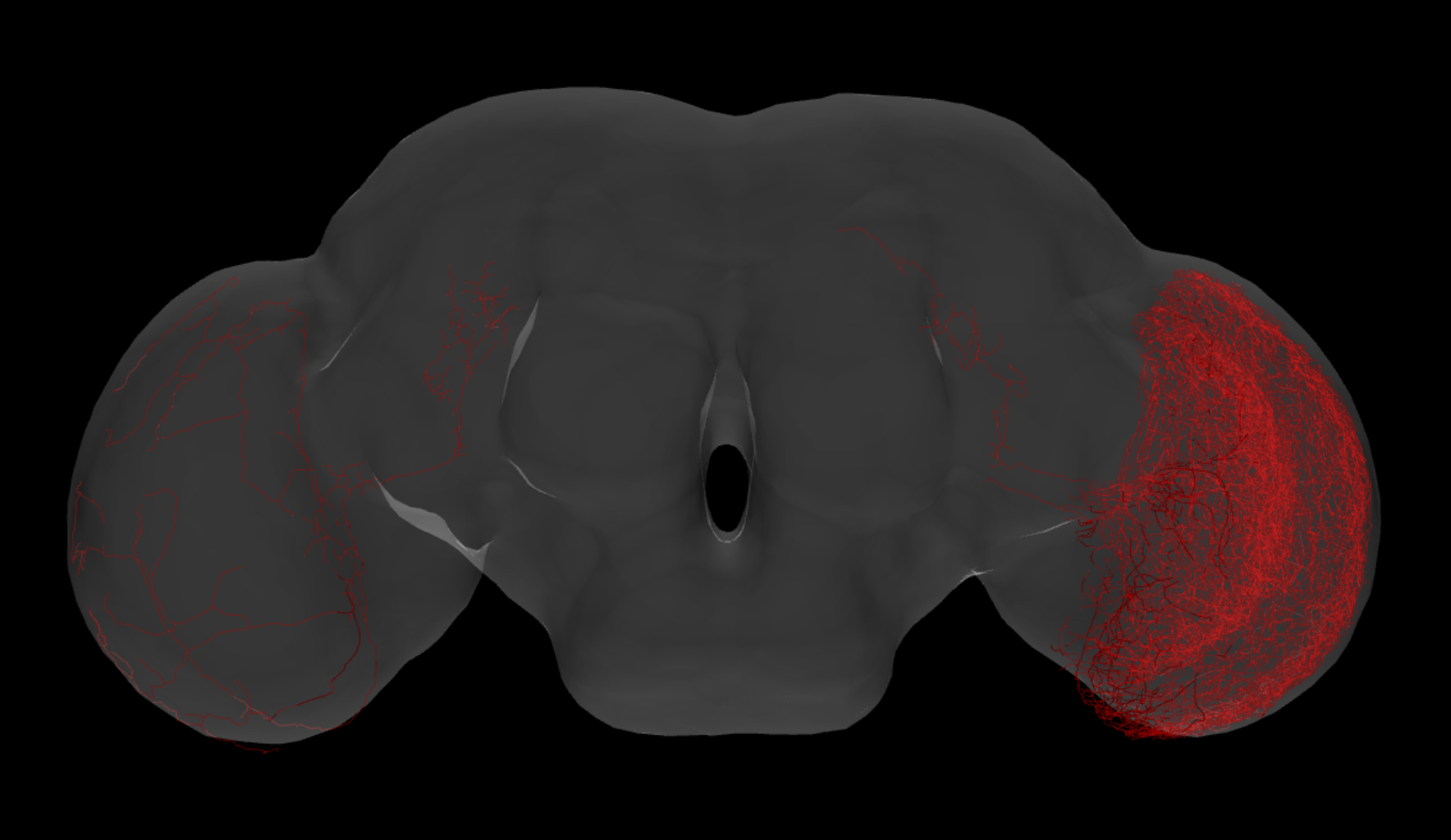} \\
     \hline

    \\ \hline

\end{tabular}
\caption{The Two-Ply structures in the Medulla of the female fly brain. }\label{MedF2Ply}
\end{center}
\end{figure}

\begin{figure}
    \centering
    \includegraphics[width=5cm]{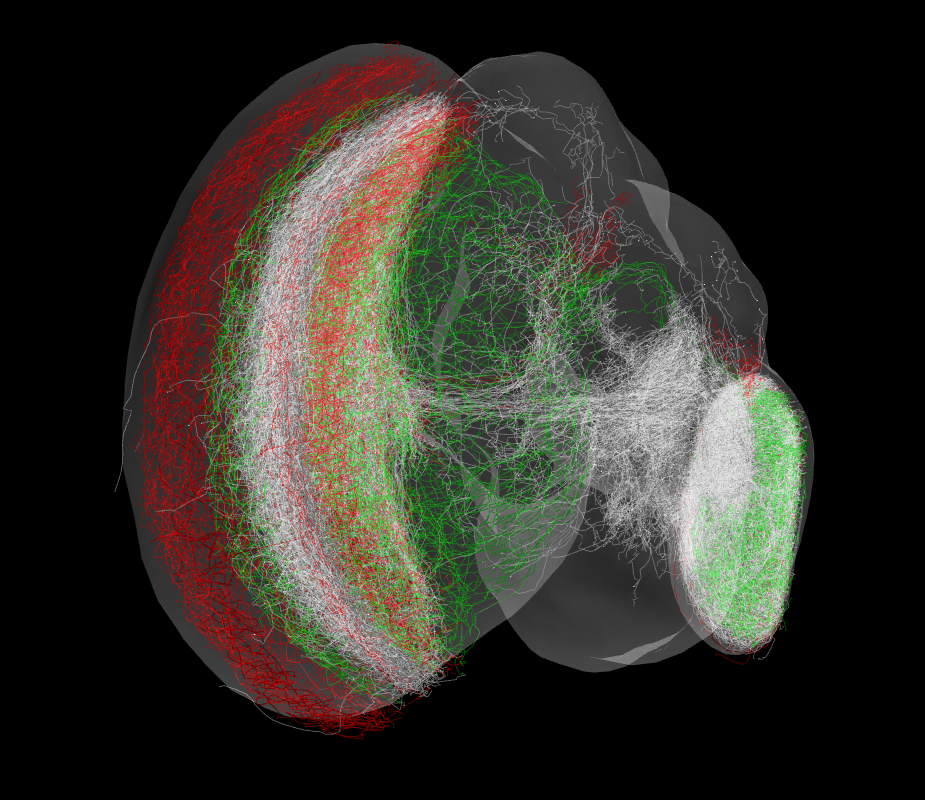}
    \includegraphics[width=5cm]{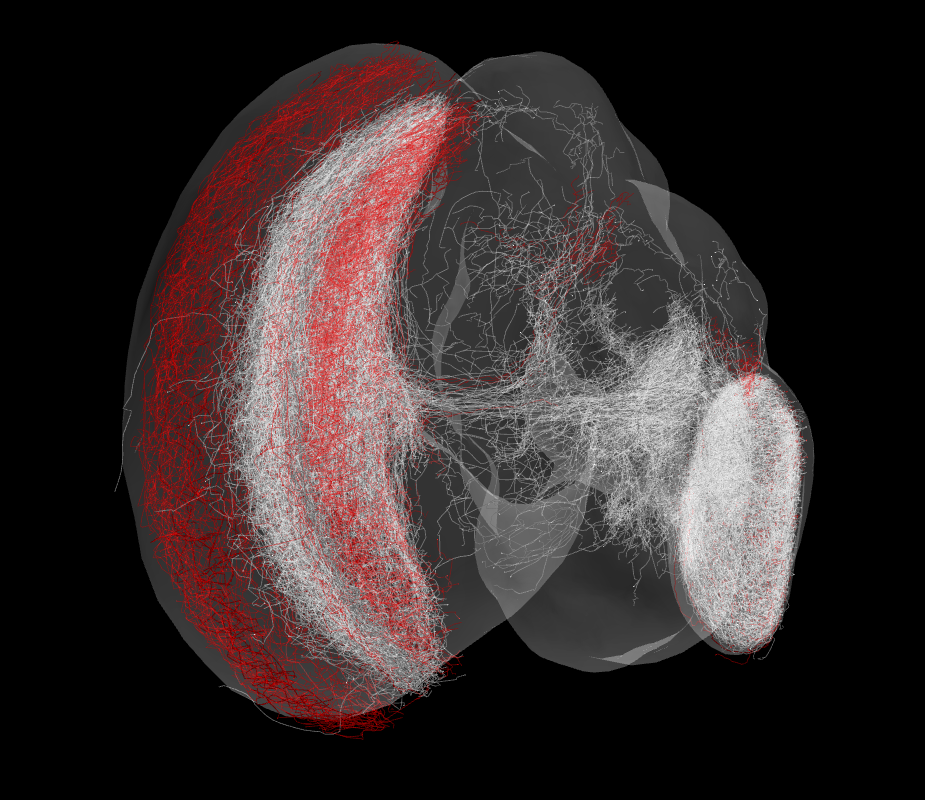}
    \includegraphics[width=5cm]{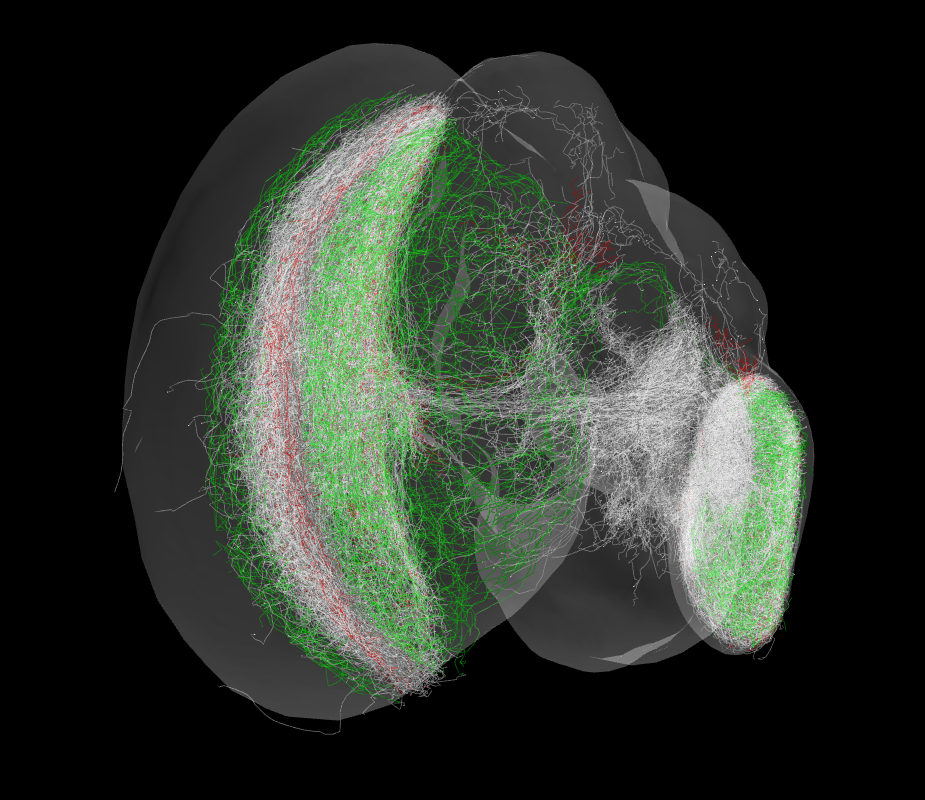}
    \caption{The two Two-Ply structures (red and green) sandwich the Base-Layer in the Medulla. }
    \label{TwoPlyBaseLayer}
\end{figure}

There are 2 Two-Ply structures found as shown in Figure~\ref{MedF2Ply}. 
Images on the first row in Figure~\ref{MedF2Ply} show the two Two-Ply structures in red and green, viewing from the front (left) and side (right). 
Each image is shown in the next row in Figure~\ref{MedF2Ply}.
Both Two-Ply structures ``sandwich'' the Base-Layer (white) shown in Figure~{\ref{TwoPlyBaseLayer}}. 

\begin{figure}
\begin{center}
\begin{tabular}{ | c | c| }
\hline 
        \includegraphics[width=7cm]{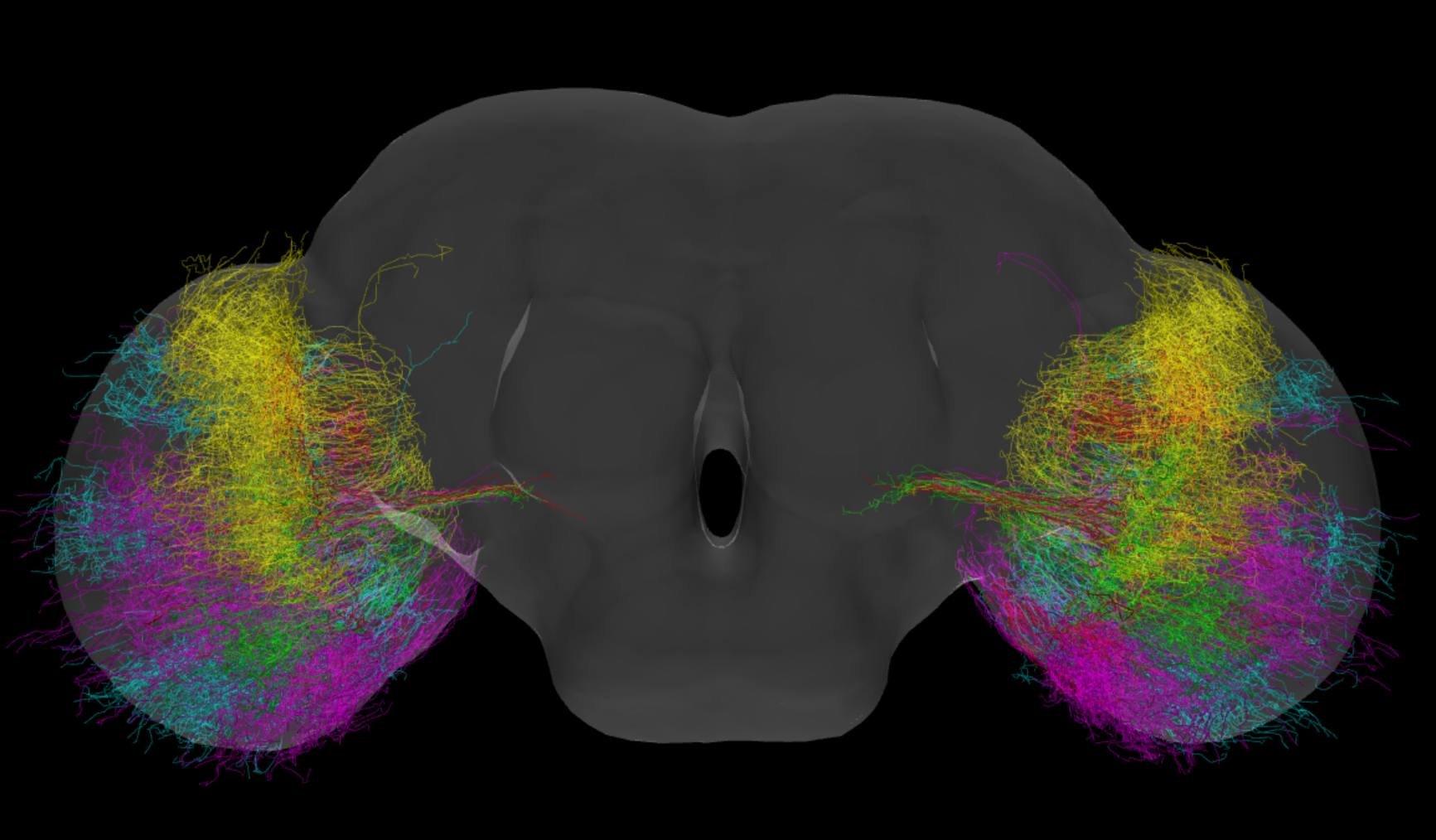} 
    &
        \includegraphics[width=7cm]{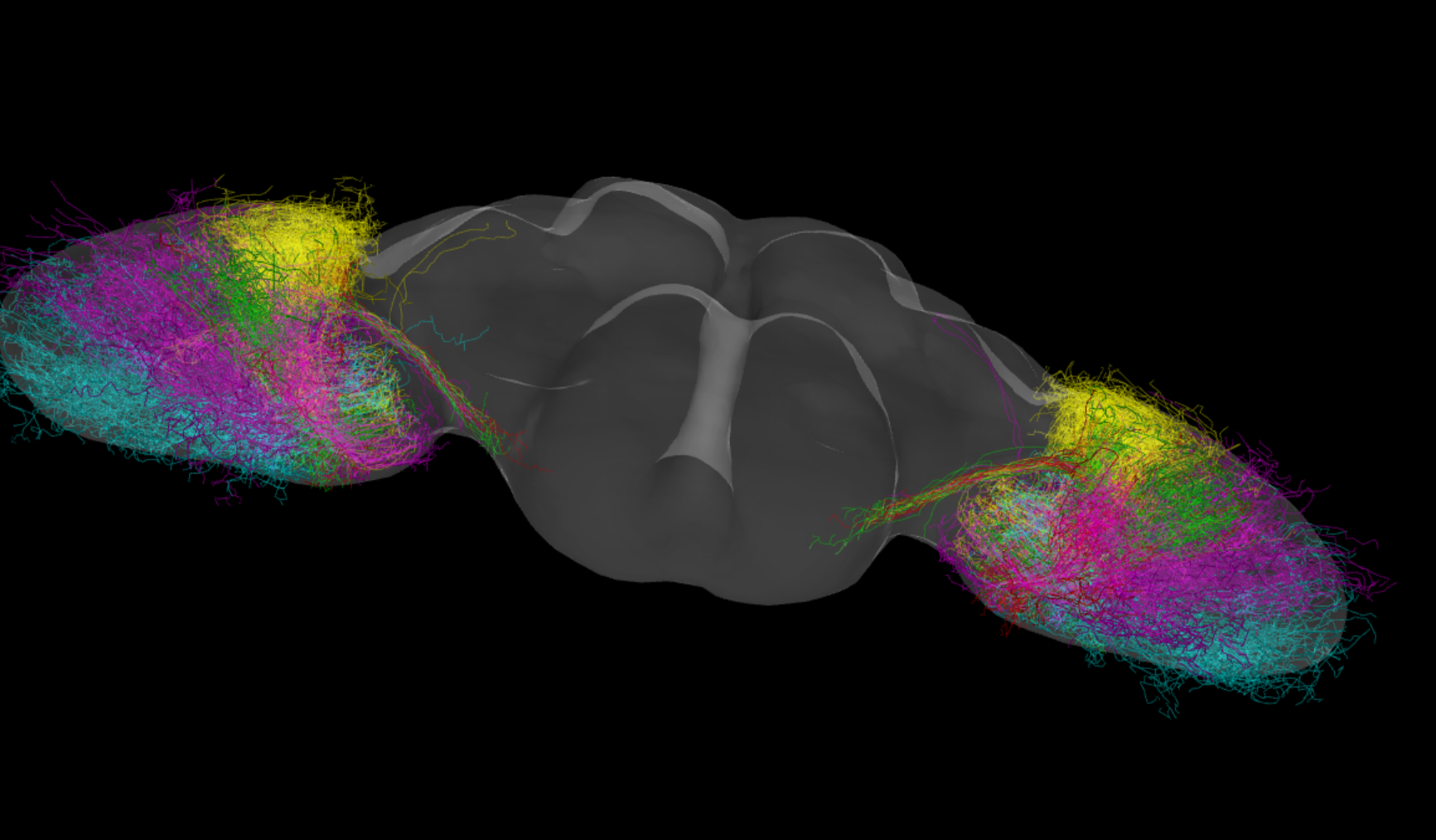} \\

        \includegraphics[width=7cm]{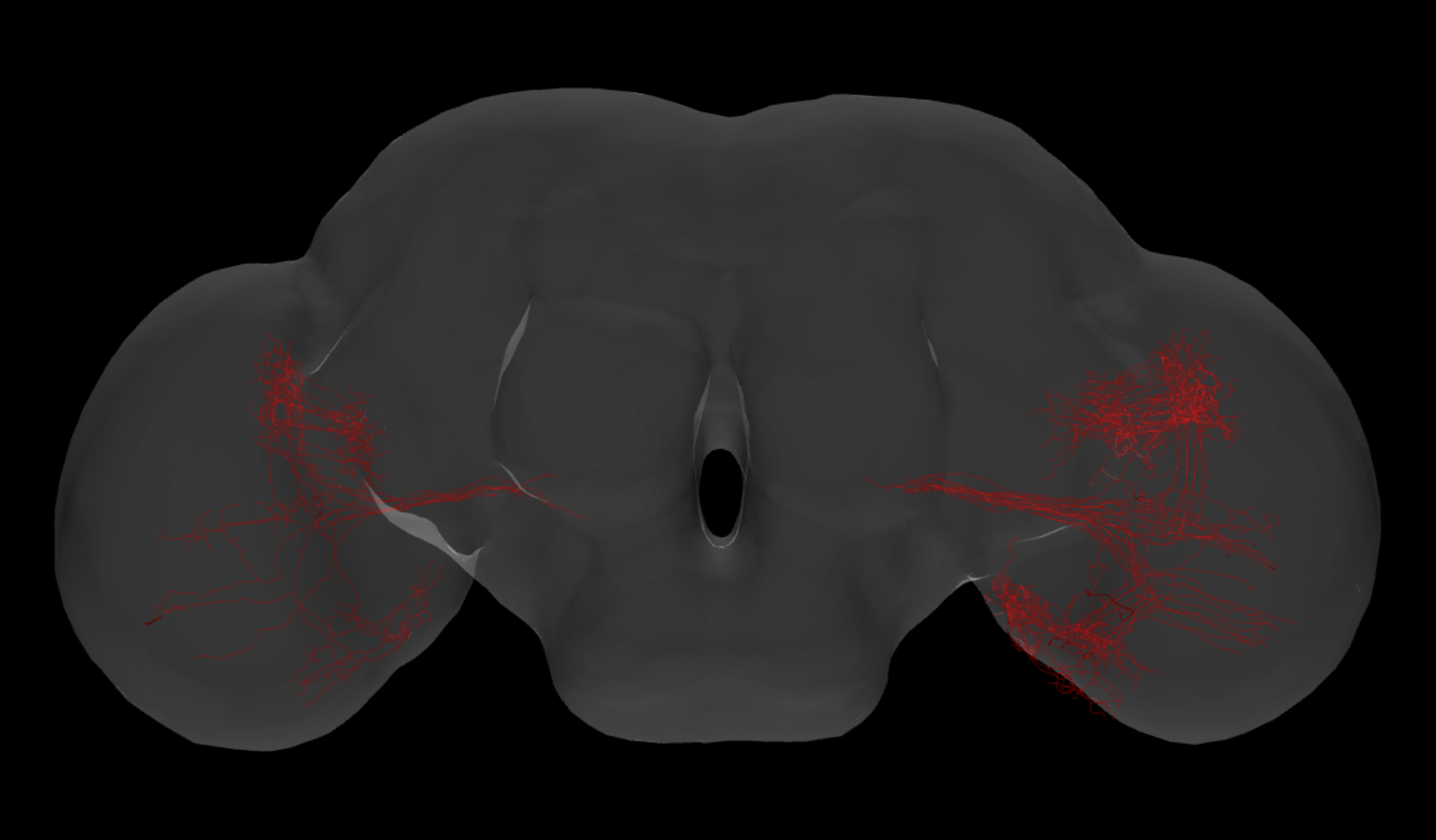} 
    & 
        \includegraphics[width=7cm]{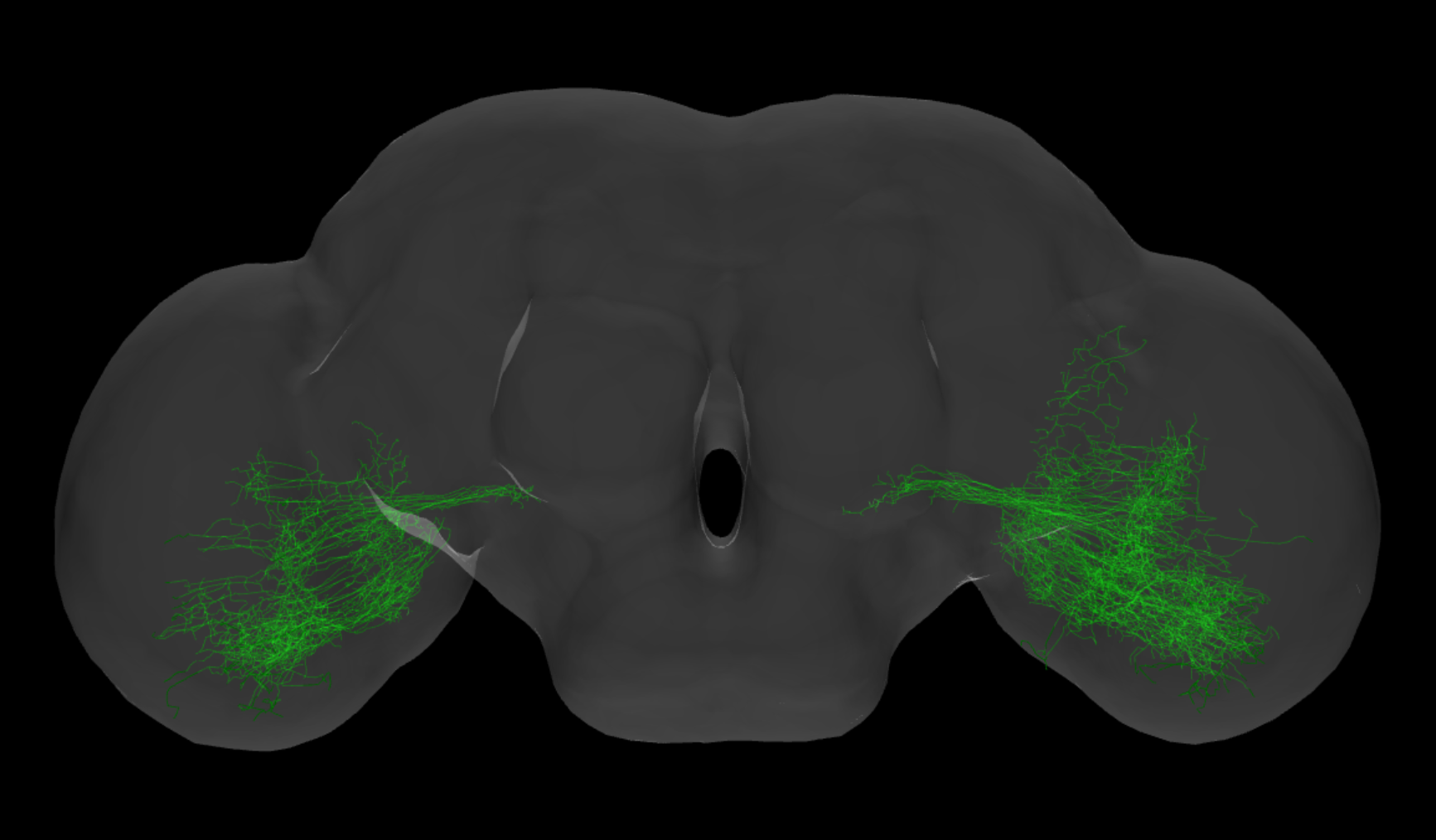} \\
     \hline
 
        \includegraphics[width=7cm]{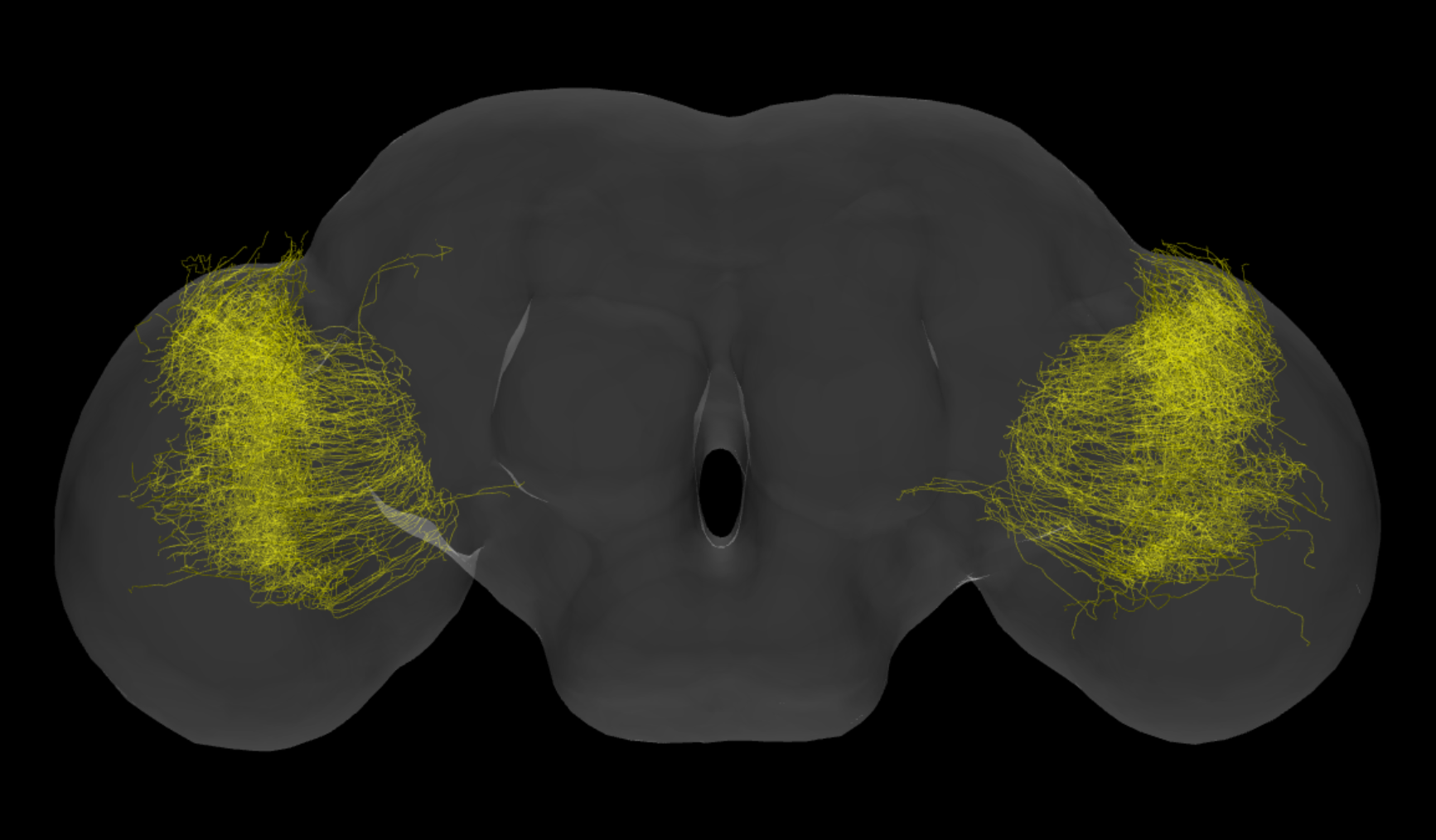} 
    & 
        \includegraphics[width=7cm]{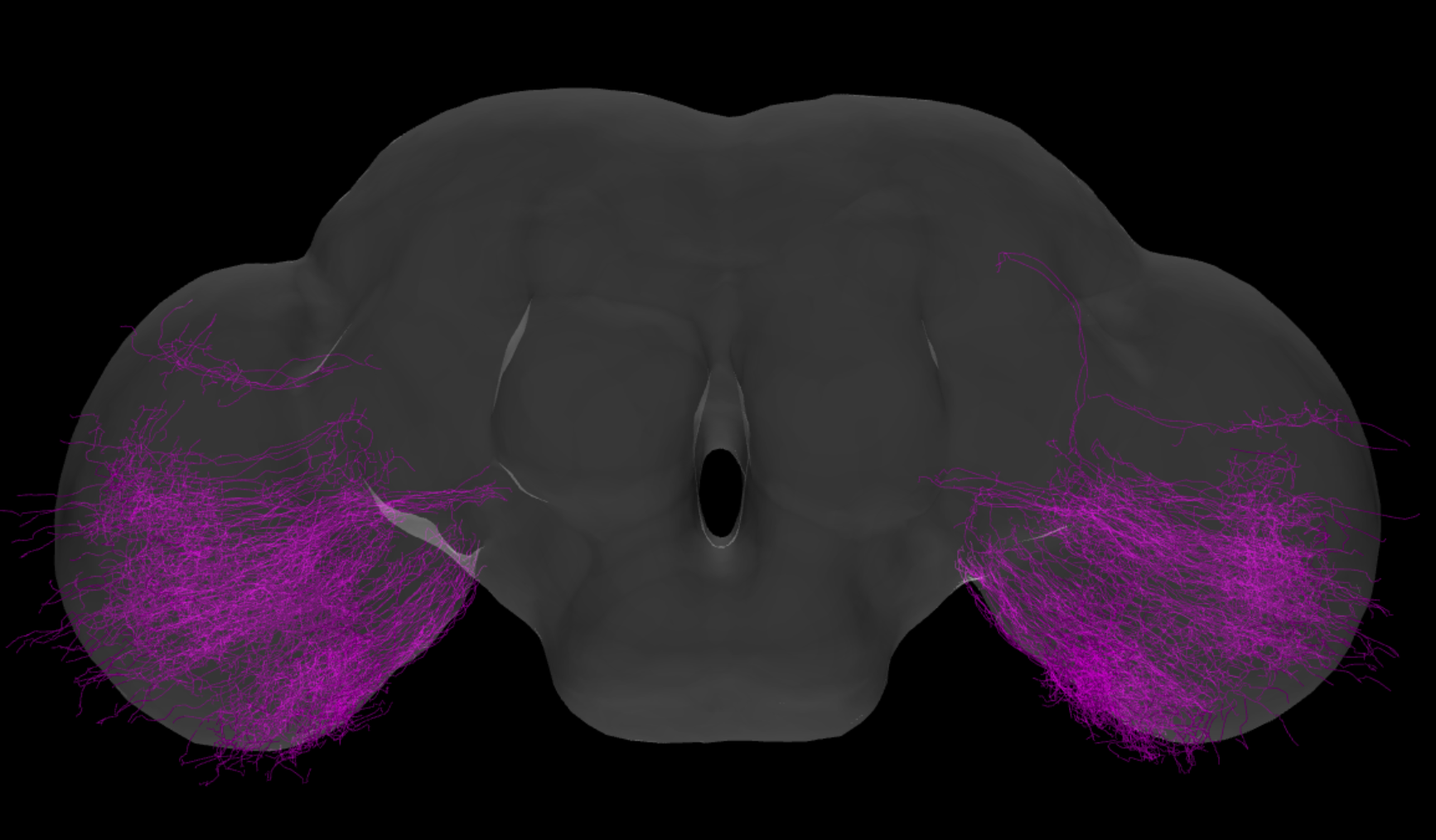} \\
     \hline
 
        \includegraphics[width=7cm]{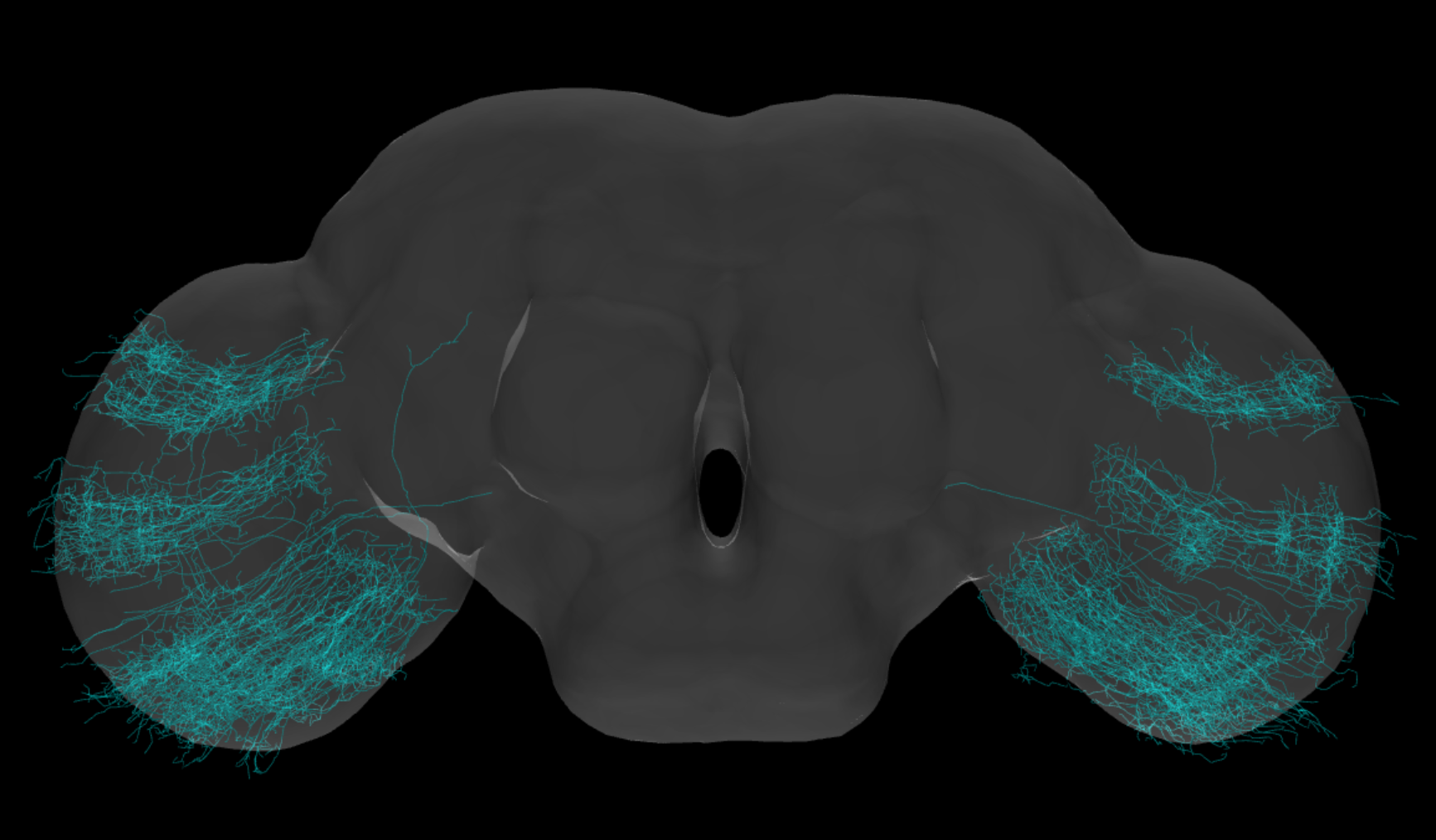} 
    & 
         \\
     \hline

    \\ \hline

\end{tabular}
\caption{The Via structures in the Medulla of the female fly brain. }\label{MedFVia}
\end{center}
\end{figure}

The Via neurons are shown in the first row in Figure~\ref{MedFVia}. 
Five Components of the Via neurons are shown in the following rows in the figure. 
These five components contain many small groups obtained by the Maximum Spanning Tree algorithm.
These small components were not connected because of the way we defined the similarity of the neurons.
These sparse components were assembled with the help of the visualization tool to get the five components. 

\begin{figure}
\begin{center}
\begin{tabular}{ | c | c| c|}
\hline 
        \includegraphics[width=5cm]{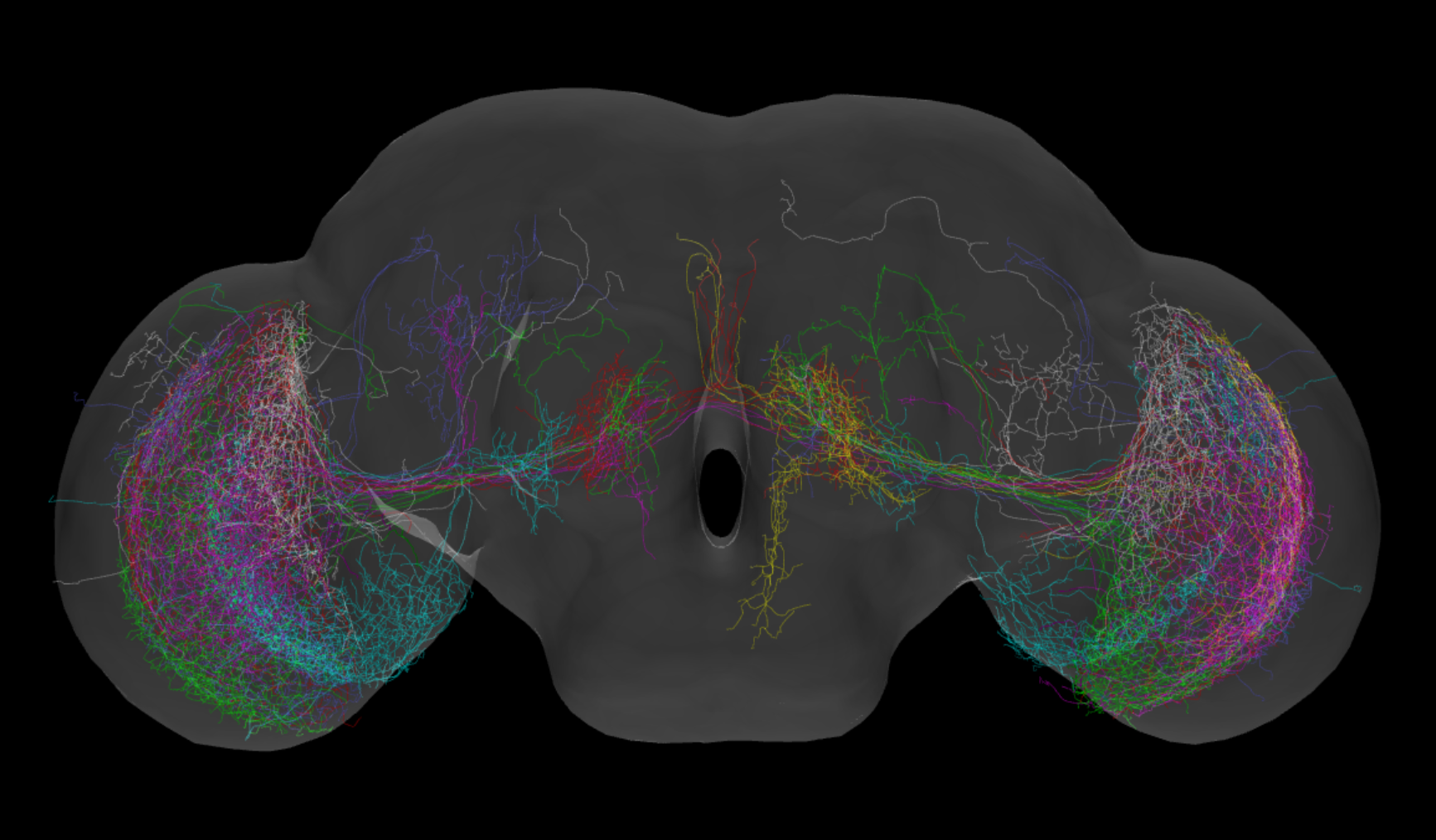} 
    &
        \includegraphics[width=5cm]{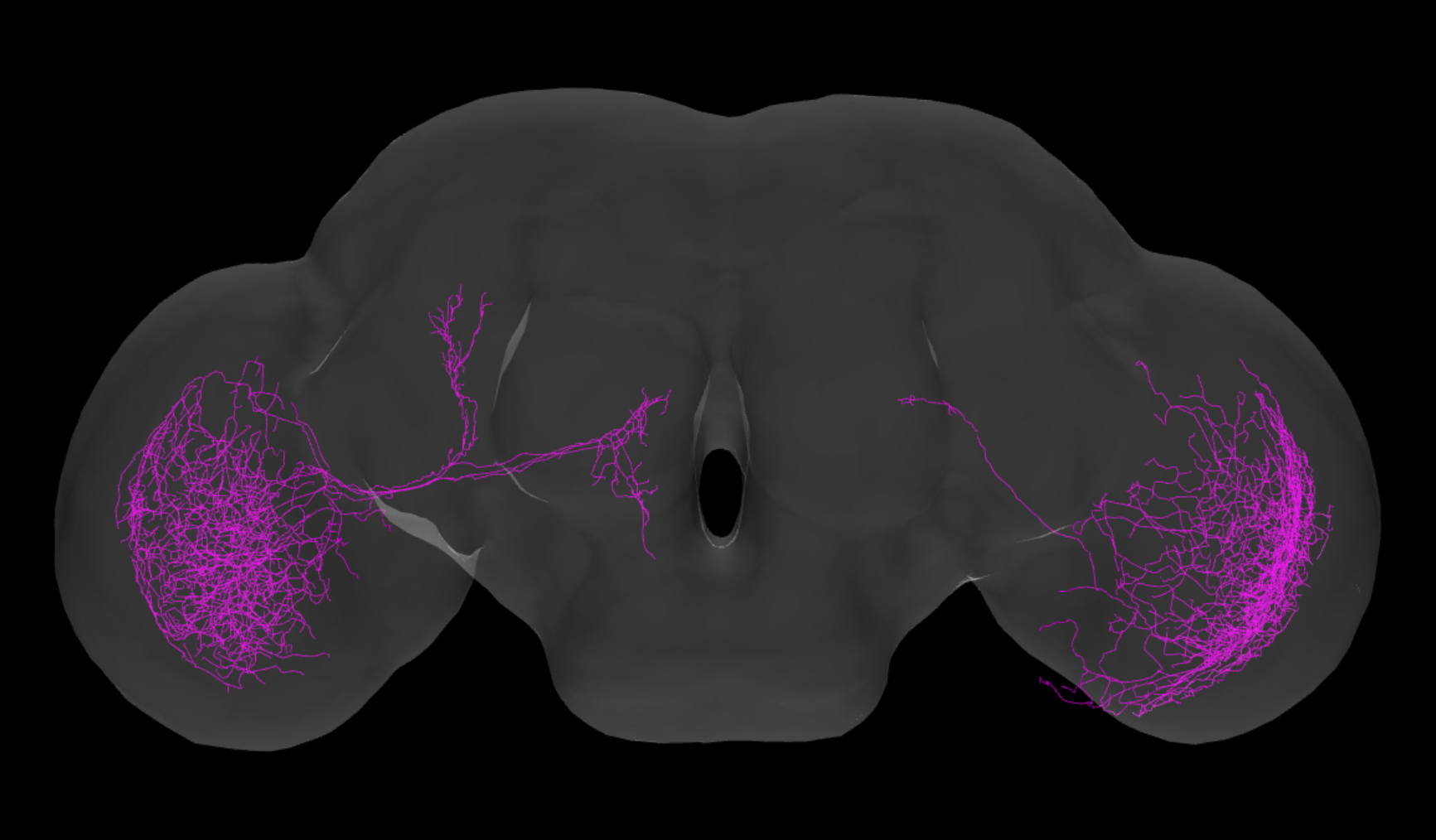} 
    &
        \includegraphics[width=5cm]{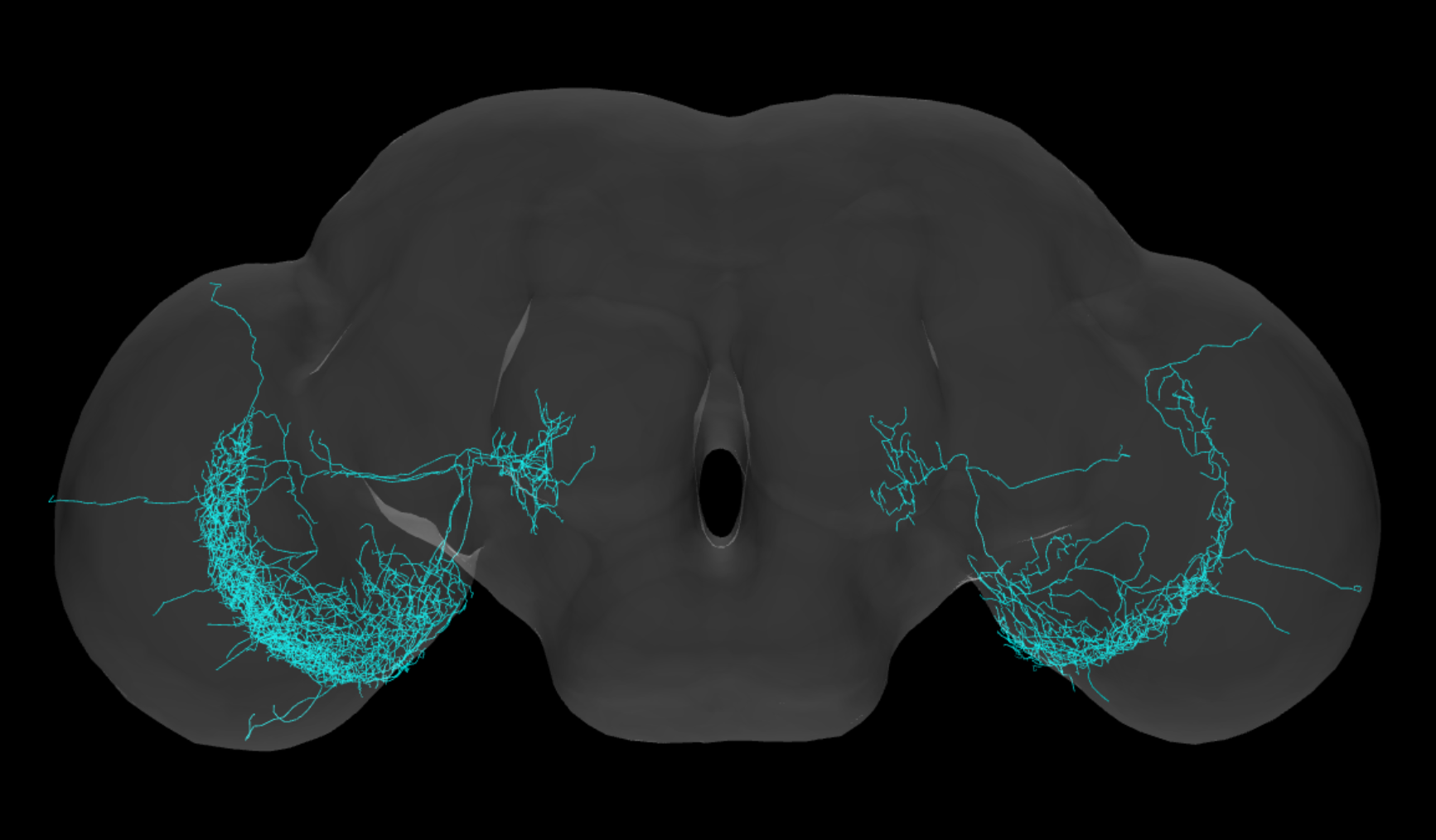} \\ \hline
     
        \includegraphics[width=5cm]{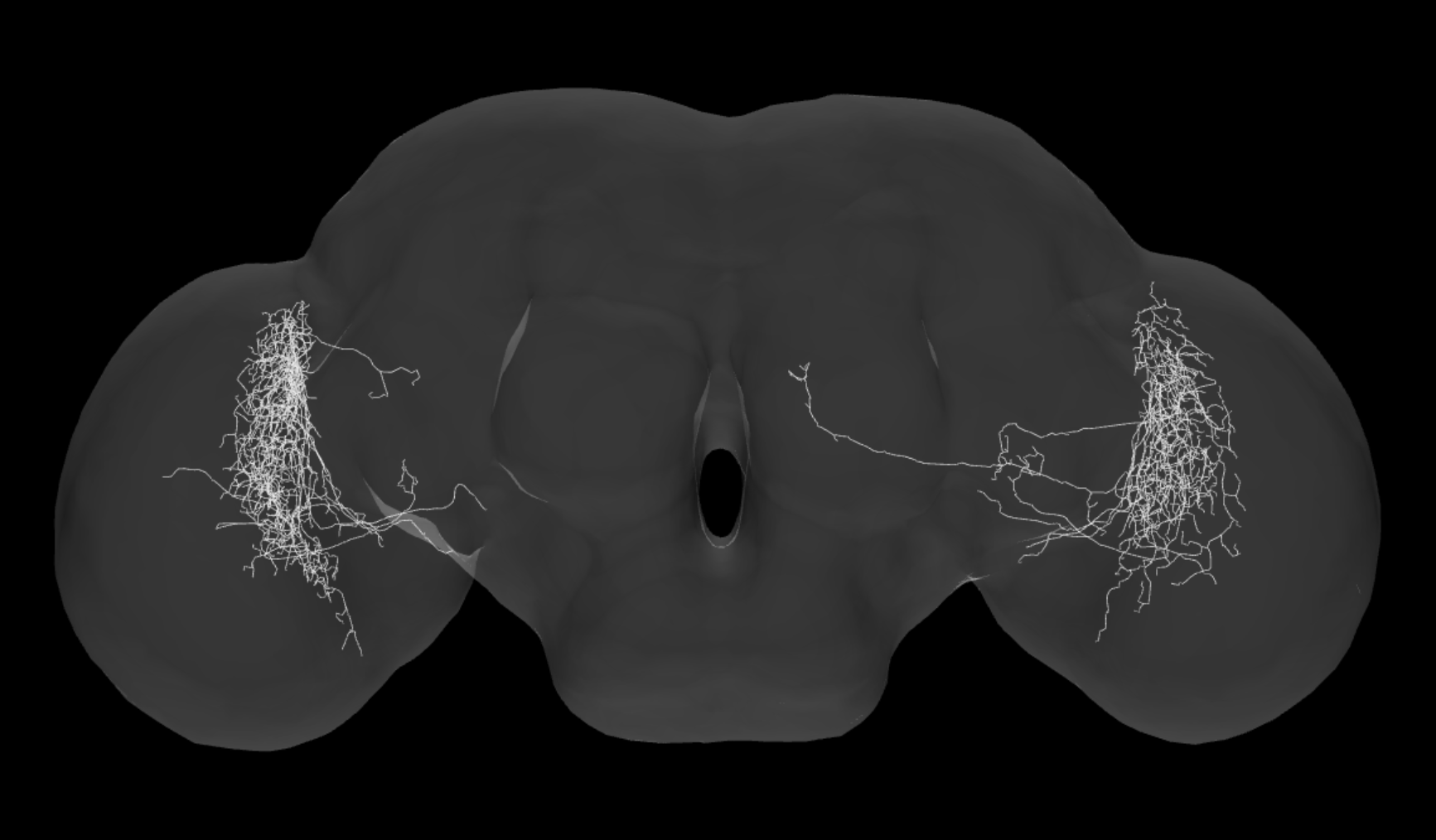} 
    &
        \includegraphics[width=5cm]{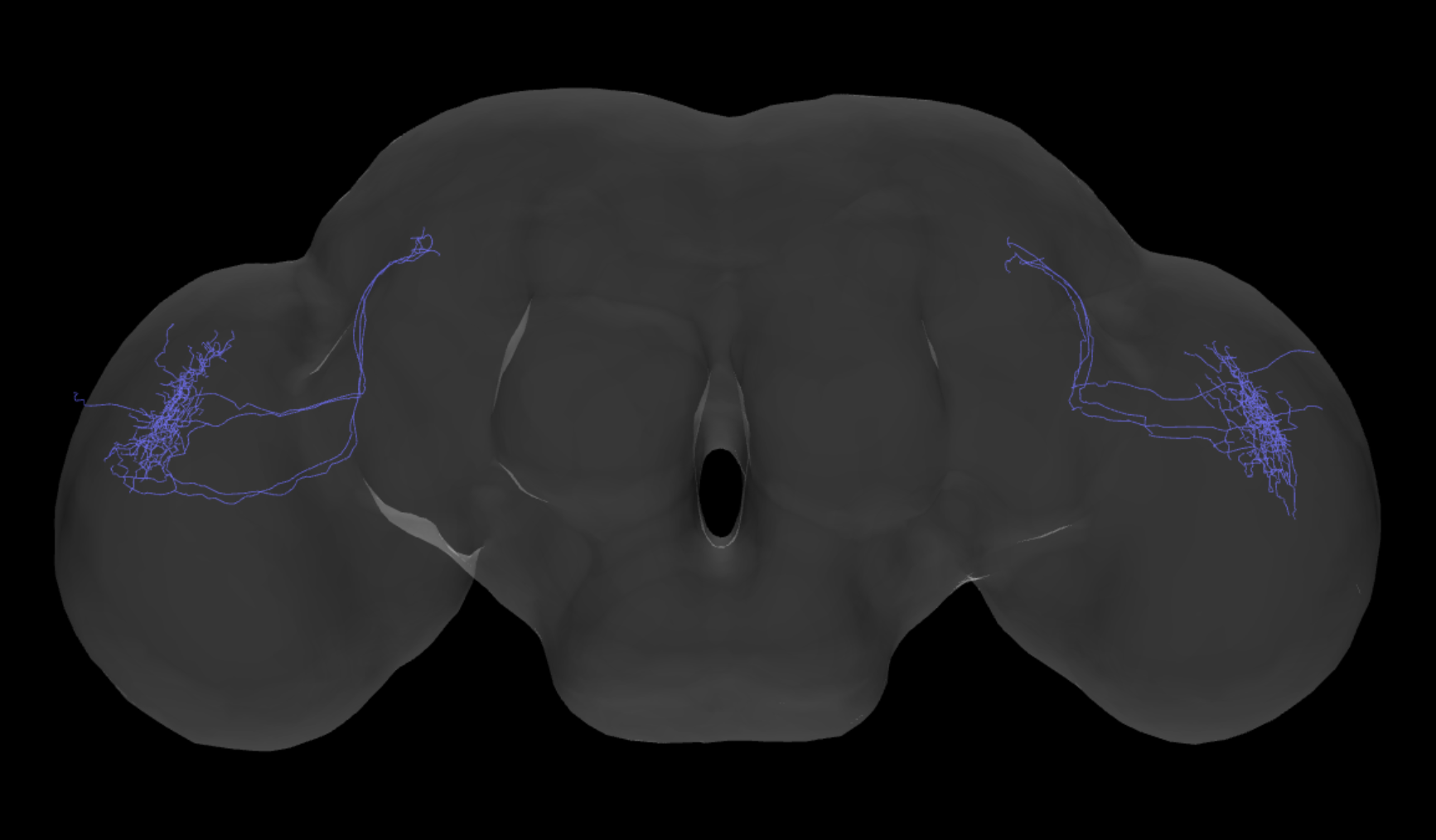} 
    & 
        \includegraphics[width=5cm]{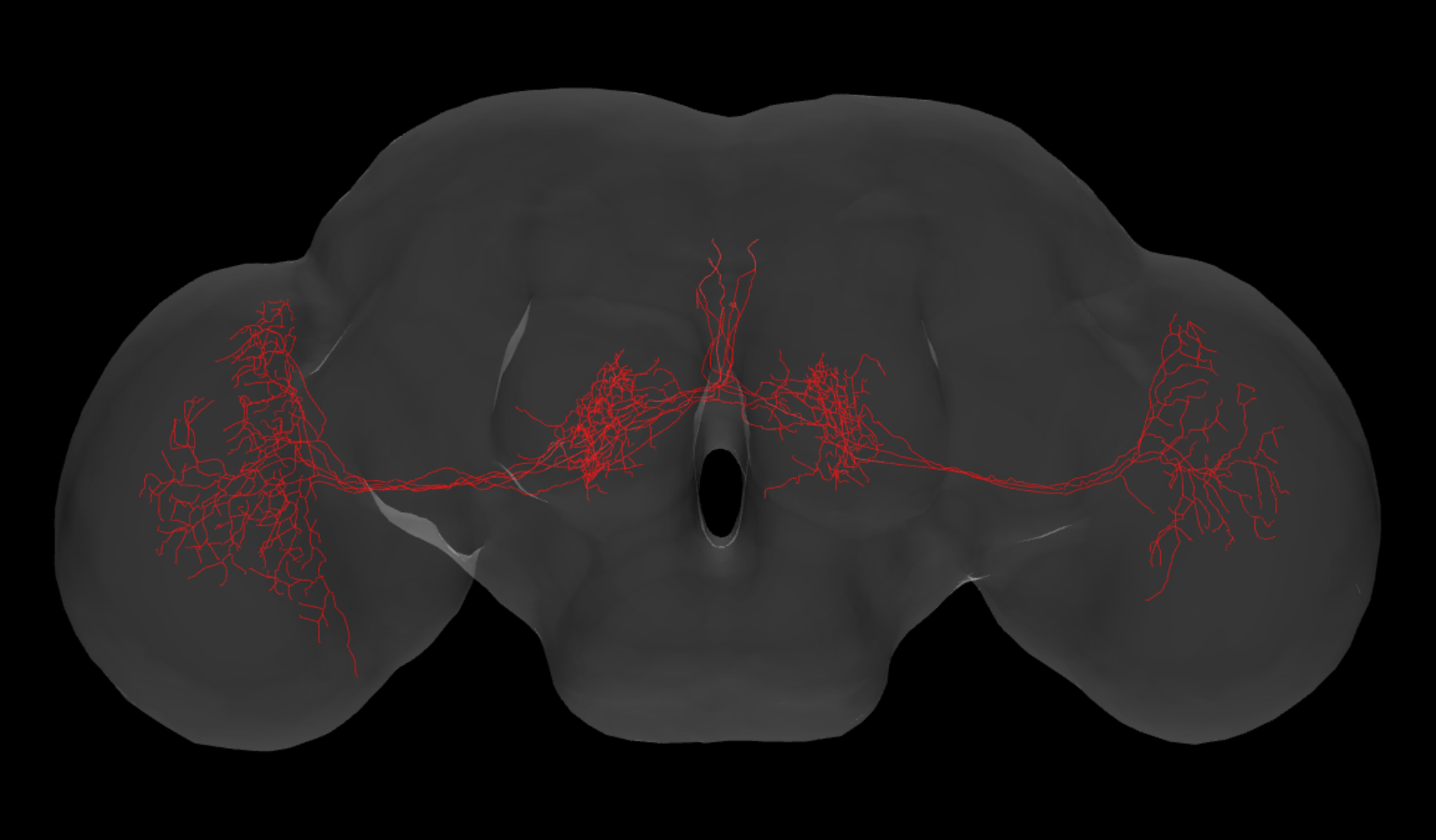} \\ \hline

        \includegraphics[width=5cm]{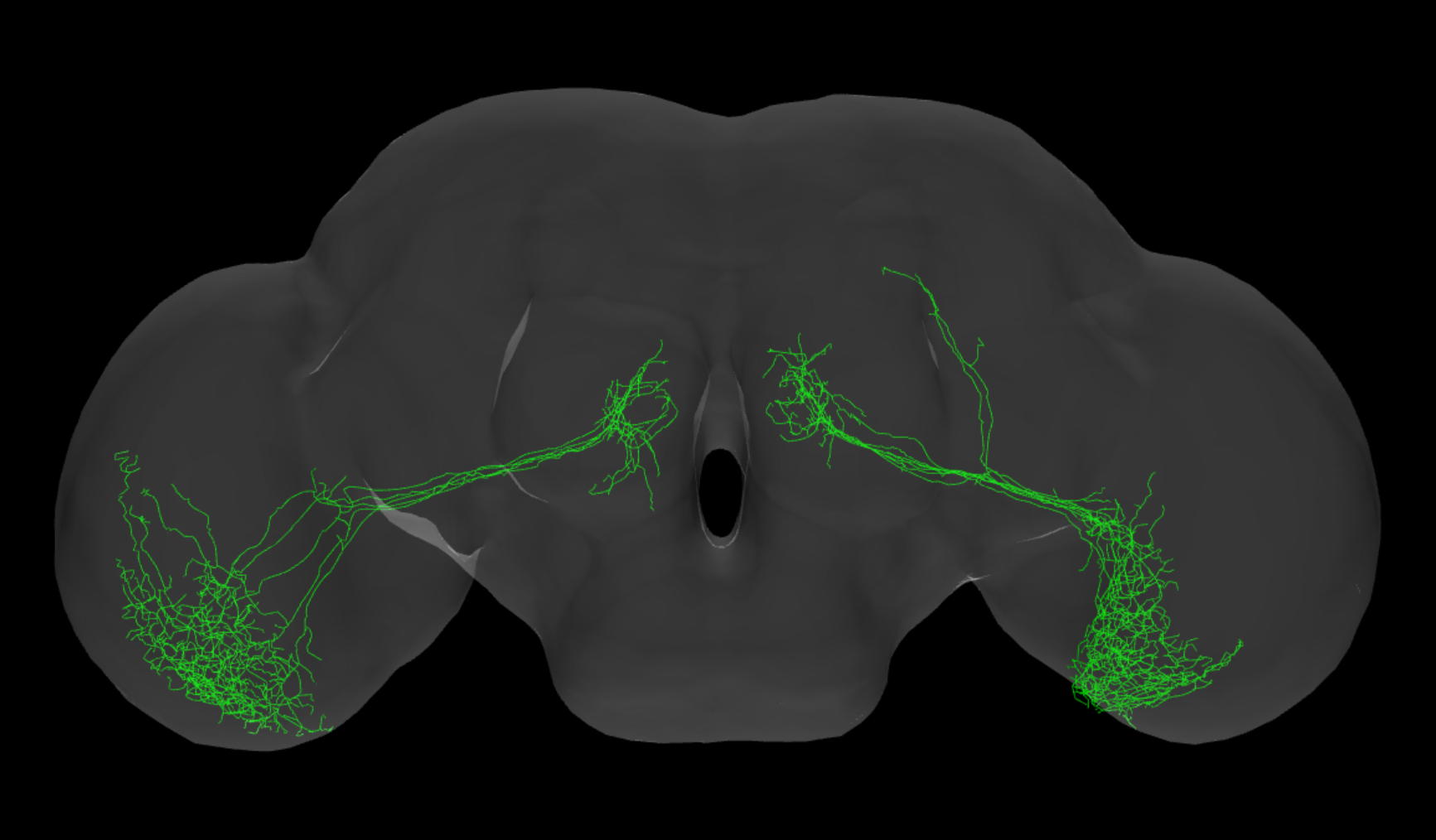} 
    &
        \includegraphics[width=5cm]{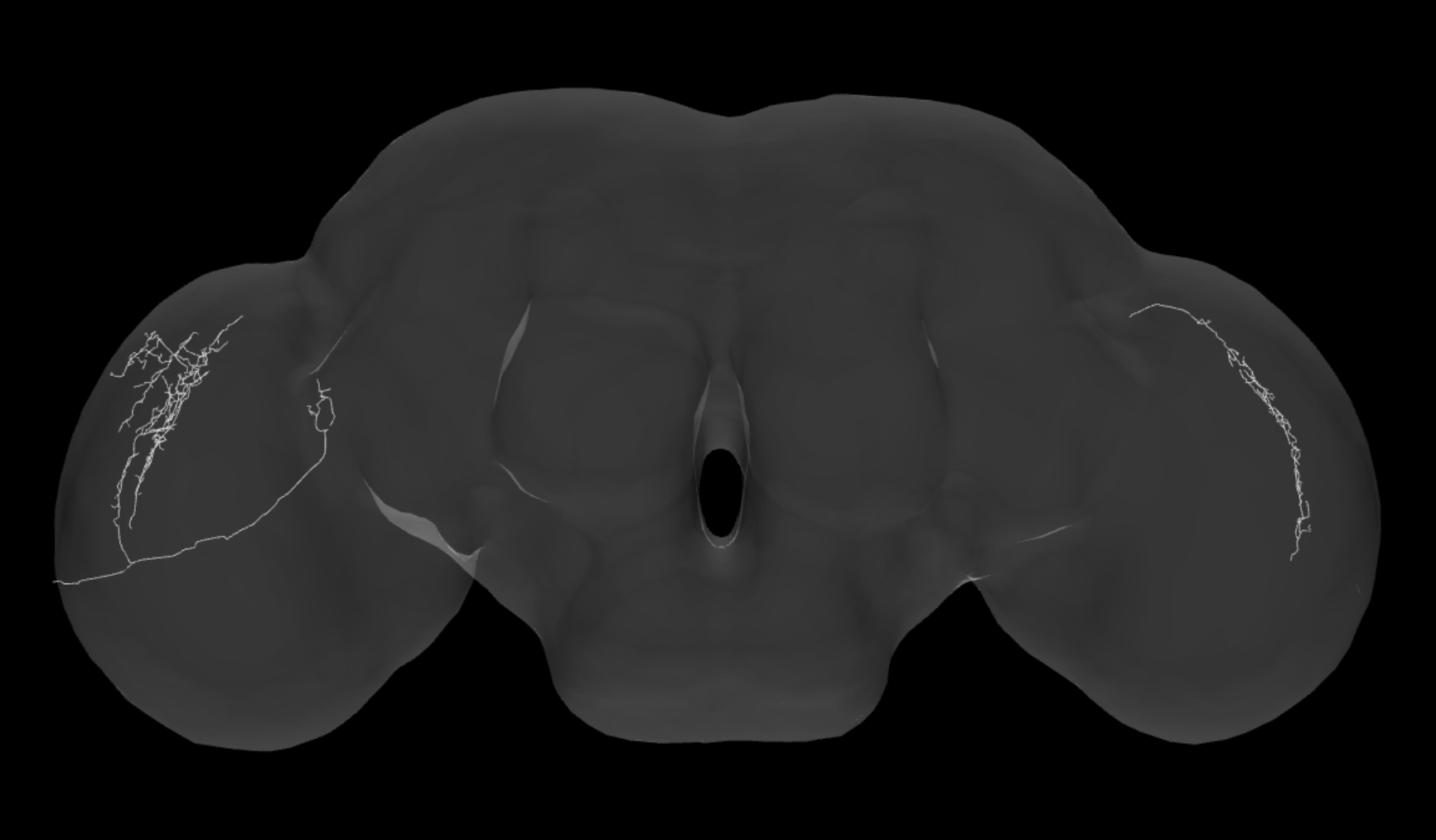} 
    & 
        \includegraphics[width=5cm]{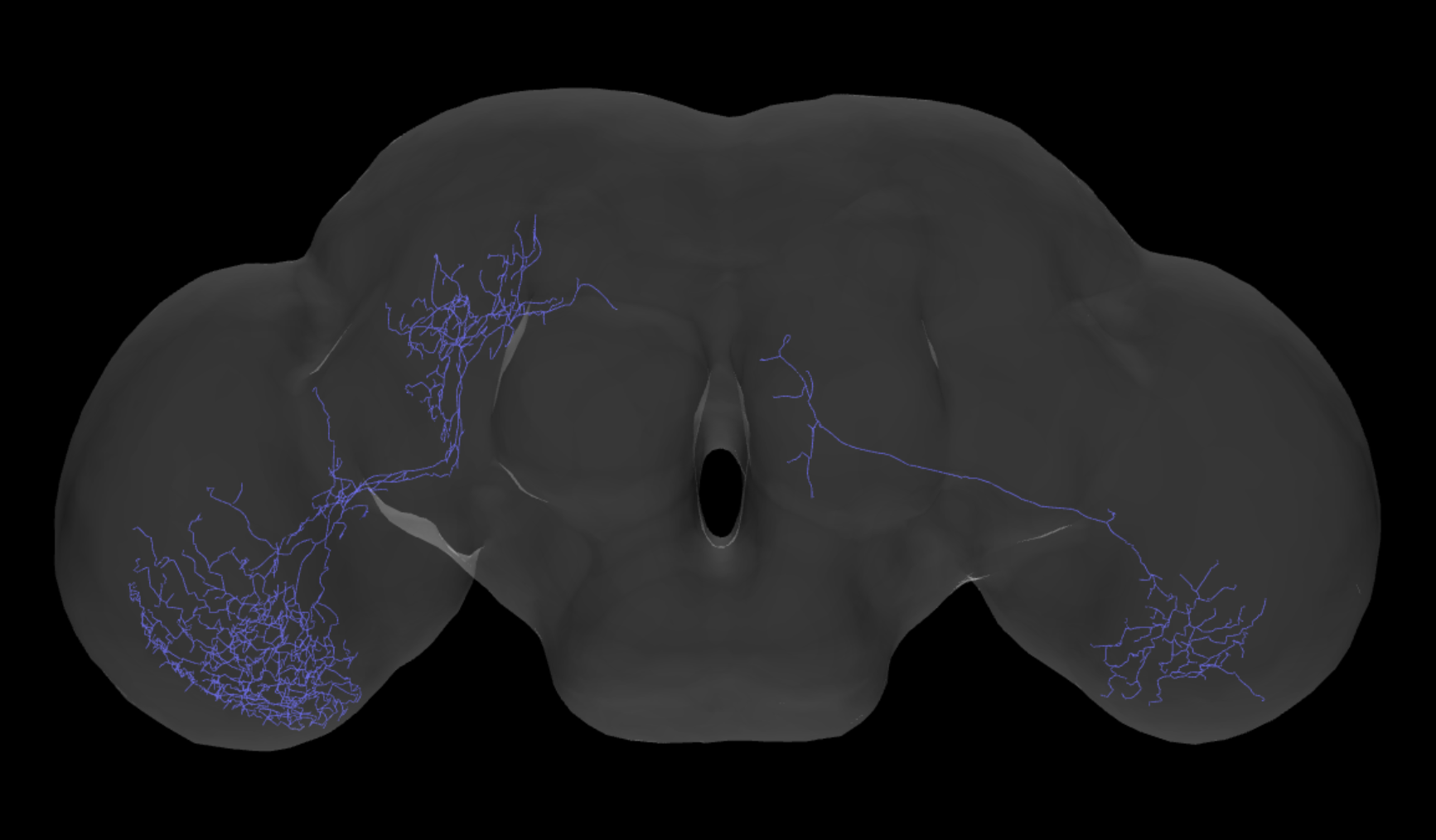} \\ \hline

        \includegraphics[width=5cm]{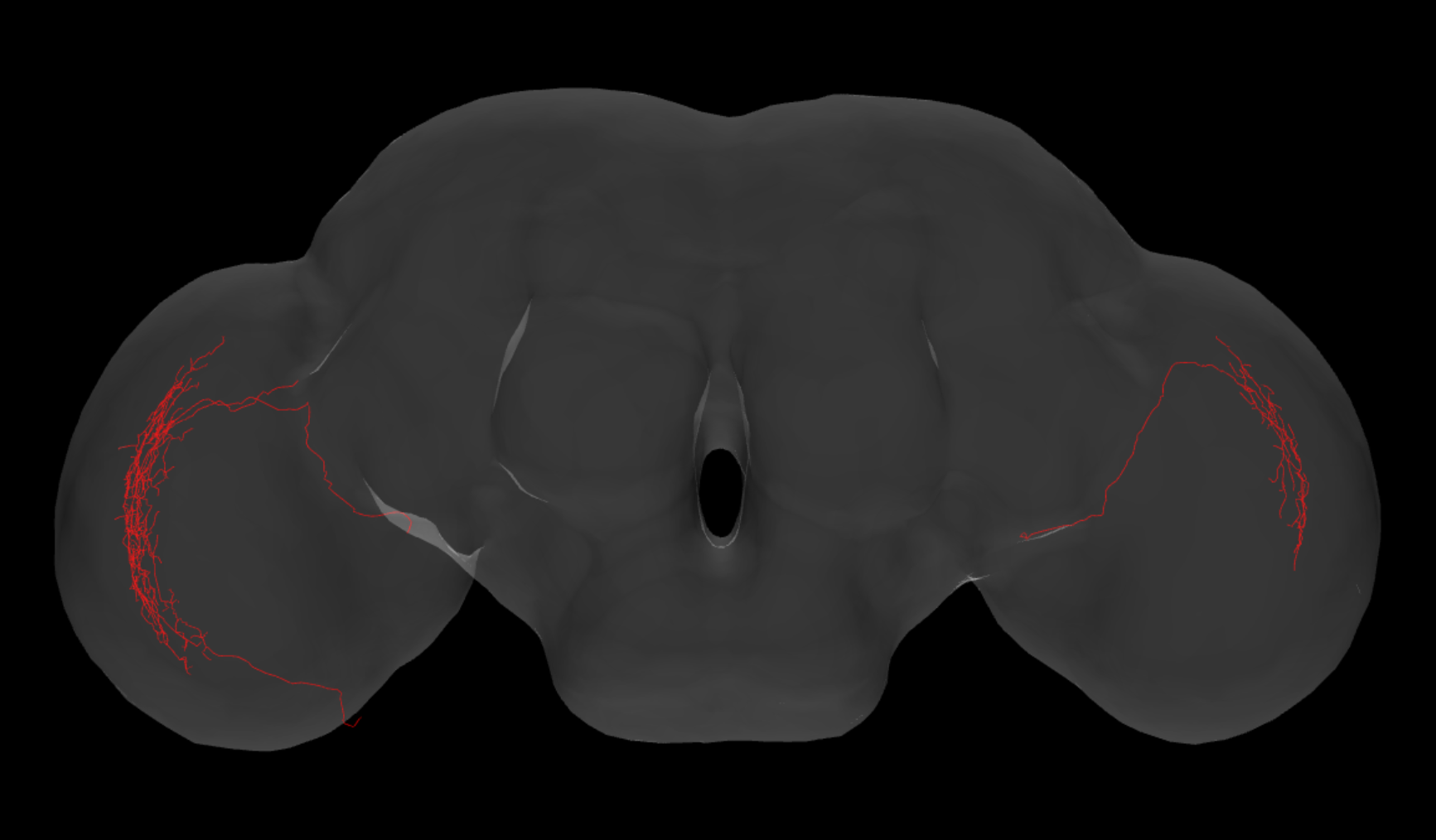} 
    &
        \includegraphics[width=5cm]{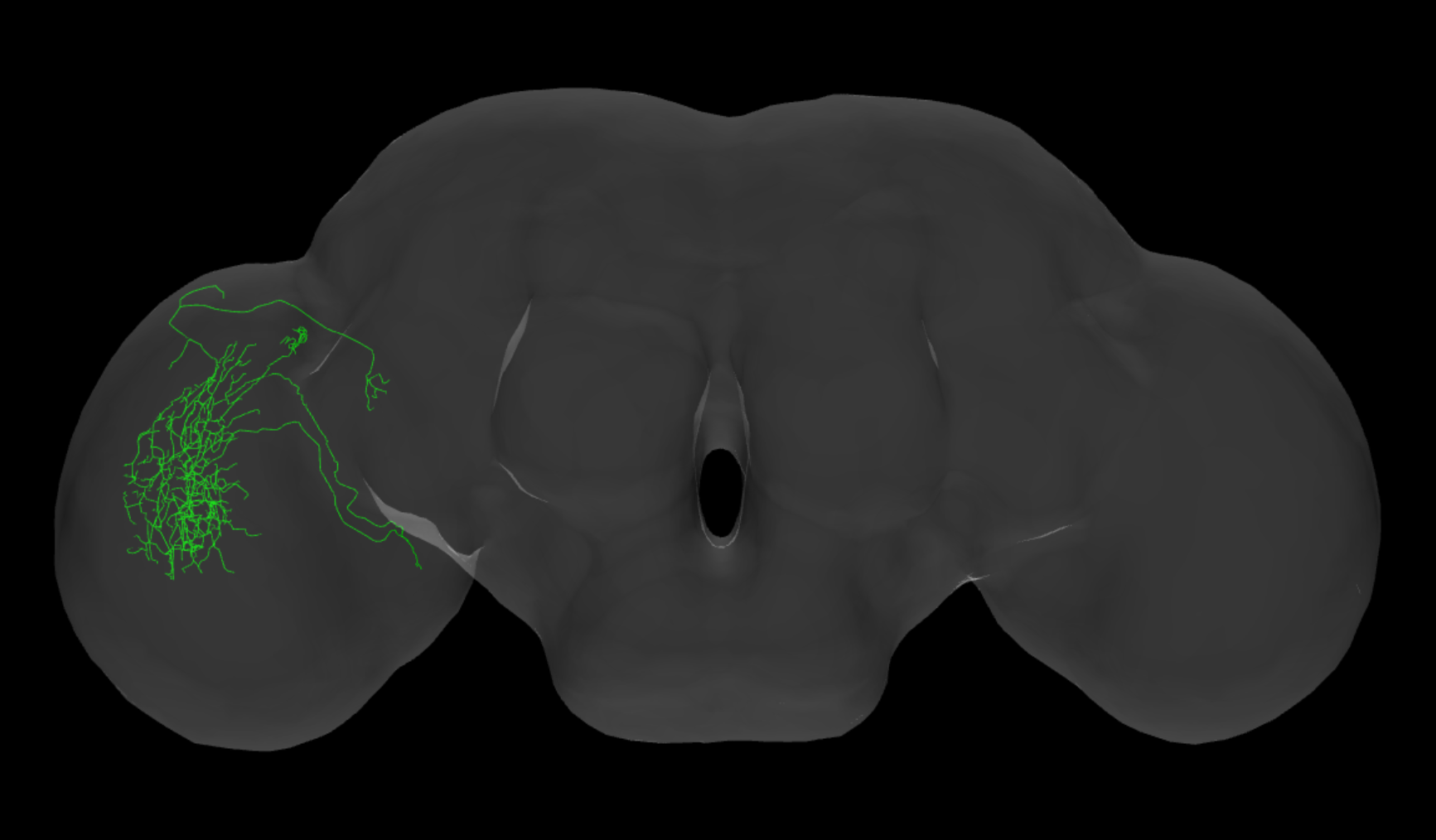} 
    & 
        \includegraphics[width=5cm]{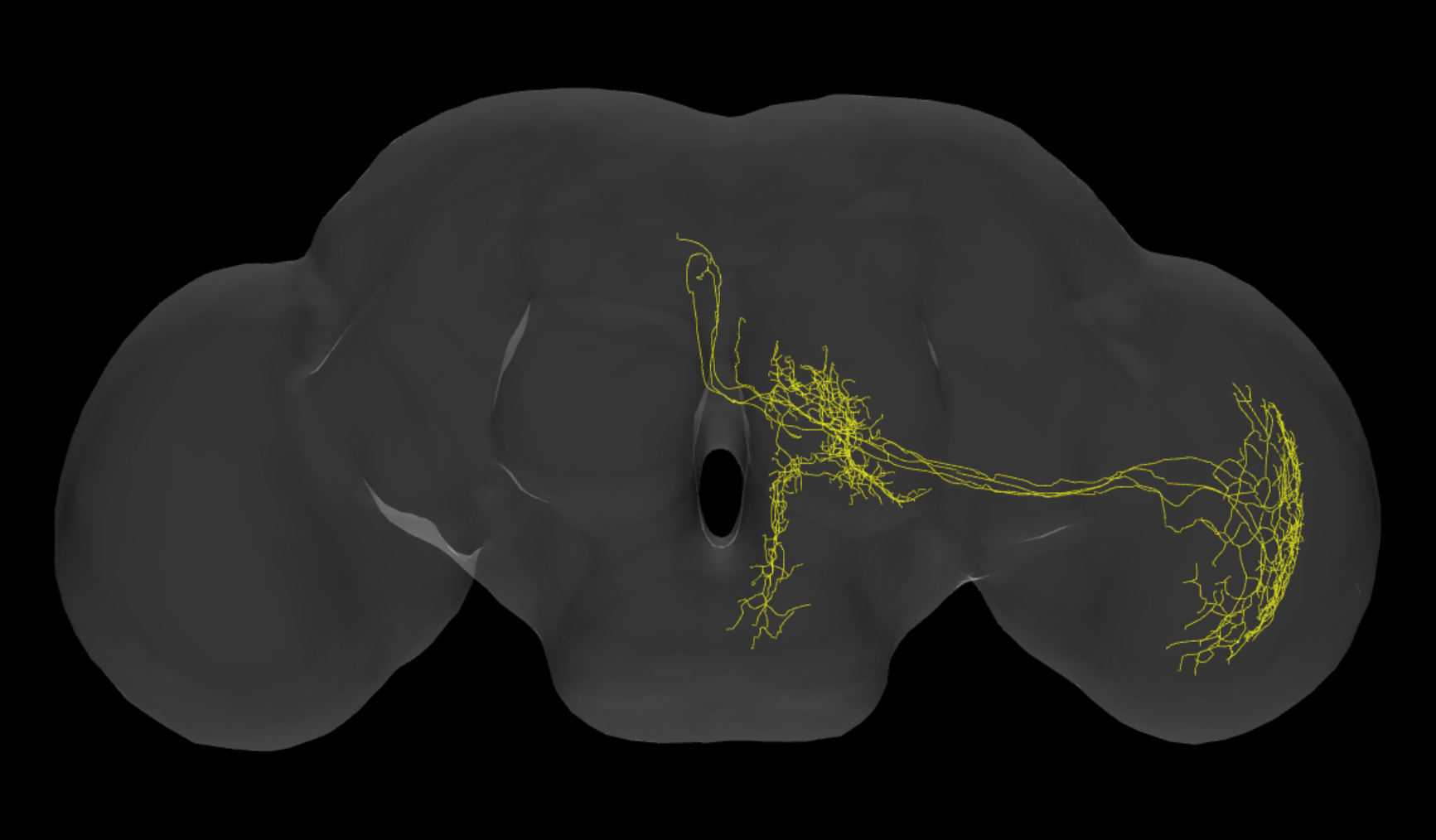} \\ \hline

        \includegraphics[width=5cm]{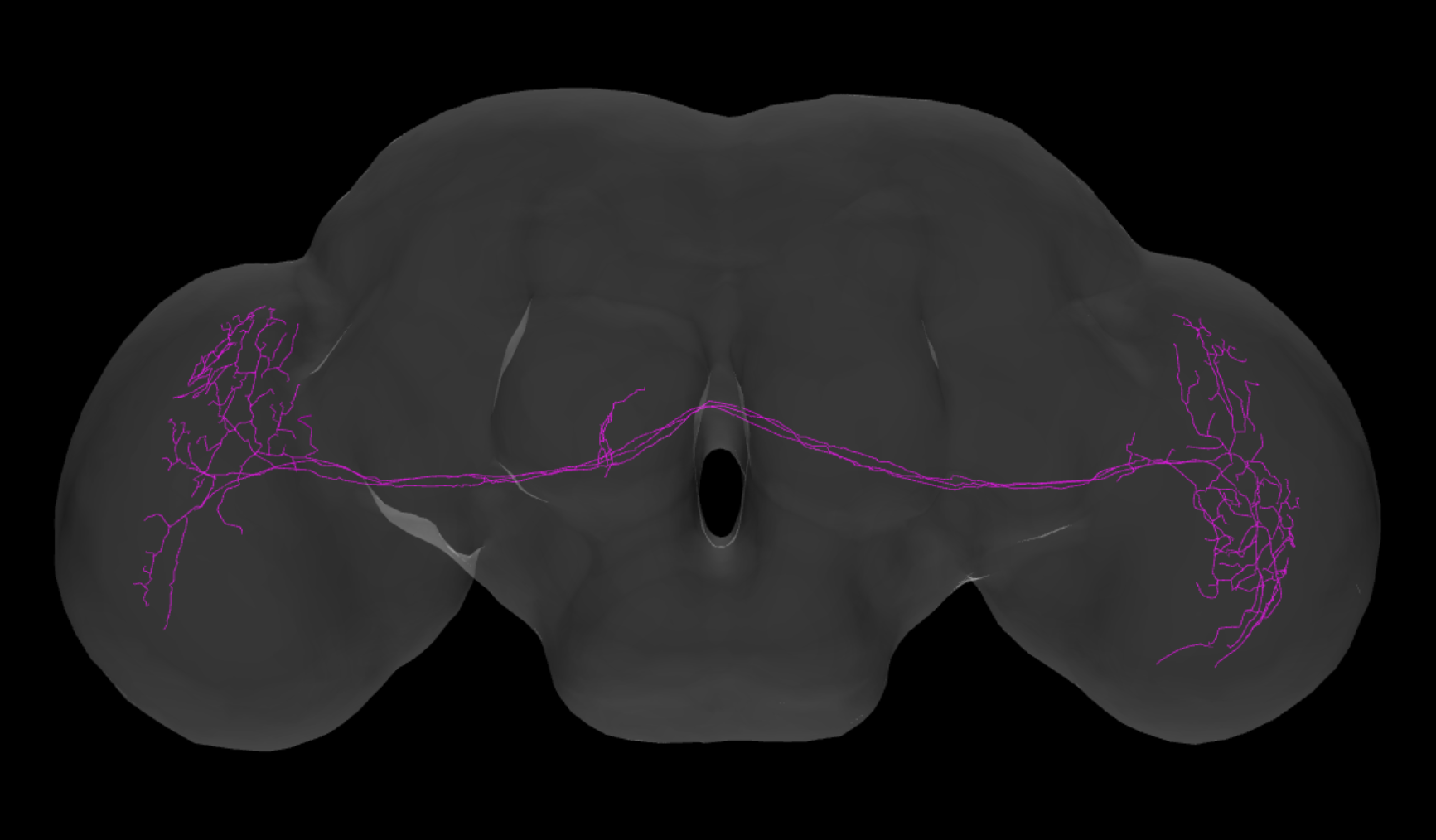} 
    &
        \includegraphics[width=5cm]{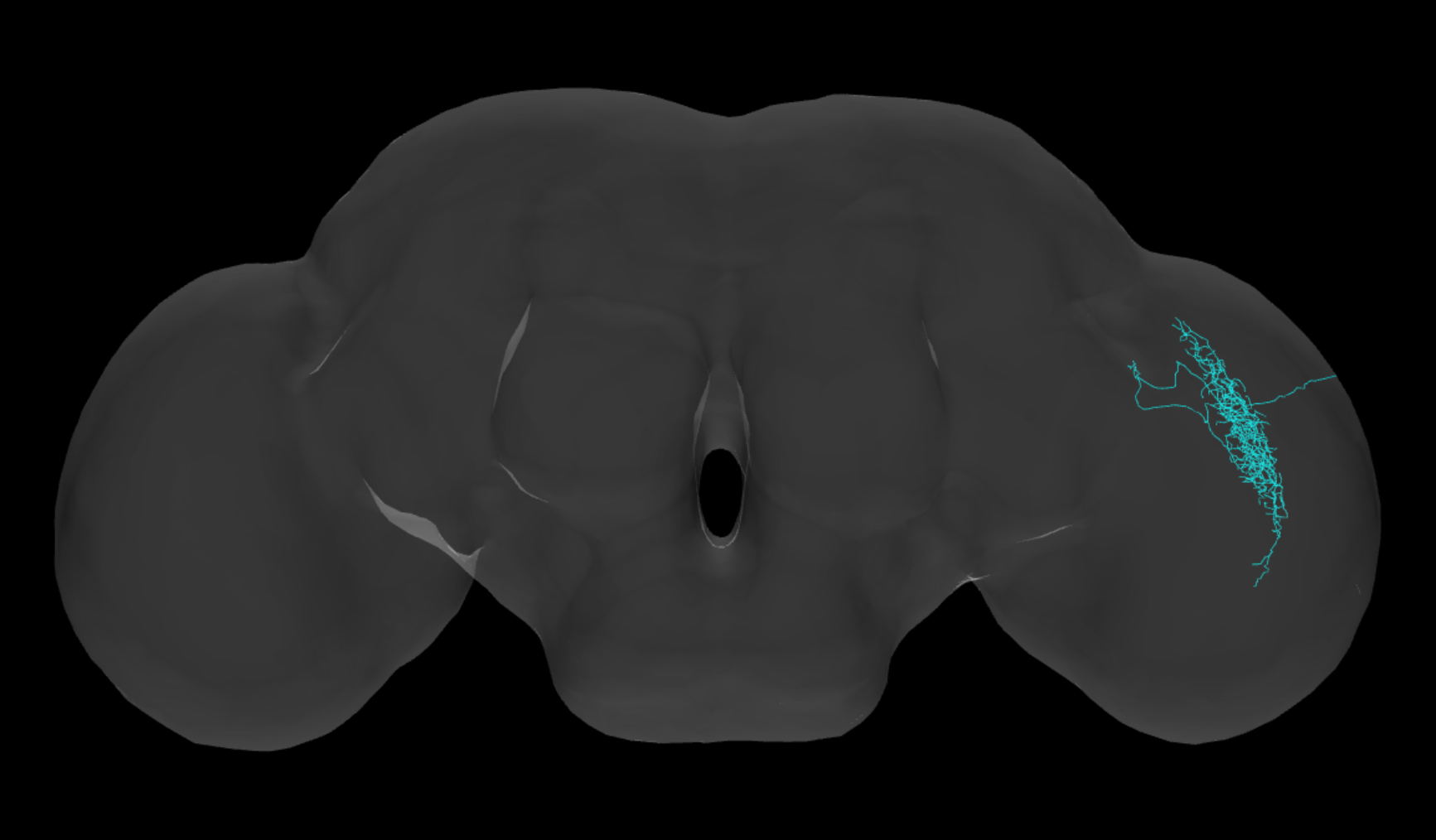} 
    & 
        \includegraphics[width=5cm]{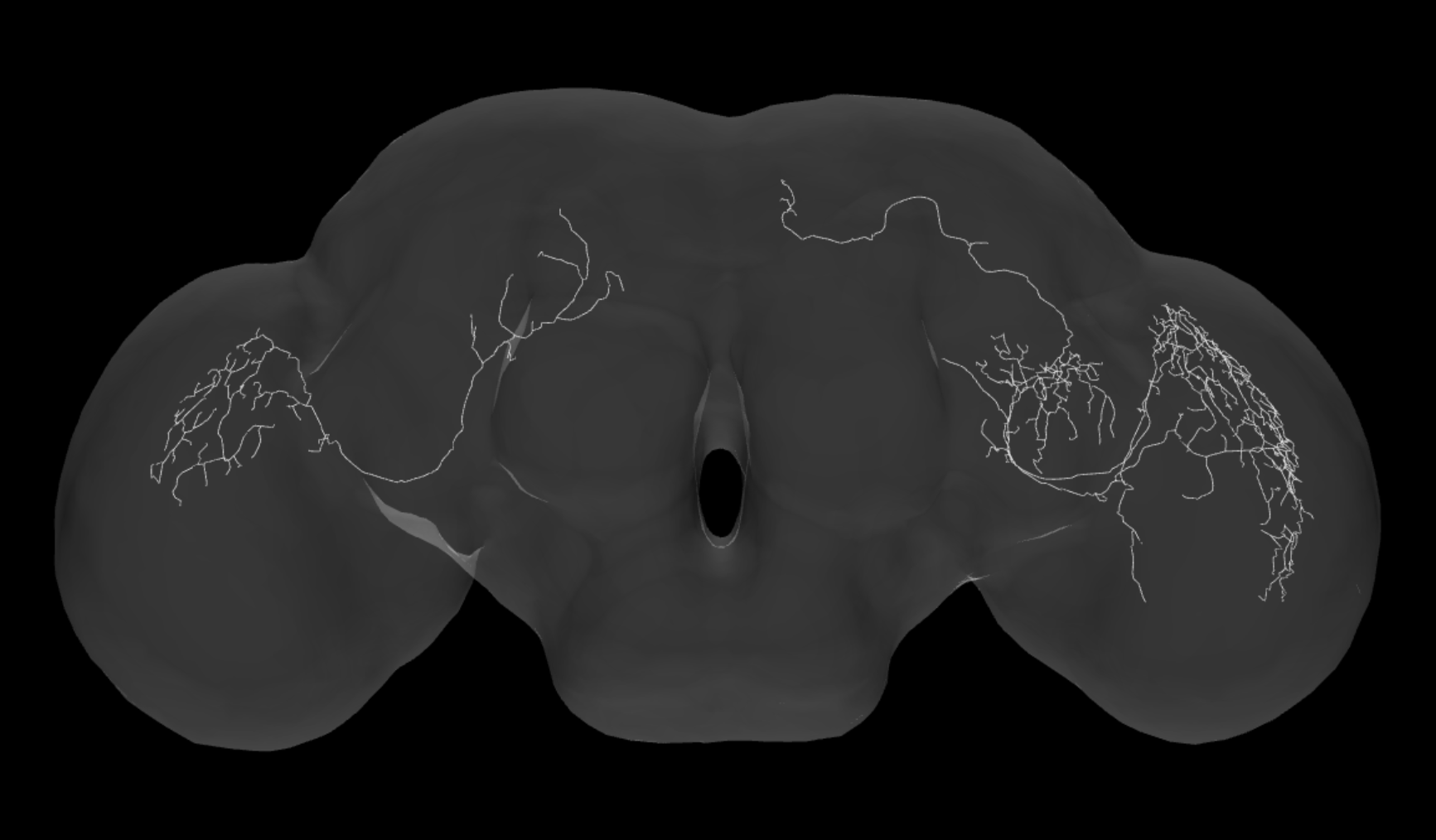} \\ \hline

        \includegraphics[width=5cm]{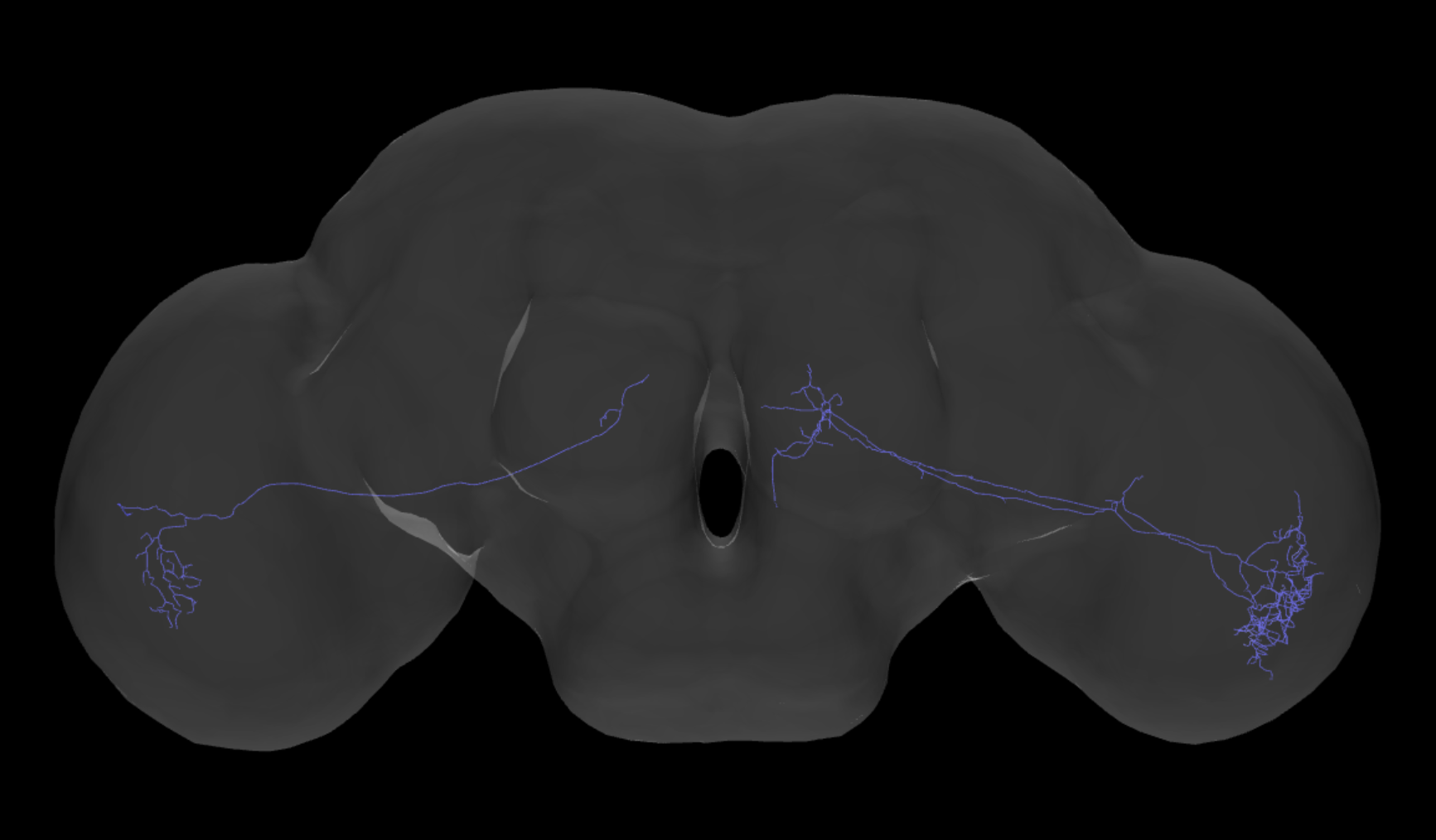} 
    &
        \includegraphics[width=5cm]{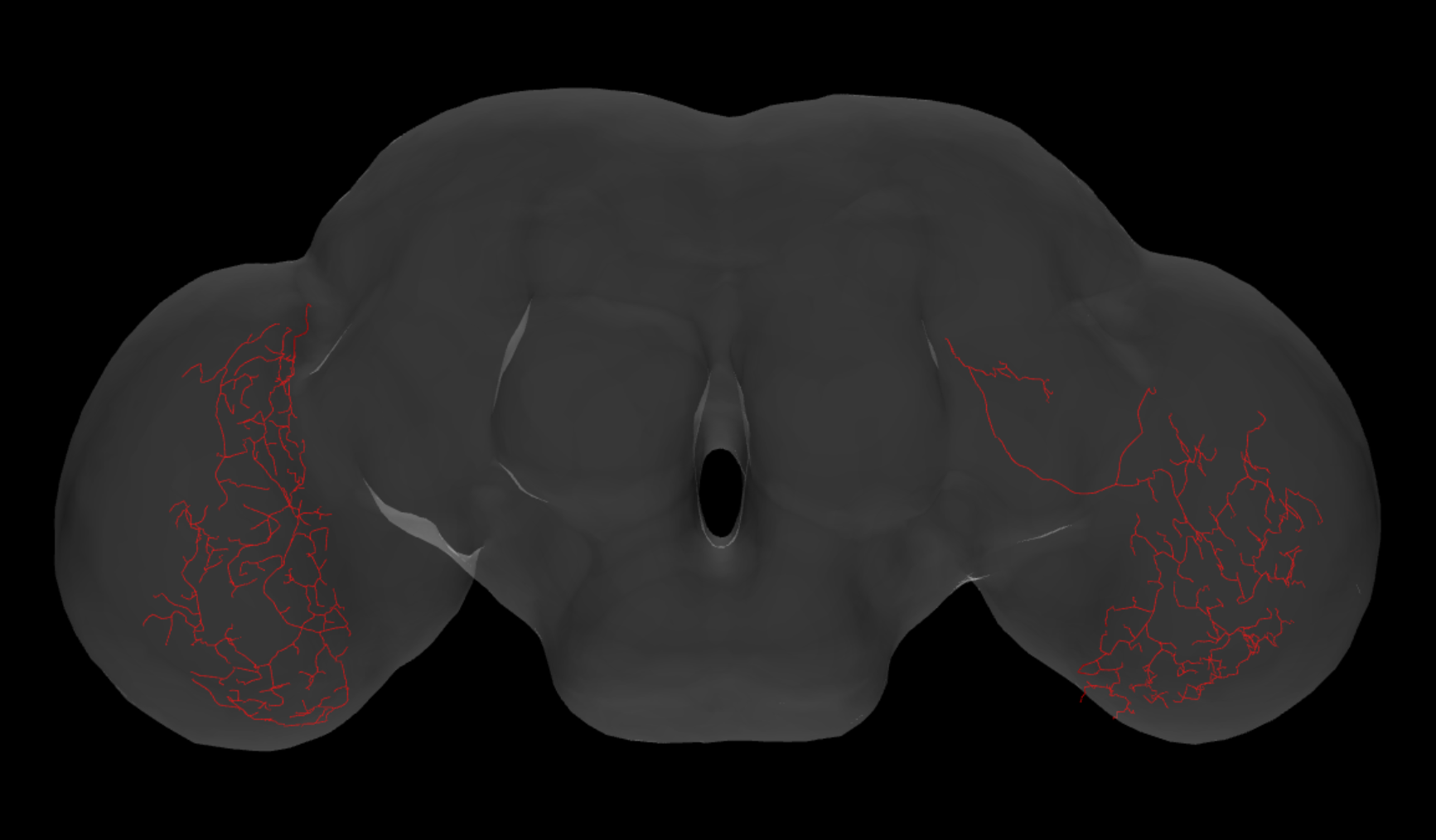} 
    & 
        \includegraphics[width=5cm]{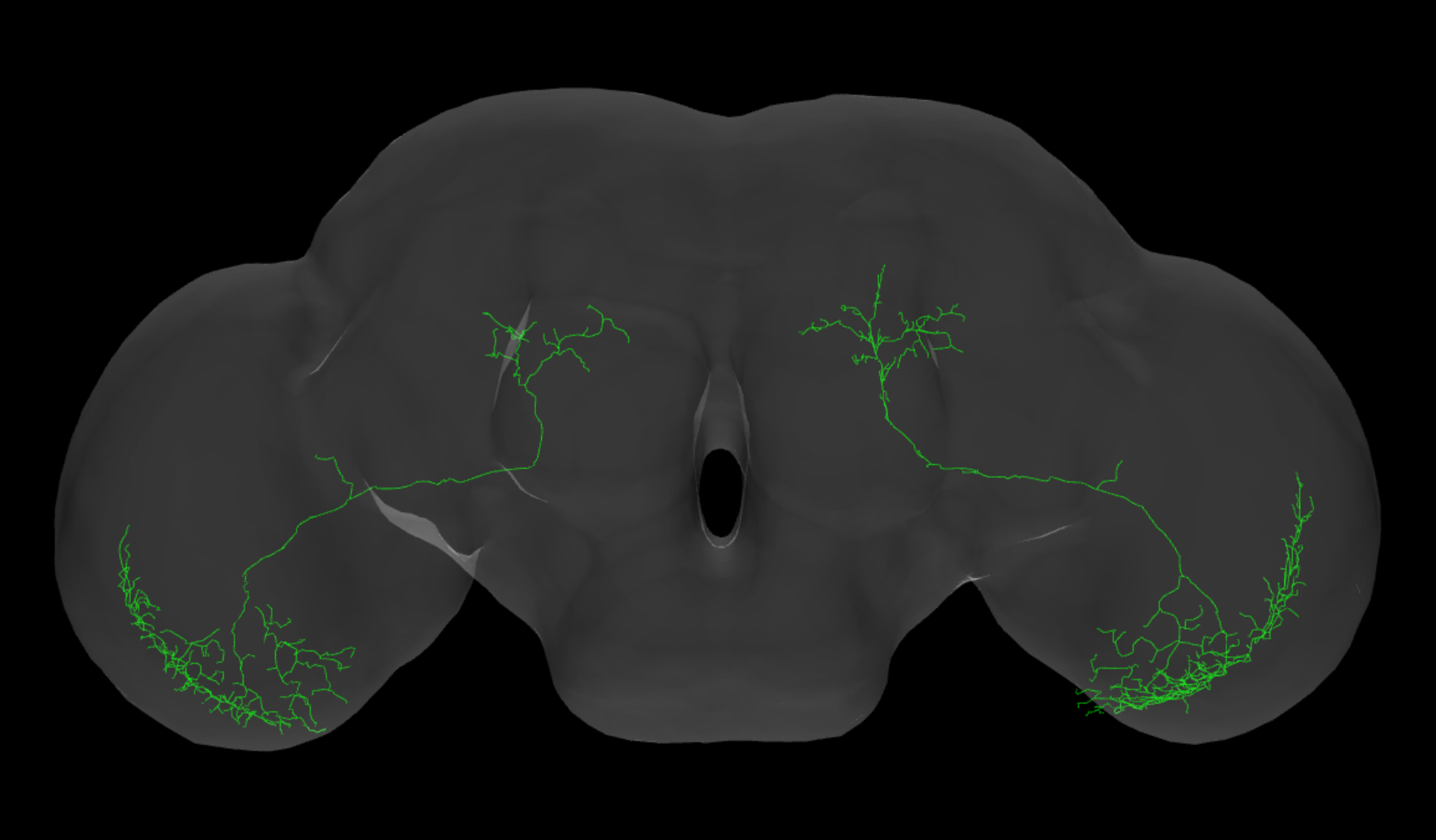} \\ \hline

     \includegraphics[width=5cm]{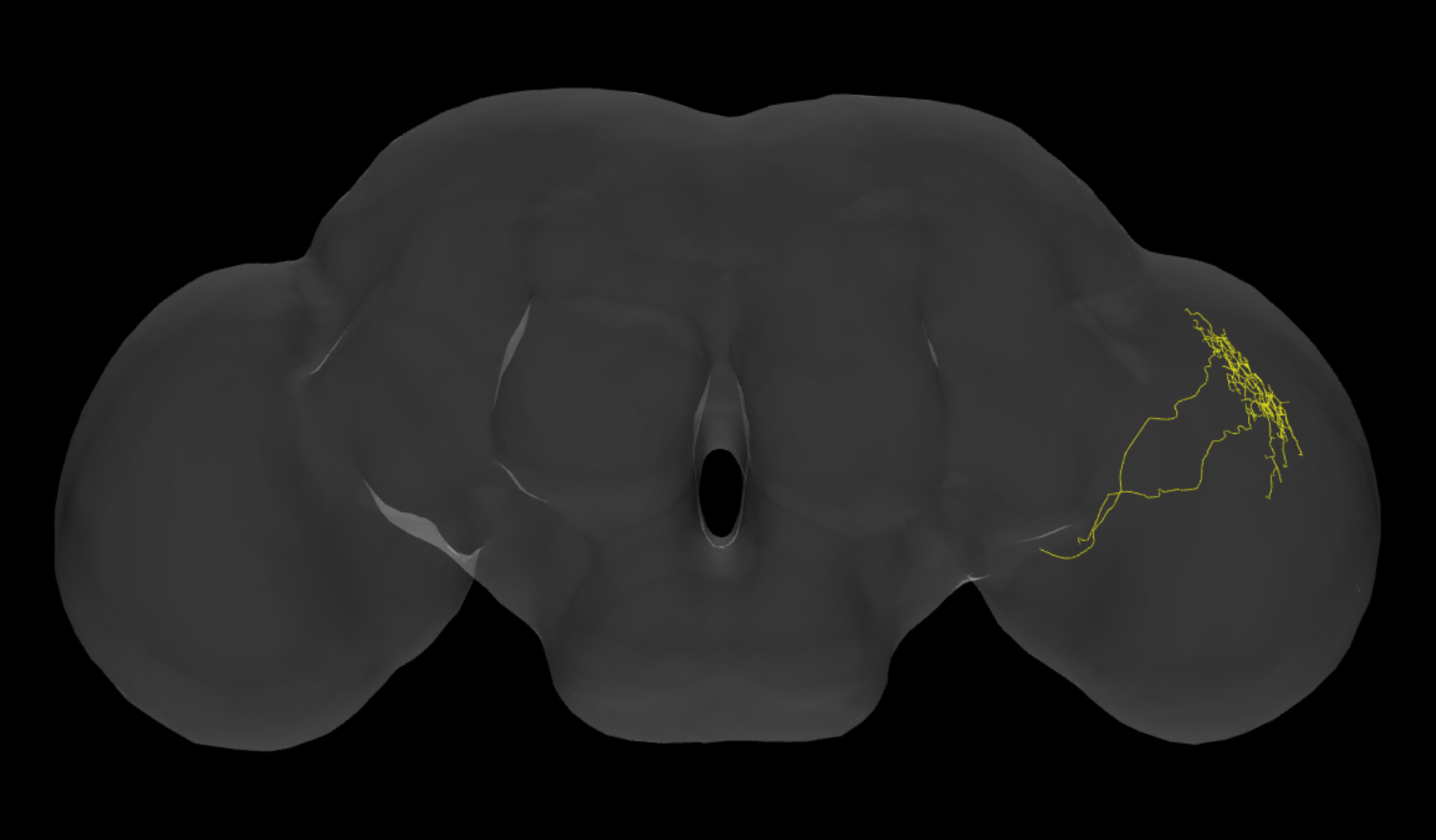} 
    &
    & 
         \\
     \hline

    \\ \hline

\end{tabular}
\caption{Many small groups of neurons in the Medulla of the female fly brain are distributed over different layers. }\label{MedFPartial}
\end{center}
\end{figure}

Finally, many small groups distribute over the layers shown in Figure~\ref{MedFPartial}.
The leftmost image on the first row depicts all the small groups in one image. 
From the distribution of the Via and the small components, it is reasonable to conclude that different areas in the Medulla are responsible for different functions. 

\section{Conclusions and Discussions\label{Conclusion}}
A neuron clustering method for reverse engineering the fly brain is presented. 
17 clusters that are potential modules were found.  
Potential usages of the clustering results are to study the structure of modules as mentioned to analyze the Medulla, and to study the neural connectivity between modules. 

\begin{figure}
    \centering
    \includegraphics[width=7cm]{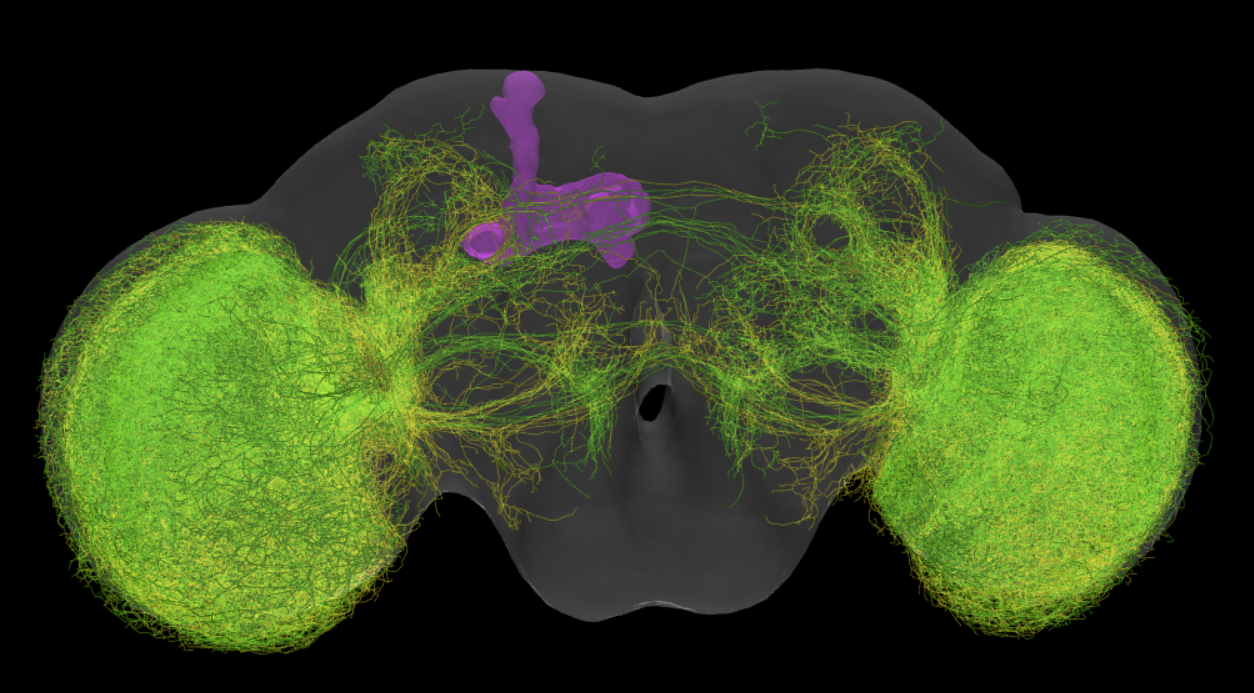}
    \caption{Groups 106 and 107 in {\tt Visual1} from tunnel-shape structure where peduncle passes through.}
    \label{MB106107}
\end{figure}

\begin{figure}
    \centering
    \includegraphics[width=7cm]{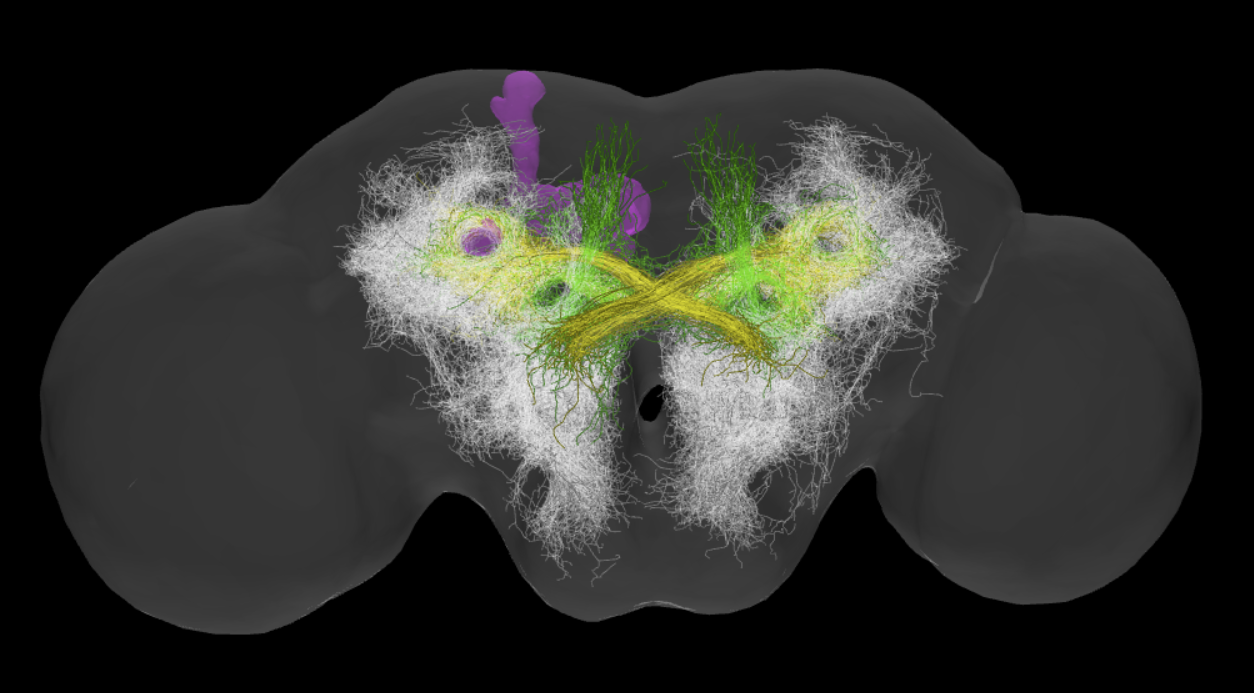}
    \caption{Groups 21, 59, 66 in {\tt Center} form tunnel-shape structure where peduncle passes through.}
    \label{MB215966}
\end{figure}

\begin{figure}
    \centering
    \includegraphics[width=7cm]{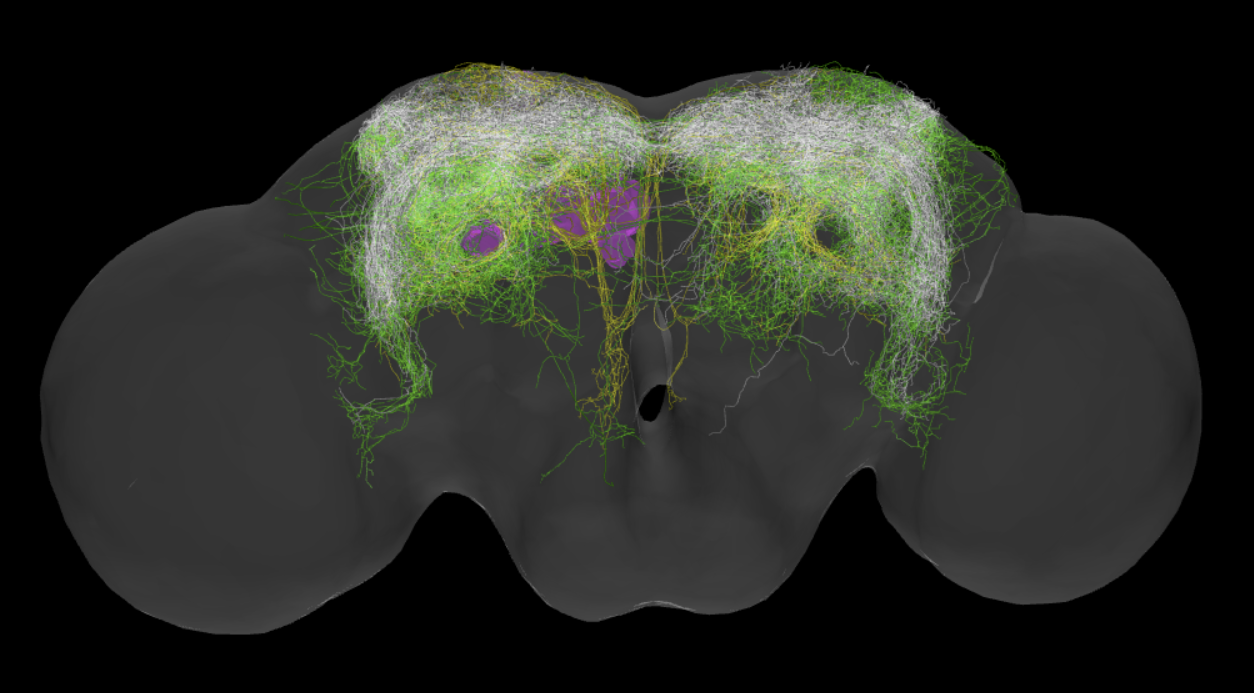}
    \caption{Groups 63, 64, 65 in {\tt Superior64} form tunnel-shape structure where peduncle passes through.}
    \label{MB636465}
\end{figure}

Intuitively, 2 groups of neurons are connected if the tips of neurons in one group are close to the tips of neurons in another group. 
Under this assumption, it is hard to explain how the peduncle of the mushroom body connects to the other groups in the fly brain because there are no observable neuron tips along the peduncle. 
However, we can use the visualization tool to identify many groups of neurons that should strongly connect to the peduncle. 
Some groups contain neurons forming a tunnel-shaped structure through which the peduncle passes. 
For examples, groups 106 and 107 in {\tt Visual1} (Figure~\ref{MB106107}), groups 21, 59, 66 in {\tt Center} (Figure~\ref{MB215966}), and groups 63, 64, 65 in {\tt Superior64} (Figure~\ref{MB636465}). 
Neurons in these groups should connect to the neurons in the mushroom body even there are no observable tips of neurons. 
The projection neurons of the antennal lobe innervate Calyx. 
Groups 106 and 107 are in the visual system. 
Neurons in group 66 innervate the antennal lobe. 
These neurons could be responsible for input information. 
If the mushroom body is considered a computer, the Calyx and the Peduncle could be the input, and the $\alpha$ and the $\beta$ arms could be the processor unit. 

Our approach lacks accurate information compared to the reversing engineering using EM data. 
On the contrary, the computation and data amount involved are manageable. 
In the fly brain connectome study, our method can be employed to create a hypothesis, and further analysis such as using the EM data then performed. 

\newpage
\printbibliography 

\newpage


\end{document}